\documentclass[aps,twocolumn,superscriptaddress,nofootinbib]{revtex4-2}
\usepackage[margin=0.7in]{geometry}
\usepackage{color}
\usepackage{babel}
\usepackage{array}
\usepackage{float}
\usepackage{multirow}
\usepackage{amsmath}
\usepackage{amsthm}
\usepackage{soul}
\usepackage{graphicx}
\usepackage[colorlinks=true,
            linkcolor=blue,
            citecolor=blue,
            urlcolor=blue]{hyperref}

\begin{document}

\title{Stochastic tensor contraction for quantum chemistry}

\author{Jiace Sun}
\affiliation{Division of Chemistry and Chemical Engineering, California Institute of Technology, Pasadena, CA 91125, USA}
\affiliation{Marcus Center for Theoretical Chemistry, California Institute of Technology, Pasadena, CA 91125, USA}

\author{Garnet Kin-Lic Chan}
\email{gkc1000@gmail.com}
\affiliation{Division of Chemistry and Chemical Engineering, California Institute of Technology, Pasadena, CA 91125, USA}
\affiliation{Marcus Center for Theoretical Chemistry, California Institute of Technology, Pasadena, CA 91125, USA}

\date{\today}

\begin{abstract}
Many computational methods in \textit{ab initio} quantum chemistry are formulated in terms of high-order tensor contractions, whose cost determines the size of system that can be studied. 
We introduce stochastic tensor contraction to perform such operations with greatly reduced cost,
and present its application to the gold-standard quantum chemistry method, coupled cluster theory with up to perturbative triples. For total energy errors more stringent than chemical accuracy, we reduce the computational scaling to that of mean-field theory, while starting to approach the mean-field absolute cost, thereby challenging the existing cost-to-accuracy landscape.
Benchmarks against state-of-the-art local correlation approximations further show that we achieve an order-of-magnitude improvement in both total computation time and error, with significantly reduced sensitivity to system dimensionality and electron delocalization. 
We conclude that stochastic tensor contraction is a powerful computational primitive to accelerate a wide range of quantum chemistry.
\end{abstract}


\maketitle

\section*{Introduction}

The objective of \emph{ab initio} quantum chemistry is to predict the behavior of molecules and materials from first-principles calculation of the electronic structure.
Commonly used systematically improvable methods compute the quantum wavefunction of the electrons by starting from a mean-field electron approximation, and then incorporating electron correlations through contributions that are theoretically justified from perturbation theory in their interactions.~\cite{MBPT_book,helgaker2013molecular,CC_review,pople1987quadratic,dreuw2005single,chen2017random,sharma2017semistochastic,shee2021regularized,dreuw2015algebraic,van2013gw} In computational form, such techniques share a mathematical structure of tensor contractions, i.e. the behavior of the electrons (the quantum amplitudes) and their interactions are represented by arrays of numbers (tensors), which are multiplied and summed over (contracted) to yield simple output observables, such as the energy and electron densities. The computational and memory cost of these operations determines the size and complexity of systems we can study today. 

A prominent example of such a method is coupled-cluster (CC) theory~\cite{MBPT_book,helgaker2013molecular,CC_review}, most commonly used in its variant known as CCSD(T)~\cite{CCSDT,pt_ref,urban1985towards} (denoting singles, doubles, and perturbative triples excitations). This is regarded as the gold standard of quantum chemistry, because it provides high accuracy in many applications~\cite{CC_review,yang2014ab,shi2025accurate}. While the mean-field starting point has an $O(N^4)$ cost, where $N$ is the number of atoms, in CCSD(T) the most complex tensor contraction has a much higher $O(N^{7})$ cost, limiting its direct application to small problems. Much effort has been directed to reducing this scaling~\cite{scuseria1999linear,schutz2001low,yang2012orbital,schutz2013orbital,neese2009efficient1,neese2009efficient2,riplinger2016sparse,liakos2019comprehensive,nagy2018optimization,nagy2019approaching,parrish2014communication,jiang2023tensor}, most notably through the use of the local approximation~\cite{saebo1993local}, which captures the observation that electron correlations can be neglected when they are widely separated. In practice, this is expressed by truncating parts of the tensors that become small when expressed in a local basis, reducing the amount of computation. 
However, due to these truncations, local correlation methods introduce additional systematic errors, implementation complexity, and a cost which still scales steeply as a function of the electron delocalization and dimensionality of the system.

Here we introduce an alternative approach to simplify the computation of many electron correlation theories through  \emph{stochastic tensor contraction} (STC). While including stochastic elements when formulating correlation theories is not new~\cite{lee2022twenty,motta2018ab,sharma2017semistochastic,austin2012quantum,hermann2020deep,PhysRevLett.99.143001,PhysRevLett.105.263004,neuhauser2013expeditious,baer2022stochastic,willow2012stochastic,damour2024stochastically}, STC introduces sampling in a different and more pervasive manner to earlier works. 
The main idea is to trade the exact evaluation of contractions for unbiased statistical estimates generated by a properly designed importance sampling of the elements of the contraction.  
We show through detailed theoretical analysis and numerical demonstration that this leads to a drastic reduction in cost, without requiring the approximation of truncating tensors in a local basis. 
For the example of CCSD(T), we describe an implementation with a one-time $O(N^{4})$ deterministic cost to set up probability tables, followed by (for a specified statistical error in the total energy) an $O(N^{2})$ stochastic cost for CCSD, and $O(N^{4})$ for (T). The scaling of this gold-standard quantum chemistry method is thus reduced to that of its mean-field starting point. 
To support these scalings, we benchmark the practical performance of STC-CCSD(T) on a wide set of molecules against a state-of-the-art local correlation implementation (using domain localized pair natural orbitals~\cite{riplinger2016sparse}), where we find an order of magnitude improvement in total computation time and total energy errors, reduced and much more consistent reaction energy errors, and a weak sensitivity to dimensionality and electron delocalization. 
Our work thus introduces stochastic tensor contraction as a simple {but powerful} computational primitive to accelerate many quantum chemistry correlation theories, and which, in practice, can already provide leading performance today.

\newpage

\section*{General theory of stochastic tensor contraction}
\begin{figure}
\centering

\includegraphics[width=\linewidth]{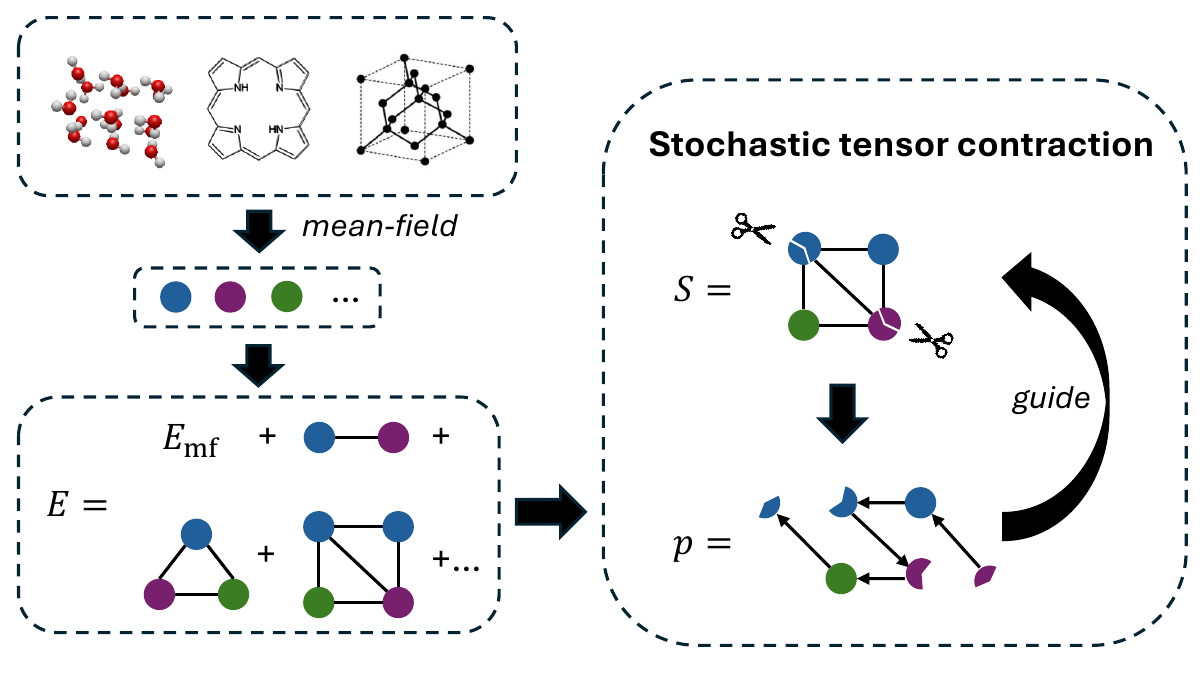}

\caption{Workflow of stochastic tensor contraction in quantum chemistry. Starting from the chemical structure and an initial mean-field calculation, input tensors are created for electronic correlation computations. The physical quantities (energy) are contractions of these tensors, typically involving loopy contractions. The loopy contractions are computed stochastically by importance sampling from a distribution constructed by breaking the loops in the contraction.}

\label{fig:workflow}
\end{figure}

\begin{figure*}[t]
\centering

\includegraphics[width=0.9\linewidth]{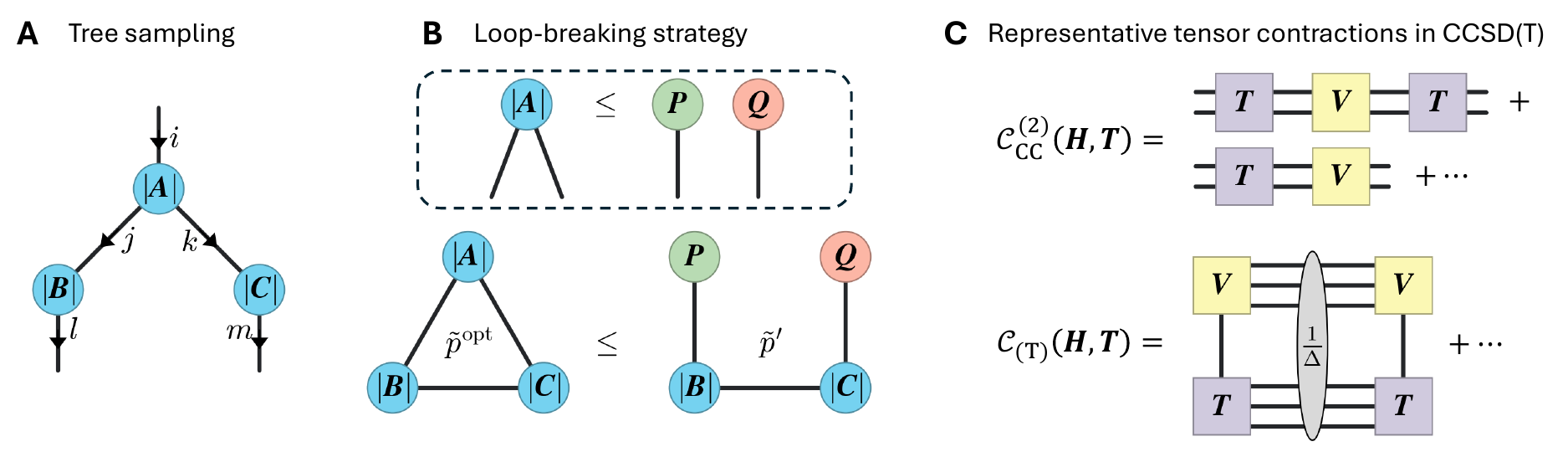}

\caption{(A) Illustration of optimal sampling for tree tensor contractions. An example $S_{ilm}=\sum_{jk}A_{ijk}B_{jl}C_{km}$ is considered. One can exactly sample the indices from the optimal probability distribution $\tilde{p}_{ijklm}^{\text{opt}}=|A_{ijk}B_{jl}C_{km}|$ by the procedure (1) sample $i$, (2) sample $j,k$ conditioned on $i$, (3) sample $l$ conditioned on $j$, (4) sample $m$ conditioned on $k$. All exact marginal and conditional probability tables can be constructed with a cost proportional to the tensor sizes. With the tables constructed, one can sample a set of indices with $O(1)$ cost. (B) Illustration of the loop breaking strategy for general loopy tensor contractions with the example $S=\sum_{ijk} A_{ik}B_{ij}C_{jk}$. We apply the approximate decomposition $|\boldsymbol{A}|\protect\leq\boldsymbol{P}\otimes\boldsymbol{Q}$ to the optimal sampling distribution $\tilde{p}_{ijk}^{\text{opt}}=|A_{ik}B_{ij}C_{jk}|$, which results in $\tilde{p}_{ijk}^{\prime}=|P_{i}B_{ij}C_{jk}Q_{k}|\protect\geq\tilde{p}_{ijk}^{\text{opt}}$. $\tilde{p}_{ijk}^{\prime}$ has a tree structure and thus can be exactly sampled by the previous strategy. (C) Two representative CCSD tensor contraction terms in $\mathcal{C}_{\mathrm{CC}}^{(2)}$ (Eq.~\ref{eq:CCSD_contraction}), and a representative (T) tensor contraction term in $\mathcal{C}_{\text{(T)}}$  (Eq.~\ref{eq:(T)_contraction}).}
\label{fig:demo}
\end{figure*}

We first introduce the general setting of tensor contractions in quantum chemistry. The overall workflow is illustrated in Fig.~\ref{fig:workflow}. The standard approach to electronic correlation starts from a mean-field Hartree--Fock (HF) wavefunction, which is the ground-state of an effective one-electron Hamiltonian, the Fock operator. The subsequent electron correlation arises from the two-electron repulsive interaction. Expressed in an orbital basis, these quantities become the Fock matrix (tensor) $F_{pq}$, and the two-electron repulsion tensor $V_{pqrs}=\langle\phi_{p}(\boldsymbol{r}_{1})\phi_{q}(\boldsymbol{r}_{2})|\frac{1}{|\boldsymbol{r}_{1}-\boldsymbol{r}_{2}|}|\phi_{r}(\boldsymbol{r}_{1})\phi_{s}(\boldsymbol{r}_{2})\rangle$, where $\phi$ is an orbital, $\boldsymbol{r}$ is the electron coordinate, and $p,q,r...$ are orbital indices, which have an $O(N)$ range. Electron correlation theories are then theories where $\boldsymbol{F}$ and $\boldsymbol{V}$ are the inputs, and the quantities of interest (such as the electronic energy) are the outputs, obtained through equations based on tensor contractions. Many theories introduce additional intermediate tensors, such as amplitude tensors $\boldsymbol{T}$ that are associated with electronic excitations, but these intermediates can be viewed as functions of \textbf{$\boldsymbol{F}$} and $\boldsymbol{V}$. 
Using bold symbols $\boldsymbol{A},\boldsymbol{B},...$ to indicate general tensors, a general tensor contraction $\boldsymbol{S}=\mathcal{C}(\boldsymbol{A},\boldsymbol{B},...)$ can be written as a partial summation over a set of common indices $I$, leaving a set of free indices $O$ in the output tensor $\boldsymbol{S}$: 
\begin{equation}
S_{O}=\mathcal{C}(\boldsymbol{A},\boldsymbol{B},...)_{O}=\sum_{I}(\boldsymbol{AB}...)_{IO}, \label{eq:contraction}
\end{equation}
{where $\boldsymbol{AB}...$ here denotes element-wise multiplication for the common indices.}
As a concrete example, we can consider the second-order Moller-Plesset (MP2) perturbation energy.~\cite{MBPT_book} In the canonical basis (in which $\boldsymbol{F}$ is diagonal), {and letting $i,j,...$ denote occupied orbitals, and $a,b,...$, virtual orbitals,} this is 
\begin{equation}
E_{\text{MP2}}=-\frac{1}{4}\sum_{ijab}\frac{\langle ij||ab\rangle^{2}}{\Delta_{ijab}^{(2)}}\label{eq:MP2_example}
\end{equation}
where $\langle ij||ab\rangle=V_{ijab}-V_{ijba}$, $\Delta_{ijab}^{(2)}=(e_{a}+e_{b}-e_{i}-e_{j})$ and $e_{i}$ and $e_{a}$ are the diagonal elements of the Fock matrix $\boldsymbol{F}$; the contraction has $O(N^{4})$ cost. In general, the complexity of the tensor contractions increases with the level of the electron correlation formulation: for example, in coupled cluster theory, considering up to $n$th order excitations in the amplitude tensor involves contractions with a cost of $O(N^{2n+2})$, while $n$th order Moller-Plesset perturbation theory has a cost $O(N^{n+3})$.~\cite{MBPT_book}

We now introduce stochastic tensor contraction as a general primitive to perform these tensor contractions. 
The principal idea is to use importance sampling to evaluate the tensor contractions. For an arbitrarily chosen probability distribution $p_{IO}$, we can express the unbiased sampling of an output tensor $\boldsymbol{S}$ (with output indices labelled by $O$) as
\begin{equation}
\boldsymbol{S}=\left\langle s_{IO}=\frac{(\boldsymbol{AB}...)_{IO}\boldsymbol{\delta}_{O}}{p_{IO}}\right\rangle _{I,O\sim p_{IO}},\label{eq:STC}
\end{equation}
where $\boldsymbol{\delta}_{O}$ is the tensor of the same dimension as $\boldsymbol{S}$, with elements equal to one at index $O$ and zero otherwise. Given $N_{\text{sample}}$ samples drawn from $p_{IO}$, the estimate is associated with a statistical error $\epsilon$; in the simple case when $\boldsymbol{S}$ is a scalar, this is $\epsilon\sim\sqrt{\mathrm{Var}(S)/N_{\text{sample}}}$, where $\mathrm{Var}$ denotes the variance, which depends on the distribution $p_{IO}$. Although this estimate is unbiased for any distribution $p_{IO}$, a careful choice is necessary to give low variance. Taking the MP2 energy as an example, we show in the SI, Sec 3.1 that uniform sampling of $i,j,a,b$ for a set of $N$ identical non-interacting molecules leads to an $O(N^{5})$ energy variance. If we desire only a relative error $\epsilon$ (i.e. $\epsilon \sim O(N)$), then we can use $N_{\text{sample}}\sim O(N^{3})$ (with $O(1)$ per-sample cost), however, if we aim for a fixed $\epsilon$ in the total energy, independent of system size, then $N_{\text{sample}} \sim O(N^{5})$, giving a cost even higher than that of the exact summation.

We can derive the optimal probability distribution $p_{IO}$ that minimizes the error of the contraction (or its $l^2$-norm, for the case of a non-scalar output), which is 
\begin{equation}
p_{IO}^{\text{opt}}=\frac{1}{Z}\tilde{p}^{\text{opt}}_{IO}=\frac{1}{Z}\left|(\boldsymbol{AB}...)_{IO}\right|,\label{eq:STC_prob}
\end{equation}
where $Z^{\text{opt}}=\sum_{IO}\tilde{p}^{\text{opt}}_{IO}$
is the ``partition function'' that normalizes the distribution, and we use the tilde quantity $\tilde{p}$ to indicate unnormalized distributions. The intuition is clear -- we wish to sample the largest elements, by magnitude, with higher probability. When the output $\boldsymbol{S}$ is a scalar, the relative variance using the optimal distribution takes a simple form (proof in SI, Sec 2.2): 
\begin{align}
\text{RelVar} & \equiv\frac{\text{Var}}{S^{2}}=\left(\frac{\mathcal{C}(|\boldsymbol{A}|,|\boldsymbol{B}|,...)}{\mathcal{C}(\boldsymbol{A},\boldsymbol{B},...)}\right)^{2}-1\label{eq:STC_var}
\end{align}
where $|\cdot|$ is the element-wise absolute value. Clearly, $\frac{\mathcal{C}(|\boldsymbol{A}|,|\boldsymbol{B}|,...)}{\mathcal{C}(\boldsymbol{A},\boldsymbol{B},...)}$ measures to what extent the tensor values have opposite signs. If all elements have the same sign, the ratio becomes unity, and the variance becomes zero. This is the case in the MP2 example, as the numerator $\langle ij||ab\rangle^{2}$ and denominator $\Delta^{(2)}_{ijab}$ are both positive, thus if one can perform importance sampling with $p_{ijab}^{\text{opt}}\propto\tilde{p}_{ijab}^{\text{opt}}=\frac{\langle ij||ab\rangle^{2}}{\Delta_{ijab}^{(2)}}$, the variance is strictly zero, and only one sample is needed to obtain the exact result. However, this still requires one to (1) sample from the given distribution $p^{\text{opt}}$, and (2) compute the partition function $Z^{\text{opt}}$, both of which can be as difficult as the original tensor contraction problem in many cases.

Nevertheless, for all tensor contractions with a tree structure (which means the graphical depiction of the contraction as in Fig.~\ref{fig:demo} has no loops), one can achieve efficient sampling by recursively sampling conditional probabilities for each tensor while traversing the tree. 
For all tensors, the probability of the children indices conditioned on the parent indices, and the partition function, can be computed with a cost proportional to the tensor sizes, and one can then perform sampling with $O(1)$ per-sample cost by the alias method.~\cite{alias} An example of sampling from $\tilde{p}_{ijklm}^{\text{opt}}=|A_{ijk}B_{jl}C_{km}|$ is shown in Fig. \ref{fig:demo}(A). The detailed step-by-step algorithm to construct the probability tables is given in the SI, Sec 2.1.

For general tensor contractions with loops, we can still construct an approximate $\tilde{p}^{\prime}\approx\tilde{p}^{\text{opt}}$ that supports efficient sampling. One well-known way to do this is through loopy belief propagation.~\cite{yedidia2003understanding,murphy2013loopy} Here, we provide a simpler ``loop-breaking'' strategy to construct $\tilde{p}^{\prime}$, which has the same one-time and per-sample cost as for a spanning tree of the contraction. We sequentially choose tensors $\boldsymbol{A}$ and replace them by the outer product $\boldsymbol{P}\otimes\boldsymbol{Q}$ whose elements satisfy
$|\boldsymbol{A}|\leq\boldsymbol{P}\otimes\boldsymbol{Q}$ (this condition is necessary for the bound below), until all loops are broken, and construct $\tilde{p}^{\prime}$ using the decomposed tensors (see Fig. \ref{fig:demo}(B)). 
Intuitively, the variance will not grow significantly if $\tilde{p}^{\prime}$ and $\tilde{p}^{\text{opt}}$ are close. Let $\text{RelVar}(\tilde{p})$ be the relative variance of the contraction using $\tilde{p}$. Then we can bound the growth of the relative variance by (see SI, Sec 2.3 for proof)
\begin{align} \label{eq:approximate_variance_bound}
\frac{\text{RelVar}(\tilde{p}^{\prime})+1}{\text{RelVar}(\tilde{p}^{\text{opt}})+1}\leq\exp(\Delta F)\equiv \frac{Z^{\prime}}{Z^{\text{opt}}}
\end{align}
where we have introduced the STC ``free energy difference'' $\Delta F$ of the sampling distributions.
We emphasize that only the importance sampling distribution is modified, the result remains unbiased, and so far everything applies to general tensor contractions.

Now we apply the above STC algorithms for tensor contractions relevant to quantum chemistry, where the tensors have additional properties that can be utilized. 
Here, the involved physical tensors are representations of operators that have some locality in real space. We assume a gapped system and an exponentially localized orthonormal basis (e.g. the localized molecular orbital basis, although the asymptotic forms do not require this basis). The fundamental tensors then exhibit the asymptotic behavior:
\begin{equation}
\begin{aligned}F_{pq} & \sim\exp(-R_{pq}),\\
V_{pqrs} & \sim\exp(-R_{pr})\exp(-R_{qs})R_{pq}^{-\alpha},
\end{aligned}
\label{eq:asymptotic}
\end{equation}
where the algebraic exponent is $\alpha=3-\delta_{pr}-\delta_{qs}$. Importantly, although Eq.~\ref{eq:asymptotic} is written in a local basis, 
even in a non-local basis, the tensors must be related to the above forms by a unitary transformation, which is a strong constraint. This reflects itself in improved variance properties of the tensor contractions regardless of the basis in which they are expressed. 

In the following, we give a variance analysis of STC for a scalar output (such as the energy) in several setups where the fundamental tensors are expressed in different bases, such as a general arbitrary basis, and special bases, including a local basis, Haar-random basis, and the canonical molecular orbital basis. For tree structure tensor contractions, the variance is completely determined by Eq.~\ref{eq:STC_var}. For loop tensor contractions, it is additionally affected by $\exp(\Delta F)$ in Eq.~\ref{eq:approximate_variance_bound}. The results for the two practically most interesting cases, the local basis and the canonical basis, are shown in Table \ref{table:scaling}, with the restriction that contractions involve only the off-diagonal part of $\boldsymbol{V}$ ($p\neq r,q\neq s$) (and other similar tensors generated from $\boldsymbol{V}$) as this is the simplest to summarize.
However, even when diagonal elements appear in the contraction, they only slightly affect the theoretical asymptotic scaling in high-dimensional systems, and in practice, this effect remains sub-dominant, as the off-diagonal elements are far more numerous than the diagonal ones.
Such effects are comprehensively discussed in the SI, Sec 3.7-3.10.

\noindent\emph{Basis independent STC variance {upper bound}}. For STC of a contraction $\mathcal{C}(\boldsymbol{A}_{1},\boldsymbol{A}_{2},...,\boldsymbol{A}_{m})$ with the optimal sampling probability, there exists a universal and basis-independent variance upper bound (proof in SI, Sec 2.4):
\begin{equation}
\text{Var}\leq\left(\prod_{i=1}^{m}\Vert\boldsymbol{A}_{i}\Vert_{2}\right)^{2},\label{eq:general_STC_var}
\end{equation}
where $\Vert\cdot\Vert_{p}$ is the element-wise tensor $l^p$-norm, i.e. the $l^p$-norm of the flattened vector. We show in the SI, Sec 3.3 that, with the diagonal elements neglected, the asymptotic $l^2$-norm is always $O(\sqrt{N})$ for all (fundamental or intermediate) tensors. Using this, the variance has an $O(N^{m})$ upper bound, thus the cost to sample to fixed error $\epsilon$ is bounded by $O(N^m/\epsilon^2)$. The cost to compute the exact contraction of $m$, $k$-index, tensors without any structure is $O(N^{O(mk)})$. Thus we obtain a large improvement when the tensors involved have many indices. 

\noindent\emph{Variance of STC in a local basis}: For STC of tensors expressed in an exponentially localized basis, such that the tensor elements follow the asymptotic form in Eq.~\ref{eq:asymptotic}, even better performance can be obtained. Consider the optimal relative variance, which is fully determined by the sign factor $\frac{\mathcal{C}(|\boldsymbol{A}|,|\boldsymbol{B}|,...)}{\mathcal{C}(\boldsymbol{A},\boldsymbol{B},...)}$ in Eq.~\ref{eq:STC_var}. The 
asymptotic form of $\boldsymbol{V}$ is the product of an exponentially decaying piece (that can be of either sign) and a positive algebraically decaying piece.
Assuming the exactness of the asymptotic forms, we show in the SI, Sec 3.7 that because of the `short-range' nature of the part of the tensor with signs, the relative variance is $O(1)$ with optimal sampling. 
We also prove in the SI, Sec 3.8 that (neglecting the diagonal contributions of $\boldsymbol{V}$) the loop-breaking strategy with a simple choice of outer product tensors results in $\exp(\Delta F)\leq O(\text{polylog}(N))$ for all loopy tensor contractions.

\noindent \emph{Variance of STC in a Haar-random basis and the canonical basis}: We can also consider STC in a real Haar-random (randomly rotated) basis and the canonical basis (i.e. $\boldsymbol{F}$ is diagonalized). Using the optimal probability distribution, {variances in both bases satisfy} the universal upper bound in Eq.~\ref{eq:general_STC_var}. However, because all tensor elements have similar magnitudes, we can obtain a more explicit estimate of $\exp(\Delta F)$ for loopy tensor contractions, and we show in the SI, Sec 3.9 that  $\exp(\Delta F)\sim O(1)$. While the canonical basis lacks easily characterized properties, we conjecture (and give an argument) that it lies between the case of the local and Haar-random bases, thus we estimate $\exp(\Delta F)$ to also be in between the corresponding estimates.

\noindent \emph{Comparison with local correlation}. Finally, we briefly compare STC with local correlation methods. Local correlation approaches achieve efficiency primarily through explicit local truncations, in which a large number of small contributions are neglected. In contrast, STC exploits locality via the implicitly favorable variance scaling, which does not introduce truncations or bias. In local correlation methods, the dominant computational cost comes from the sheer number of small but non-negligible contributions, which becomes particularly severe in systems with a high electron density and delocalized orbitals. In STC, however, such contributions are naturally suppressed in the variance due to their small amplitudes together with short-range origin of the sign factors.
As a result, we expect the performance of STC to be significantly less sensitive to the details of the system.

\begin{table}[t!]

\caption{Relative variance scaling of scalar properties in STC in a local basis and in a canonical basis for quantum chemistry tensor contractions. Two topologies are considered: tree tensor contractions and loopy tensor contractions. $\boldsymbol{m}$ is the number of tensors in the contraction. {For simplicity, these results do not include the contribution from diagonal elements of $\boldsymbol{V}$, but the scalings are affected only for high-dimensional systems, and only in a minor fashion even then. A comprehensive summary can be found in Table 1 of the SI, Sec 3.10.}}

\begin{tabular}{|c|c|c|}
\hline 
Relative variance scaling  & STC (local basis) & STC (canonical basis)\tabularnewline
\hline 
\hline 
Tree tensor contractions  & $O(1)$  & $O(N^{m-2})$\tabularnewline
\hline 
Loopy tensor contractions  & $\leq O(\text{polylog}N)$ & $^{*} \leq O(N^{m-2}\text{polylog}N)$\tabularnewline
\hline 
\multicolumn{3}{l}{$*$: Obtained as a non-rigorous estimate.}
\end{tabular}

\label{table:scaling}
\end{table}

\section*{Stochastic tensor contraction applied to coupled cluster theory}

\begin{table*}
\centering

\caption{Computational scaling of CCSD(T) by STC and exact computation. The STC cost includes a deterministic part and a sampling part. The scaling of the sampling cost is shown for two target error settings: a fixed target total energy error and a fixed target relative energy error. The shown STC-CCSD cost is for a single iteration. Note that to converge the CCSD equations requires $\boldsymbol{N_{\text{sample}}}\protect\geq\boldsymbol{N_{\text{critical}}}$. The scaling of $\boldsymbol{N_{\text{critical}}}$ is not included in the cost. For simplicity, we only show the polynomial part of the scaling, and neglect $\boldsymbol{O(\log N)}$ factors. The STC sampling cost also accounts for tensor contractions where diagonal elements of $\boldsymbol{V}$ contribute (see Eq.~\ref{eq:asymptotic} and subsequent discussion).}

\begin{tabular}{|c|c|c|c|c|}
\hline 
\multirow{3}{*}{Problems} & \multicolumn{3}{c|}{STC} & \multirow{3}{*}{Exact computation}\tabularnewline
\cline{2-4}
 & \multirow{2}{*}{Deterministic cost } & \multicolumn{2}{c|}{Sampling cost} & \tabularnewline
\cline{3-4}
 &  & fixed total error  & fixed relative error  & \tabularnewline
\hline 
CCSD, local basis  & $O(N^{4})$  & $O(N^{2})$  & $O(1)$  & $O(N^{6})$\tabularnewline
\hline 
CCSD, canonical basis  & $O(N^{4})$  & $O(N^{4})$  & $O(N^{2})$  & $O(N^{6})$\tabularnewline
\hline 
(T), canonical basis  & $O(N^{4})$  & $\leq O(N^{4})$  & $\leq O(N^{2})$  & $O(N^{7})$\tabularnewline
\hline 
\end{tabular}

\label{table:CC_scaling} 
\end{table*}

We now describe the application of stochastic tensor contraction to the gold-standard quantum chemistry method, CCSD(T). For this, we first give a brief overview of CCSD(T) and focus on the required tensor contractions. We present expressions in the general spin orbital basis for simplicity. In CC theory, the wavefunction is parameterized by an excitation operator $\hat{T}=\sum_{l=1}\hat{T}_{l}$ as $|\psi_{\text{CC}}\rangle=e^{\hat{T}}|\psi_{\text{HF}}\rangle$. The operators $\hat{T}_{1},\hat{T}_{2},...$ are associated with coefficient (amplitude) tensors $T_{ia},T_{ijab},...$, and these amplitudes are obtained through the iterative solution of the CC equations 
\begin{equation}
\begin{aligned}\boldsymbol{\Delta}^{(l)}\boldsymbol{T}^{(l)} & =-\mathcal{C}_{\mathrm{CC}}^{(l)}(\boldsymbol{H},\boldsymbol{T}),\end{aligned}
l=1,2,...\label{eq:CC_eq}
\end{equation}
where $\boldsymbol{H}=\{\boldsymbol{F},\boldsymbol{V}\}$, $\boldsymbol{T}=\{\boldsymbol{T}^{(l)}\}$, $\mathcal{C}_{\mathrm{CC}}^{(l)}(\boldsymbol{H},\boldsymbol{T})$ is shorthand for the sum of a number of tensor contractions between these tensors, and $\boldsymbol{\Delta}^{(l)}$ is a diagonal tensor with elements $\Delta_{ia}^{(1)}=(F_{aa}-F_{ii})$, $\Delta_{ijab}^{(2)}=(F_{aa}+F_{bb}-F_{ii}-F_{jj})$, etc. The correlation energy is then obtained as

\begin{equation}
E_{\mathrm{CC}}(\boldsymbol{T})=\frac{1}{4}\sum_{ijab}\langle ij||ab\rangle\left(T_{ijab}+T_{ia}T_{jb}-T_{ib}T_{ja}\right).\label{eq:cc_energy}
\end{equation}

In CCSD, both the excitation operators $\hat{T}$ and the CC equations in Eq.~\ref{eq:CC_eq} are truncated to $l=1,2$. One can further correct the CCSD energy by perturbatively including the contribution from the triple excitations $\hat{T}_{3}$, which is denoted ``(T)''. The (T) energy correction is also the sum of a number of tensor contractions, and we write it as 
\begin{equation}
E_{\text{(T)}}=\mathcal{C}_{\text{(T)}}(\boldsymbol{H},\boldsymbol{T}).\label{eq:(T)_eq}
\end{equation}

For illustration, we show a few representative terms of the tensor contractions in $\mathcal{C}_{\mathrm{CC}}^{(l=1,2)}$ and $\mathcal{C}_{(T)}$:
\begin{equation}
\mathcal{C}_{\text{CC}}^{(2)}(\boldsymbol{H},\boldsymbol{T})_{ijab}=\frac{1}{2}\sum_{cd}V_{abcd}T_{ijcd}+\frac{1}{4}\sum_{cd}V_{klcd}T_{ijcd}T_{klab}+...\label{eq:CCSD_contraction}
\end{equation}
\begin{equation}
\mathcal{C}_{\text{(T)}}(\boldsymbol{H},\boldsymbol{T})=-\frac{1}{4}\sum_{ijkabc}\frac{1}{\Delta_{ijkabc}^{(3)}}\left[\left(\sum_{f}T_{ijaf}V_{fkbc}\right)^{2}+...\right]\label{eq:(T)_contraction}
\end{equation}
where the given $\mathcal{C}_{(T)}$ expression is only valid in the canonical basis. Interested readers can refer to Ref.~\cite{MBPT_book,CCSDT} for the full expressions. Exact evaluation of the terms in $\mathcal{C}_{\text{CC}}^{(2)}(\boldsymbol{H},\boldsymbol{T})$ scales as $O(N^{6})$, and for the terms in (T), like $O(N^{7})$.

\noindent\emph{STC-CCSD}: We follow a ``greedy'' strategy to design an STC sampling that minimizes the energy variance in the next CCSD iteration, given the $\boldsymbol{T}^{(l)}$ tensors in the current iteration. Thus we expand the next-iteration energy $E_{\mathrm{CC}}(\{\mathcal{C}_{\mathrm{CC}}^{(l)}(\boldsymbol{H},\boldsymbol{T})/\boldsymbol{\Delta}^{(l)}\})$ into a sum of scalar-output tensor contractions, and apply the loop-breaking strategy to construct the sampling probabilities. Note that we can use the same set of samples to construct the unbiased updated $\boldsymbol{T}^{(l)}$ tensors. In practice, we also use a density-fitting decomposition of $\boldsymbol{V}$ to reduce memory requirements. The implementation details are discussed in the SI, Sec 4.1.

Working in the canonical basis, the updated energy expression involves contraction of at most four tensors (with some simple preprocessing, see SI, Sec 4.1), which results in a $O(N^{4})$ variance scaling for STC. {In a local basis (see SI, Sec 4.2 for details)  the energy variance is rigorously $O(N^{2})$ (up to possible $\mathrm{polylog}(N)$ factors, as described previously).

Since the CCSD equations are solved iteratively, it is important to understand their convergence behavior in the STC setting. In the SI, Sec 4.3, we show in a simple single amplitude model that the variance converges as long as the number of samples $N_{\text{sample}}$ exceeds a critical number $N_{\text{critical}}$. The amplitude estimate is formally biased if the equations involve a non-linear tensor contraction (i.e. involving the same tensor more than once), but the magnitude of the bias is $\text{bias}\sim\left(\text{statistical error}\right)^{2}$. Since we will always target the high-accuracy regime, we expect this bias to be much smaller than the statistical error in any practical application. Currently, we lack a theoretical understanding of $N_{\text{critical}}$, so we leave further discussion of this quantity to the numerical results.

\begin{figure*}[!t]
\centering

\includegraphics[scale=0.4]{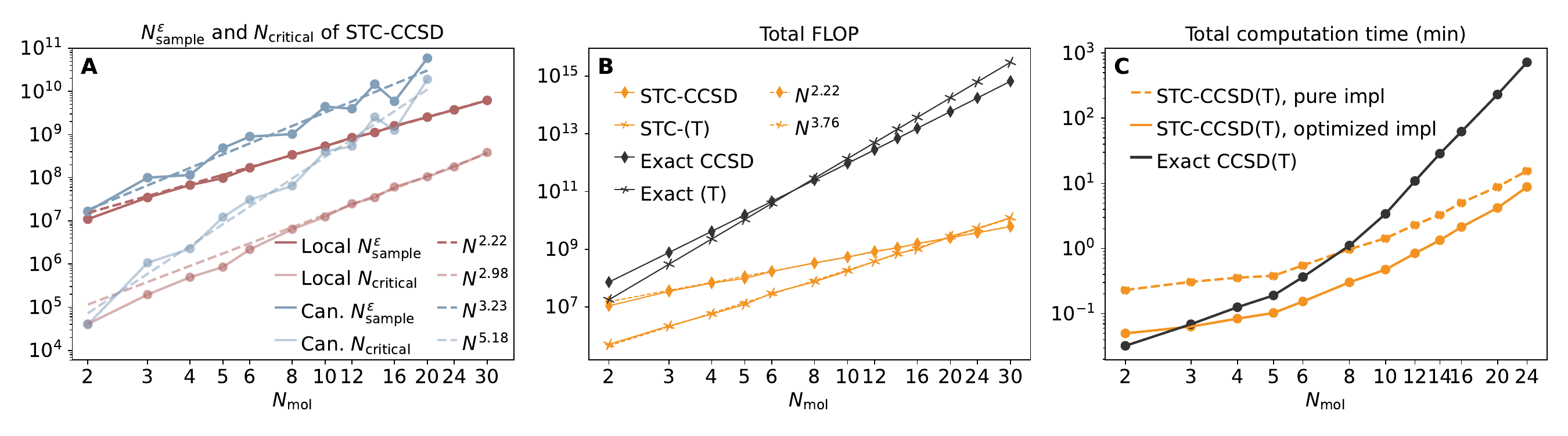}

\caption{Numerical scalings and computation time of STC-CCSD(T) and comparison with exact CCSD(T) demonstrated on water clusters with $2\sim30$ water molecules. (A) Scaling of $N_{\text{sample}}^{\epsilon}$ and $N_{\text{critical}}$ for STC-CCSD with local and canonical (labeled as "Can.") bases. (B) Scaling of the number of floating-point operations (FLOP) for STC and exact CCSD and (T). (C) Total computation time of the complete STC and exact CCSD(T) calculation. The reported number of samples or floating point operations of CCSD is for each iteration. In STC calculations, the target total error is fixed to be 0.2m$E_{h}$ for both CCSD and (T). All structures are lowest-energy structures, and all calculations use the 6-31G basis and are performed on the same CPU with 8 cores.}

\label{fig:scaling}
\end{figure*}

\noindent\emph{STC-(T)}: For simplicity, we discuss the treatment of only the first term in Eq.~\ref{eq:(T)_contraction}, and the generalization to the full expressions can be found in the SI, Sec 4.5. Following the STC procedure, one can directly write the optimal sampling probability $p^{\text{opt}}$ for the 8 indices appearing in Eq.~\ref{eq:(T)_contraction}. We can apply the loop-breaking strategy (with some simple preprocessing, see SI, Sec 4.4), which leads to an $O(N^{4})$ energy variance (using the conjectured behavior of the $\exp(\Delta F)$ factor in the canonical basis discussed above). However, utilizing more of the structure of the (T) expression, we can design an improved strategy that we use to rigorously prove  $O(N^{3})$ energy variance of the main expressions with a $O(N)$ per-sample cost. 

Instead of viewing $\boldsymbol{T}$ and $\boldsymbol{V}$ as separate tensors, we implicitly view the contracted $\boldsymbol{T}\cdot\boldsymbol{V}$ as a single 6-index tensor, with elements $(\boldsymbol{T}\cdot\boldsymbol{V})_{ijkabc}=\sum_{f}T_{ijaf}V_{fkbc}$, and the energy is now a contraction involving only 6 indices. We construct the approximate sampling probability 
\begin{equation}
\tilde{p}_{ijkabc}^{\prime}=N(\boldsymbol{T}^{2}\cdot\boldsymbol{V}^{2})_{ijkabc}=N\sum_{f}T_{ijaf}^{2}V_{fkbc}^{2}
\end{equation}
which takes a tree contraction form by viewing $\boldsymbol{T}^{2}$ and $\boldsymbol{V}^{2}$ as input tensors (the superscript 2 means element-wise square), and therefore supports efficient exact sampling with again an $O(N^{4})$ one-time cost. In the SI, Sec 4.4, we prove that this leads to $\exp(\Delta F)\sim O(N)$ assuming the asymptotic forms in Eq.~\ref{eq:asymptotic}. Thus, the final energy relative and absolute variances are bounded by $O(N)$ and $O(N^{3})$, respectively. The additional $O(N)$ per-sample cost comes from computing $(\boldsymbol{T}\cdot\boldsymbol{V})_{ijkabc}$ and $\tilde{p}_{ijkabc}^{\prime}$ for each sampled $ijkabc$.

A summary of the computational scaling of the full STC-CCSD(T) implementation and the comparison with exact computation can be found in Table \ref{table:CC_scaling}.

\section*{Numerical experiments}
We now carry out numerical experiments to verify our theoretical analysis and establish the performance of stochastic tensor contraction. We have implemented the closed-shell STC-CCSD(T) method in Python with PyTorch~\cite{paszke2019pytorch}, interfaced with the PySCF code~\cite{sun2018pyscf,sun2020recent}.
The code is available at \url{https://github.com/SUSYUSTC/stc_qc_paper}. 
Full details of the implementation can be found in the SI. 
Some contractions in CCSD that are subleading in the exact contraction cost are, in practice, cheaper to perform deterministically for the system sizes of interest.

Therefore, for STC-CCSD, we provide both a ``pure'' implementation which applies STC to all tensor contractions with an exact contraction cost higher than $O(N^{4})$ (such that the deterministic cost is strictly $O(N^{4})$), and an optimized implementation that includes some $O(N^{5})$ deterministic steps (from computing some contractions exactly and employing an improved iterative solver) which yields
better performance at the system sizes of interest. Further, motivated by the theoretical variance bounds, when using localized orbitals we obtain them by minimizing the tensor norm of the fundamental input tensors.

Using the above implementations, we first verify the sample requirements for the stochastic tensor contractions, to converge the amplitude equations, and to obtain unbiased estimators. Next, we analyze STC-CCSD(T) compared to the leading competing strategy for large molecules, domain localized pair natural orbital (DLPNO) CCSD(T)~\cite{riplinger2016sparse}, as available through the ORCA package~\cite{neese2020orca}. We study the cost and accuracy of DLPNO-CCSD(T) and STC-CCSD(T) as a function of electronic locality and dimensionality, and provide a real-world test of cost and accuracy across a set of molecules with realistic computational settings.

\noindent \emph{Computational scaling.} In both STC-CCSD and STC-(T), the computational cost is composed of a deterministic part and a sampling part controlled by the sampling variance, target error, and (in the case of the iterative amplitude equations) the critical number of samples $N_\text{critical}$.
As discussed above, the deterministic scaling in the ``pure'' implementation is fixed at $O(N^4)$ by design, thus 
we numerically study the scaling of only the sampling part and observe the relation with the theoretical scalings in Table~\ref{table:CC_scaling}.
We will focus on targeting a fixed absolute error $\epsilon$, as the less stringent case of fixed relative error can be obtained by dividing all costs by $N^{2}$. {The required number of samples to reach the $\epsilon$ error is denoted as $N_{\text{sample}}^{\epsilon}$.}

We use a series of clusters composed of 2-30 water molecules (6-31G basis~\cite{hehre1972self}) and set $\epsilon = 0.2$~m$E_h$ $\sim$ 0.1~kcal/mol {for both CCSD and (T)}.
While we mainly present data for the pure implementation to examine the ideal theoretical scaling, we note that the optimized implementation always requires fewer samples. The numerical scalings of $N_{\text{sample}}^{\epsilon}$ and $N_{\text{critical}}$ for STC-CCSD with local and canonical bases are shown in Fig. \ref{fig:scaling}(A). We first consider the scaling of $N_{\text{sample}}^{\epsilon}$. For local STC-CCSD, the numerical scaling is $N_{\text{sample}}^{\epsilon}\sim O(N^{2.22})$, which is slightly higher than the theoretical $O(N^2)$ scaling, possibly due to the small size of the clusters and the effect of $\log N$ factors. For canonical STC-CCSD, the numerical scaling is $N_{\text{sample}}^{\epsilon}\sim O(N^{3.23})$, which is much lower than the theoretical $O(N^4)$ scaling. This arises from the competition between scaling and perturbation order. The CCSD contractions involving two $\boldsymbol{T}^{(2)}$ tensors have higher variance scaling but describe contributions from higher perturbation orders, thus the variance is in practice small, while the contractions involving {one} $\boldsymbol{T}^{(2)}$ tensor dominate in practice with $O(N^{3+1/3})$ theoretical scaling (see SI), which matches the numerical scaling.

Next we consider the scaling of $N_{\text{critical}}$. For local STC-CCSD, we find $N_{\text{critical}}\sim O(N^{2.98})$, which is slightly higher than the scaling of $N_{\text{sample}}^{\epsilon}$. However, the absolute numbers are much lower than $N_{\text{sample}}^{\epsilon}$ for the given $\epsilon$ and range of system sizes studied. For canonical STC-CCSD, we find $N_{\text{critical}}\sim O(N^{5.18})$, which is much higher, and the values of $N_{\text{critical}}$ also approach $N_{\text{sample}}^{\epsilon}$ for the large clusters. Clearly, STC-CCSD in a local basis has a better scaling and smaller  $N_{\text{sample}}^{\epsilon}$ and $N_{\text{critical}}$. Thus we focus on STC-CCSD in the local basis in later numerical experiments.

We also show the estimated number of floating point operations (FLOP) of the sampling part required for STC-CCSD(T) and exact CCSD(T) in the \textsc{PySCF} implementation in Fig.~\ref{fig:scaling}(B). The $O(N^{2.22})$ FLOP complexity for STC-CCSD matches $N_{\text{sample}}^{\epsilon}$, since only one floating point operation is performed for each sample (we do not count FLOP coming from the {random number generation}). The observed $O(N^{3.76})$ complexity for STC-(T) is below the $O(N^{4})$ theoretical upper bound. Both of these are much better than the exact scalings of $O(N^{6})$ for CCSD and $O(N^{7})$ for (T). For 30 waters, this gives around 5 orders of magnitude reduction in FLOP for both CCSD and (T).

Finally, we show the total computational wall time of STC-CCSD(T) (including the deterministic parts) and exact CCSD(T) in Fig. \ref{fig:scaling}(C), and also compare the practical performance of the pure and optimized implementations. 
Due to the much better scaling, both implementations exhibit significant speedups when the system size grows. For all studied system sizes, the optimized implementation is faster than the pure implementation. We also note that STC-CCSD(T) is already faster than the exact CCSD(T) implementation starting from 3 waters, where both the STC and exact computations take only a few seconds.

\begin{figure}
\centering

\includegraphics[width=1.0\columnwidth]{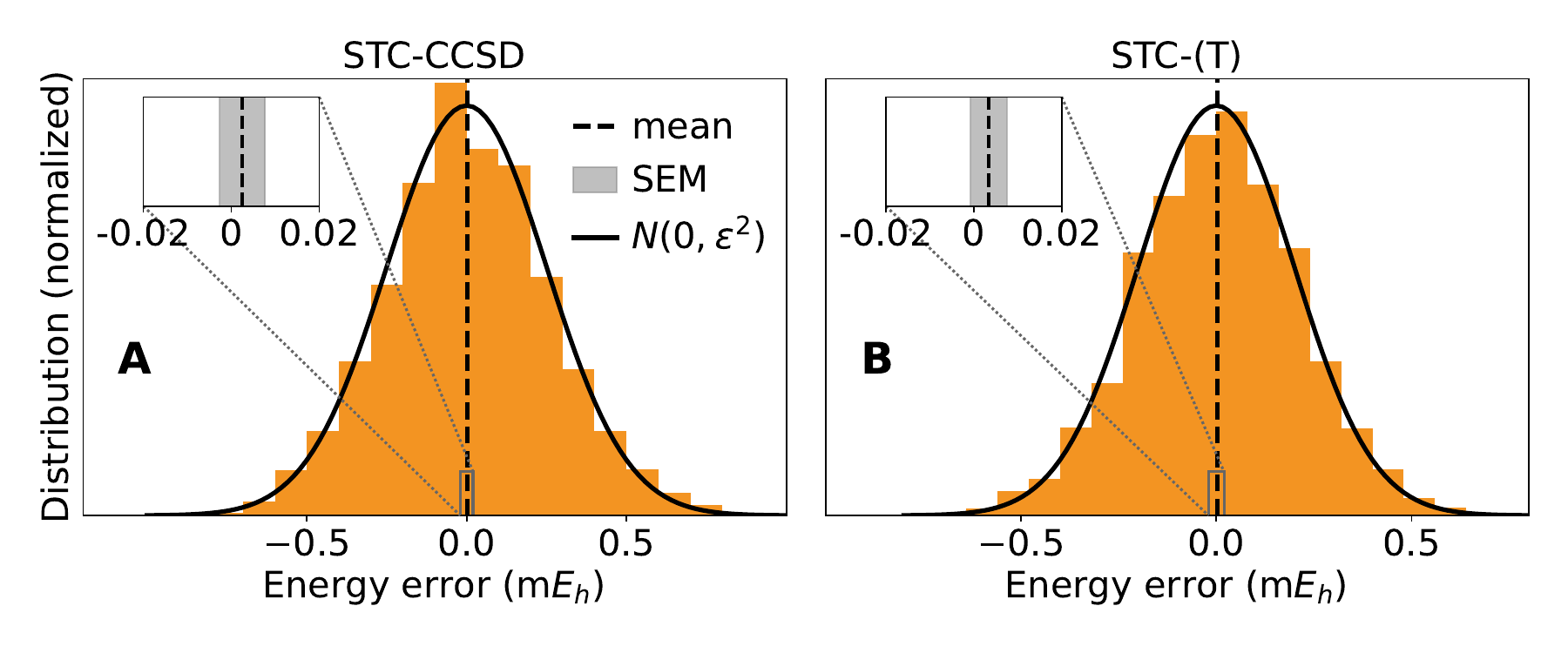}

\caption{Statistical histograms of (A) STC-CCSD and (B) STC-(T) energy errors from $n=2500$ independent STC-CCSD(T) calculations for a single benzene molecule in the cc-pVTZ basis, relative to exact CCSD(T) energies. The target per-sample statistical error is set to $\epsilon=0.25\text{m}E_{h}$ for STC-CCSD and $\epsilon=0.2\text{m}E_{h}$ for STC-(T) with input tensors from STC-CCSD. Black vertical dashed lines indicate the sample means of the 2500 calculations, which are $2.53\mu E_{h}$ for STC-CCSD and $3.40\mu E_{h}$ for STC-(T). Black solid lines are the predicted error distributions $N(0, \epsilon^2)$ from the target error $\epsilon$. The sample standard deviations, $\sigma=$0.257 m$E_{h}$ for STC-CCSD and $\sigma=$0.206 m$E_{h}$ for STC-(T), are consistent with the target errors. Gray shaded regions denote the standard error of the mean (SEM), given by $\sigma/\sqrt{n}$, corresponding to $5\mu E_{h}$ and $4\mu E_{h}$ for STC-CCSD and STC-(T), respectively. In both cases, zero lies within the SEM, indicating the absence of detectable bias in the $\mu E_{h}$ range.}

\label{fig:unbiaseness}
\end{figure}

\begin{figure*}[!hbtp]
\centering

\includegraphics[width=\linewidth]{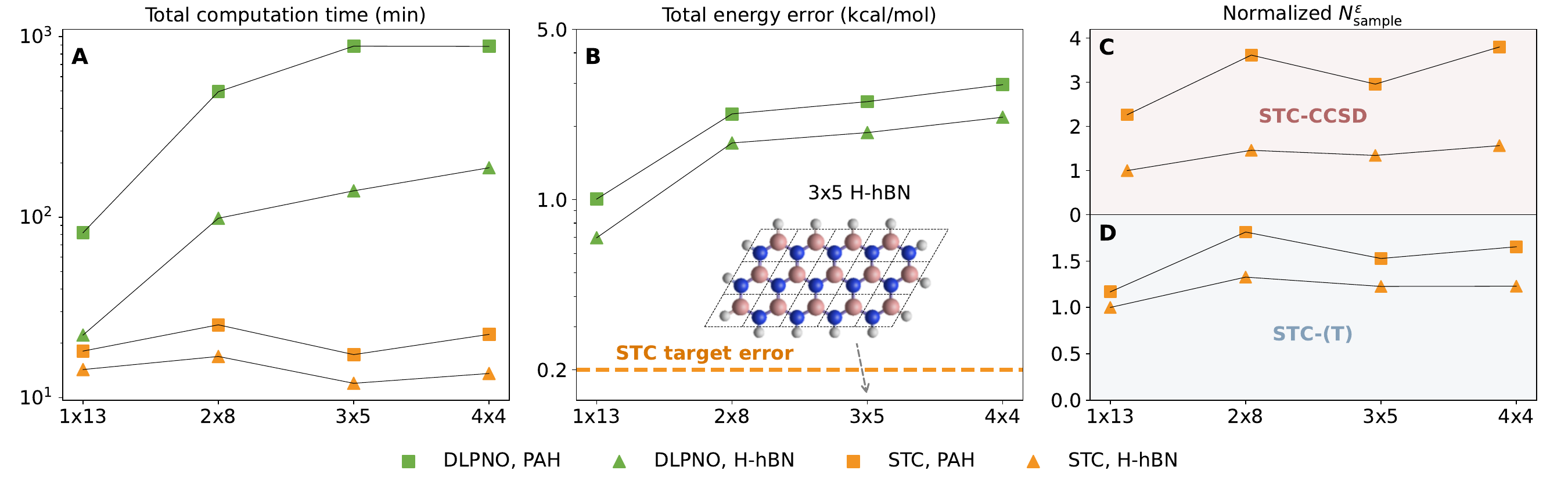}

\caption{Dependence of STC-CCSD(T) on system dimensionality and orbital delocalization compared with DLPNO-CCSD(T), demonstrated on finite hydrogen-terminated h-BN (H-hBN) and polycyclic aromatic hydrocarbon (PAH) clusters. Four geometries, 1\texttimes 13, 2\texttimes 8, 3\texttimes 5, and 4\texttimes 4, are used to capture the transition from quasi-one-dimensional to two-dimensional geometries, thereby varying the effective dimensionality. The Hartree-Fock gaps of H-hBN and PAH are $\sim 0.5$ and $\sim 0.2$ $E_{h}$ for the 4 geometries, respectively, yielding different levels of delocalization. The STC-CCSD(T) target error is set to be $0.2$ kcal/mol, while DLPNO-CCSD(T) calculations use the TightPNO setting. (A) Total computation time of STC and DLPNO CCSD(T) for H-hBN and PAH. (B) Total energy error of STC and DLPNO CCSD(T) for H-hBN and PAH. (C,D) Normalized number of samples in (C) iterative STC-CCSD and (D) STC-(T) for H-hBN and PAH calculations. An example of the 3x5 H-hBN geometry is shown in Fig.~\ref{fig:locality}(B). All computations are performed with the frozen-core approximation, cc-pVDZ basis and 8 CPU cores. In all STC-CCSD(T) calculations, the true energy errors compared to the exact reference are consistent with the target error.} 
\label{fig:locality}
\end{figure*}

\noindent \emph{Unbiased nature of errors {and predictable error statistics.}} As established above, STC is strictly unbiased at the level of a single multi-tensor contraction, but bias theoretically appears in STC-CCSD due to the iterative computation, as well as in STC-(T) when the amplitude tensors are obtained from STC-CCSD. This bias is quadratic in the relative statistical error of the intermediate tensor(s) in the contraction. 
For typical target accuracies, the relative statistical error is on the order of $10^{-3}\sim10^{-4}$, implying that the induced bias should be suppressed by an additional factor of $10^{-3}\sim10^{-4}$ compared to the statistical uncertainty and can be safely neglected.
The central limit theorem allows us to accurately estimate $N_{\text{sample}}^{\epsilon}$ needed to reach statistical error $\epsilon$, using the variance estimated from a small number of samples. Therefore, we expect the true error statistics to be in good agreement with the specified target error.

To verify this, we consider a single benzene molecule in the cc-pVTZ basis~\cite{dunning1989gaussian} and perform $n=2500$ independent STC-CCSD(T) calculations using the pure implementation and different random seeds, comparing the resulting energies against exact CCSD(T) values. 
The target statistical error is set to $\sigma=0.25\text{m}E_{h}$ for STC-CCSD and $\sigma=0.2\text{m}E_{h}$ for STC-(T) to achieve a total target error of $\sqrt{0.25^2 + 0.2^2}$~m$E_h$ $\sim$ 0.2~kcal/mol. 
The total correlation energy is approximately 1.12~$E_{h}$, so the prescribed absolute errors correspond to relative statistical uncertainties in the $10^{-3}\sim10^{-4}$ range. The histograms of the 2500 CCSD and (T) energy errors are plotted in Fig. \ref{fig:unbiaseness}. 
For both CCSD and (T), the energy errors follow a normal distribution with mean values in the $\mu E_h$ range.
Further, the measured sample standard deviations, 0.257 m$E_{h}$ for STC-CCSD and 0.206 m$E_{h}$ for STC-(T), are in good agreement with the specified target errors, confirming the reliability of the error estimates.
Thus, we confirm that the bias induced by STC is {negligible} compared to the statistical uncertainty under practical accuracy settings, and the target statistical error is consistent with the true statistical error.

\begin{figure}
\centering
\includegraphics[width=0.7\linewidth]{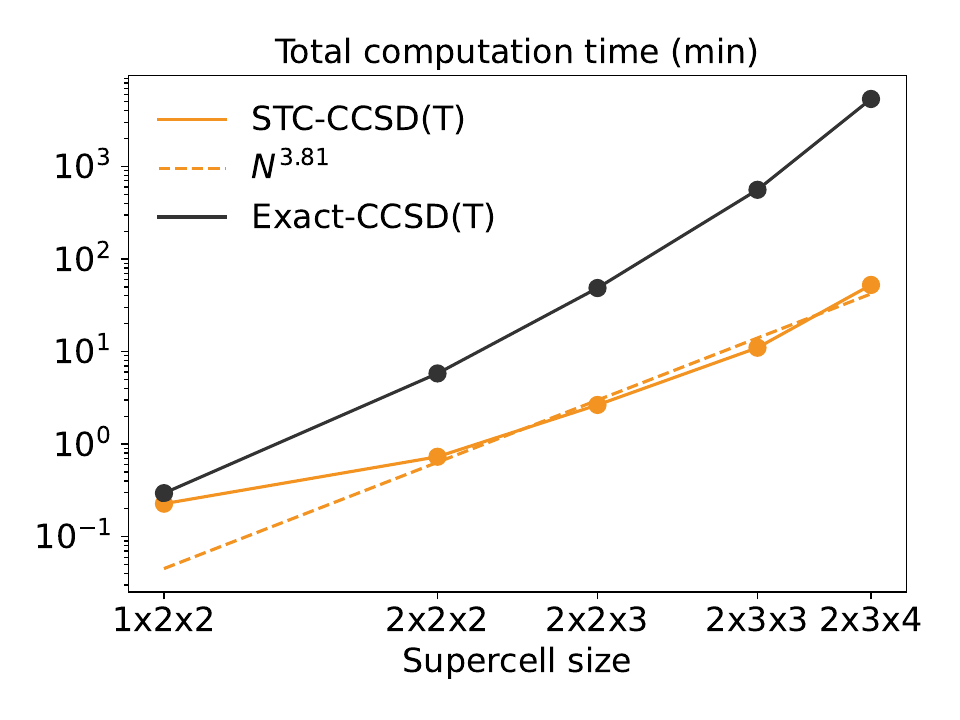}

\caption{Total computation time of STC and exact CCSD(T) on single-Si-doped diamond crystals. Each primitive cell contains two atoms. The STC-CCSD(T) target total energy error is set to be 24~meV for the whole supercell. Both STC and exact CCSD(T) calculations are performed with the GTH-DZVP basis and 32 CPU cores. The true energy errors compared to the exact reference are consistent with the target error in all the five calculations.}

\label{fig:solid}
\end{figure}

\noindent \emph{Dependence of STC on dimensionality and delocalization.}
We now analyze the performance of STC-CCSD(T) (with the optimized implementation {of STC-CCSD} in the local basis) for systems of different dimensionality and electronic delocalization, and compare to the {state-of-the-art}  DLPNO-CCSD(T) local correlation implementation in ORCA. 

We consider a series of hydrogen-terminated finite hexagonal boron nitride (H-hBN) and polycyclic aromatic hydrocarbon (PAH) clusters. The clusters are configured to have roughly the same number of {electrons}, and we consider $1\times13$, $2\times8$, $3\times5$, and $4\times4$ clusters to mimic a transition from a 1D chain to a 2D lattice. An example of the $3\times5$ H-hBN structure is shown in Fig \ref{fig:locality}(B). In the infinite size limit, PAH becomes graphene, which has a vanishing gap (Dirac points), while H-hBN has a finite system gap regardless of size. Thus, comparing H-hBN to PAH clusters allows us to compare systems with different degrees of delocalization. As an indication, their HF gaps in the cc-pVDZ basis~\cite{dunning1989gaussian} are 0.486$\sim$0.507 $E_{h}$ for H-hBN, and 0.192$\sim$0.242 $E_{h}$ for PAH. 

In DLPNO-CCSD(T), truncation is performed by constructing pair natural virtual orbitals for each occupied orbital pair and discarding those with small occupation numbers, thereby restricting correlation to localized pair domains. The truncation level is controlled by a series of thresholds; in the ORCA implementation, default thresholds are provided, referred to as NormalPNO, TightPNO, etc. Here we choose the TightPNO setting in the DLPNO-CCSD(T) calculations. In STC-CCSD(T) calculations, we fix the target energy error to 0.2 kcal/mol, and all the observed STC-CCSD(T) errors are consistent with this. All calculations were performed in a cc-pVDZ basis and using the same CPU with 8 cores.

The total computational time and total energy error of both STC-CCSD(T) and DLPNO-CCSD(T) are shown in Fig.~\ref{fig:locality}(A) and Fig.~\ref{fig:locality}(B), respectively, while the number of (iterative) STC-CCSD and STC-(T) samples are shown in Fig.~\ref{fig:locality}(C).
We first analyze the dependence on dimensionality. For DLPNO-CCSD(T), both the computation time and energy error grow monotonically as we cross from one-dimensional clusters to two-dimensional geometries. {Consistently for both H-hBN and PAH}, compared to the $1\times13$ structure, the computation time of the $4\times4$ structure is larger by a factor $\sim10$, and the energy error increases by a factor $\sim3$.
In contrast, the number of samples for STC-CCSD and STC-(T) increases only by a factor of $\sim1.6$ and $\sim1.3$, respectively. The total computation time changes by an even smaller factor, once we account for the (essentially) fixed deterministic cost across all clusters.

\begin{figure}[t!]
\centering

\includegraphics[width=0.96\linewidth]{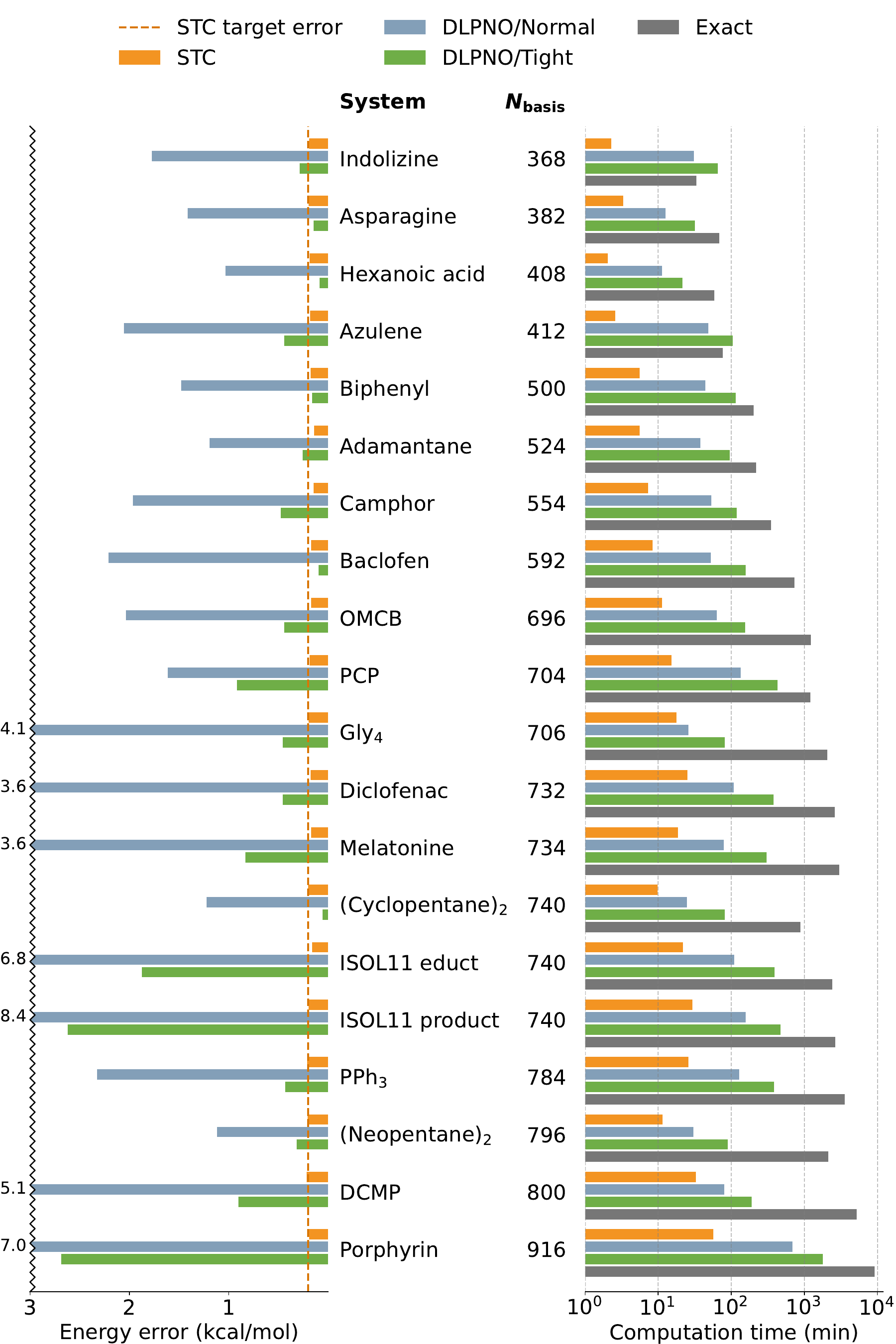}

\caption{Total energy error and computation time for STC-CCSD(T), DLPNO-CCSD(T), and exact CCSD(T) calculations on a set of 20 realistic molecules. Three abbreviated names are used, including OMCB for octamethylcyclobutane, PCP for {[}2.2{]}paracyclophane, and DCMP for deoxycytidine-monophosphate. The systems in the figure are ordered by the number of orbitals from top to bottom. We use the aug-cc-pVTZ basis~\cite{kendall1992electron} for non-hydrogen atoms in the two interacting dimer systems, and the cc-pVTZ basis elsewhere. In all STC-CCSD(T) calculations, the target energy error is set to be 0.2 kcal/mol for STC-CCSD(T), and the mean absolute error (MAE) of 25 independent calculations is reported as the energy error in the left panel. The DLPNO-CCSD(T) results are reported with two truncation settings NormalPNO and TightPNO.
A summary of the four CCSD(T) results, as well as two DFT results with different functionals under the same basis, is shown in Table.~\ref{table:benchmark_stats}.
All computations are performed with the frozen-core approximation (except DFT) and 8 CPU cores.}

\label{fig:benchmarking}
\end{figure}

We next analyze the dependence on the electronic delocalization and gap. {Since the relative behaviors of the total computation time, energy error, and number of samples are consistent across different structures,} we use the $4\times4$ H-hBN and PAH structures as an example. For DLPNO-CCSD(T), we observe a $4.7\times$ and $1.4\times$ increase in total computation time and energy error from H-hBN to PAH, respectively. For STC-CCSD(T), the number of samples increases by $2.4\times$ and $1.4\times$ for STC-CCSD and STC-(T), respectively, and the total computation time increases only by a factor of $1.7$ for fixed error, showing the weak dependence on delocalization and gap. We also see that STC-(T) (which is formulated in the canonical basis) has an even weaker dependence on dimensionality and the gap than STC-CCSD in the local basis, in accordance with our theoretical analysis.

The insensitivity of STC to dimensionality and delocalization suggests it will be useful in formulating electronic structure methods in materials, where the density of atoms is typically higher, and the delocalization greater, than in molecules. As a preliminary demonstration, we present STC-CCSD(T) calculations on doped diamond structures. We construct diamond cells by repeating a 2 atom diamond supercell, up to a $2\times3\times4$ lattice (48 atom supercell), replace a single atom with a silicon dopant, {and relax the cell structure using density functional theory.}
As DLPNO-CCSD(T) is not available for materials, we compare to exact CCSD(T) calculations, and in the STC-CCSD(T) calculations, we set a target total energy error of the supercell of 24 meV (to yield 0.5~meV per-atom error for the largest $2\times3\times4$ supercell). All calculations are performed with the frozen-core approximation, GTH-DZVP basis~\cite{vandevondele2007gaussian}, and 32 CPU cores.
The results are shown in Fig. \ref{fig:solid}. As we see, despite the high density of atoms, STC-CCSD(T) is already faster than the exact implementation even for the smallest system (8 atoms) and further has a much better scaling (empirically $O(N^{3.81})$, versus $O(N^{7})$). This illustrates the potential of STC formulations in materials simulations.

\begin{table}[t!]

\caption{Summary of averaged absolute energy error and geometrically averaged computation time of DLPNO-CCSD(T), STC-CCSD(T), exact CCSD(T), and additionally DFT with two functionals B3LYP~\cite{b3lyp} and $\boldsymbol{\omega}$B97M-V\cite{wb97m-v}, over 20 realistic molecules in Fig. \ref{fig:benchmarking}. The computational details of all CCSD(T) calculations can be found in the caption of Fig. \ref{fig:benchmarking}, and DFT calculations are performed with the same basis and hardware.}

\setlength{\tabcolsep}{2.5pt}

\begin{tabular}{|c|c|c|c|c|c|c|}
\hline 
\multirow{3}{*}{Average} & \multicolumn{2}{c|}{DFT} & \multicolumn{4}{c|}{CCSD(T)}\tabularnewline
\cline{2-7}
 & \multirow{2}{*}{B3LYP} & \multirow{2}{*}{$\omega$B97M-V} & \multirow{2}{*}{STC} & \multicolumn{2}{c|}{DLPNO} & \multirow{2}{*}{Exact}\tabularnewline
\cline{5-6}
 &  &  &  & Normal & Tight & \tabularnewline
\hline 
\hline 
Error (kcal/mol) & N/A & N/A & 0.183 & 3.00 & 0.70 & 0\tabularnewline
\hline 
Time (min) & 0.42 & 0.86 & 10.7 & 58 & 159 & 773\tabularnewline
\hline 
\end{tabular}

\label{table:benchmark_stats}
\end{table}

\begin{figure}[hbt]
\centering

\includegraphics[width=0.96\linewidth]{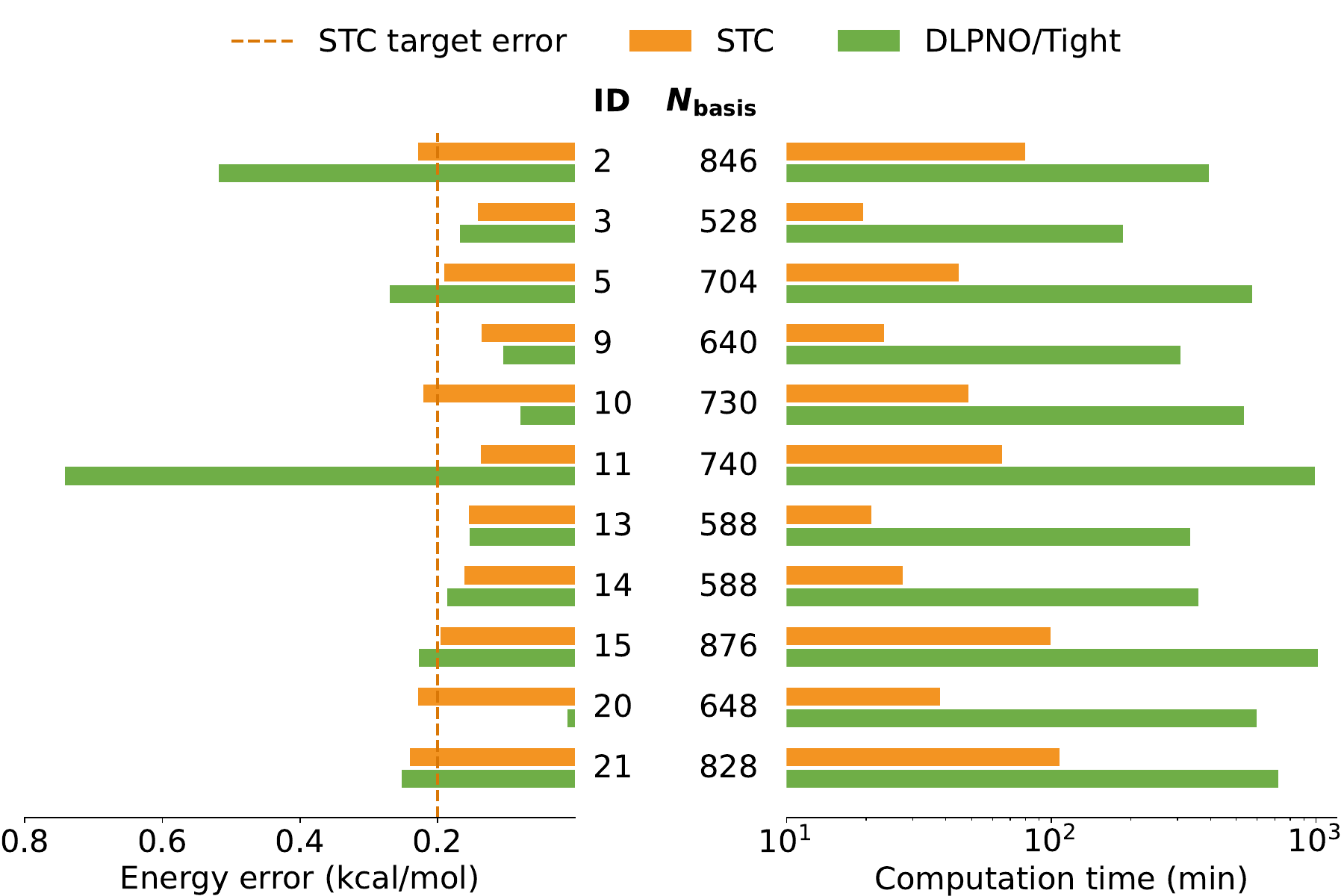}

\caption{Reaction energy error and computation time obtained from the STC-CCSD(T) and Tight DLPNO-CCSD(T) calculations of 11 reactions in the ISOL24 reaction dataset for which exact CCSD(T) computations are affordable. The STC target error for each individual reactant or product is set to be $0.14$ kcal/mol, so that the reaction target error is $0.14\times\sqrt{2}\approx 0.2$ kcal/mol. The MAE of 25 independent STC calculations is reported as the energy error in the left panel. For STC-CCSD(T) and Tight DLPNO-CCSD(T), the averaged absolute reaction energy errors are 0.185 and 0.246
kcal/mol, and the geometrically averaged (i.e., average on a log scale) computation times are 44 and 487 minutes, respectively. All computations are performed with a cc-pVTZ basis, the frozen-core approximation, and 8 CPU cores.}

\label{fig:reaction}
\end{figure}

\noindent \emph{Performance on benchmark molecular sets.}
Next, we benchmark the performance of our current STC-CCSD(T) implementation on a set of 20 realistic molecular systems with 16-38 atoms and 368-916 basis functions against DLPNO-CCSD(T) and exact CCSD(T). Note that these are not the largest systems treatable by our STC-CCSD(T) implementation but are ones where exact CCSD(T) calculations are practical. The benchmark set spans a range of molecular systems that have previously been used to benchmark the performance of local correlation methods in the literature~\cite{neese2009efficient2,riplinger2016sparse,datta2016analytic,nagy2018optimization}, ensuring that the test cases are not biased in favor of STC-CCSD(T).

In all the STC-CCSD(T) calculations, the target statistical error is fixed at 0.2 kcal/mol (the reported error is the mean absolute error (MAE) from 25 independent runs).
For DLPNO-CCSD(T), results obtained with the NormalPNO and TightPNO truncation settings are shown for comparison. All calculations were performed on the same CPU with 8 cores. For simplicity we refer to the three methods as STC, DLPNO/Normal and DLPNO/Tight. The comprehensive results are shown in Fig.~\ref{fig:benchmarking}, and a summary is shown in Table~\ref{table:benchmark_stats}.

The STC energy errors are smaller than all DLPNO/Normal and 15 of 20 DLPNO/Tight errors; on average, the absolute STC errors are $16\times$ and $3.8\times$ smaller than those from DLPNO/Normal and DLPNO/Tight. Additionally, the STC MAEs are $<0.22$ kcal/mol for all systems, matching the target error, while the DLPNO energy error varies significantly even amongst systems of similar size. Thus STC yields a more predictable and controllable error behavior. 

In terms of computation time, both STC and DLPNO are faster than exact calculations for most systems, but STC shows significant speedup for all system sizes, while DLPNO shows a large speedup only for large systems. Indeed STC is faster than DLPNO for every system, by a factor ranging from $2.5\times$ (Neopentane dimer, NormalPNO) to $32\times$ (Porphyrin, TightPNO). Averaged over the set, STC is approximately one order of magnitude faster (5.4$\times$ versus NormalPNO, 14.9$\times$ versus TightPNO), with {simultaneously} one order of magnitude smaller errors (16.4$\times$ versus NormalPNO, 3.8$\times$ versus TightPNO), than DLPNO. For additional context, the STC cost is $\sim$10-25$\times$ that of a calculation using  modern density functionals
(using the same hardware, with density fitting, and appropriate grids) as opposed to 1000-2000 times using exact CCSD(T) (see Table~\ref{table:benchmark_stats}). Thus with STC, gold-standard coupled cluster calculations not only reach the scaling of mean-field calculations, but start to approach them in absolute cost.

We further benchmark the performance of STC-CCSD(T) on 11 reactions from the ISOL24 isomerization reaction dataset~\cite{isol24} against DLPNO-CCSD(T). The 11 reactions were chosen by the criterion of $N_\text{occ}\times N_\text{vir} \leq 40000$ ($N_\text{occ}$, $N_\text{vir}$ are the number of occupied and virtual orbitals, respectively), such that exact CCSD(T) computations are feasible. The total STC target error is set to be 0.2 kcal/mol, and MAE of 25 independent calculations are reported. The reaction energy errors and total computation time are shown in Fig.~\ref{fig:reaction}.

For reaction tasks, the energy errors induced by the local correlation approximations partly cancel, leading to smaller errors; while the errors of STC in principle increase by a factor of $\sqrt{2}$, since the errors are independently Gaussian distributed. Nevertheless, the STC-CCSD(T) energy errors are lower than the TightPNO errors on average (0.185 versus 0.246 kcal/mol), and remain consistent around the target error, while TightPNO gives unexpectedly large errors for some reactions: for example, for reaction 11, the TightPNO error (0.74~kcal/mol) is more than 5 times that from STC (0.14~kcal/mol). Simultaneously, STC-CCSD(T) is on average 11$\times$ faster than DLPNO-CCSD(T) (44 versus 487 minutes).

We note that the ORCA DLPNO-CCSD(T) implementation has been optimized over more than a decade, while our STC-CCSD(T) implementation is still a prototype. Thus, the relative improvement comes from the strong theoretical foundations of the STC theory. At the same time, STC is an orthogonal strategy to local correlation truncations and can be combined with them in large systems. We leave such considerations to future work.

\section*{Conclusion}
We have shown that stochastic tensor contraction is a simple and flexible way to accelerate quantum chemistry methods that can be formulated in terms of tensor contractions. Since this encompasses many computational methods, the technique has a wide range of future applications. As shown in our theoretical analysis and numerical demonstrations in the case of the gold-standard coupled cluster theory, stochastic tensor contraction enables such calculations to be carried out (for statistical errors more stringent than chemical accuracy) with the scaling of mean-field theory, and, even in our preliminary implementation, only an order of magnitude greater absolute cost.

Quantum chemistry has generally advanced through the combination of three types of innovation: the formulation of theoretical equations in place of the full quantum many-body problem, the development of new numerical approaches to solve those equations, and the utilization of new computer hardware. While our application of stochastic tensor contraction has focused on the second setting, it also opens up  directions for formulating new theories based on the new achievable complexities. Further, the efficient implementation of this computational primitive on modern hardware, such as graphical processing units, is an obvious avenue to explore. Overall, we see a bright future for stochastic tensor contraction in the quantum many-body problems of chemistry.

\section*{Data availability}
The data and code repository is available on GitHub: \url{https://github.com/SUSYUSTC/stc_qc_paper}. The repository contains all numerical data used to generate figures in the main text and supporting information, and the code to reproduce all numerical results.

\section*{Acknowledgement}
This work was supported by the U.S. Department of Energy, Office of Science, Basic Energy Sciences, through Award No. DE-SC0018140. We also thank Prof. Lixue Cheng, Prof. Sandeep Sharma, Dr. Xing Zhang, Dr. Junjie Yang, Yuhang Ai, Dr. Huanchen Zhai, Dr. Chenghan Li, Dr. Johnnie Gray, and Jielun Chen for helpful discussions and comments.


\bibliography{main}

\newpage
\pagebreak

\end{document}


\date{}
\maketitle
\noindent $^{*}$Corresponding author: gkc1000@gmail.com

\tableofcontents{}

\section{Notation}

For an arbitrary vector $\boldsymbol{v}$, we define its $p$-norm (also named the $l^{p}$-norm) as 
\begin{equation}
\Vert\boldsymbol{v}\Vert_{p}=\left(\sum_{i}|v_{i}|^{p}\right)^{1/p}.\label{eq:vec_norm}
\end{equation}
For an arbitrary tensor $\boldsymbol{A}$, we define its $p$-norm as the vector $p$-norm of the flattened vector i.e. 
\begin{equation}
\Vert\boldsymbol{A}\Vert_{p}=\left(\sum_{ijk...}|A_{ijk...}|^{p}\right)^{1/p},\label{eq:tensor_norm}
\end{equation}
where $ijk...$ loops over all tensor indices. Note that $\Vert\cdot\Vert_{p}$ in the literature is sometimes used for the operator norm, which is different from the tensor norm here. The operator norm of a matrix $\boldsymbol{A}$ is 
\begin{equation}
\Vert\boldsymbol{A}\Vert_{\text{op},p}=\max_{\boldsymbol{v}}\frac{\Vert\boldsymbol{A}\boldsymbol{v}\Vert_{p}}{\Vert\boldsymbol{v}\Vert_{p}}\label{eq:op_norm}
\end{equation}
where $\boldsymbol{v}$ is a vector, and $\boldsymbol{A}\boldsymbol{v}$ corresponds to matrix-vector multiplication.

For a tensor $\boldsymbol{A}$, we define \textbf{$|\boldsymbol{A}|$}, or sometimes $\bar{\boldsymbol{A}}$, as the tensor after taking the element-wise absolute value of $\boldsymbol{A}$, i.e. 
\begin{equation}
\left(|\boldsymbol{A}|\right)_{ijk...}=\left(\bar{\boldsymbol{A}}\right)_{ijk...}=\left|\left(\boldsymbol{A}\right)_{ijk...}\right|
\end{equation}
A general tensor contraction $\boldsymbol{S}=\mathcal{C}(\boldsymbol{A}_{1},\boldsymbol{A}_{2},...,\boldsymbol{A}_{m})$ of tensors $\boldsymbol{A}_{1},\boldsymbol{A}_{2},...,\boldsymbol{A}_{m}$ sums over some common indices $I$, and leaves some free indices $O$, in the output tensor $\boldsymbol{S}$: 
\[
S_{O}=\sum_{I}\left(\boldsymbol{A}_{1}\boldsymbol{A}_{2}...\boldsymbol{A}_{m}\right)_{IO},
\]
where $\boldsymbol{A}_{1}\boldsymbol{A}_{2}...\boldsymbol{A}_{m}$ corresponds to element-wise multiplication along the common indices. We say that a tensor contraction satisfies the Einstein summation rule (or simply is Einstein), if each common index appears exactly twice, and each free index appears exactly once.

Note that we use notations like $\boldsymbol{A}\boldsymbol{B}$ by default always for element-wise multiplication, and for matrix product only when we explicitly specify.

\section{STC for general tensor contractions}

\subsection{Algorithm for tree tensor contractions }\label{subsec:tree}

\begin{figure}[tbh]
\centering

\includegraphics[width=0.6\columnwidth]{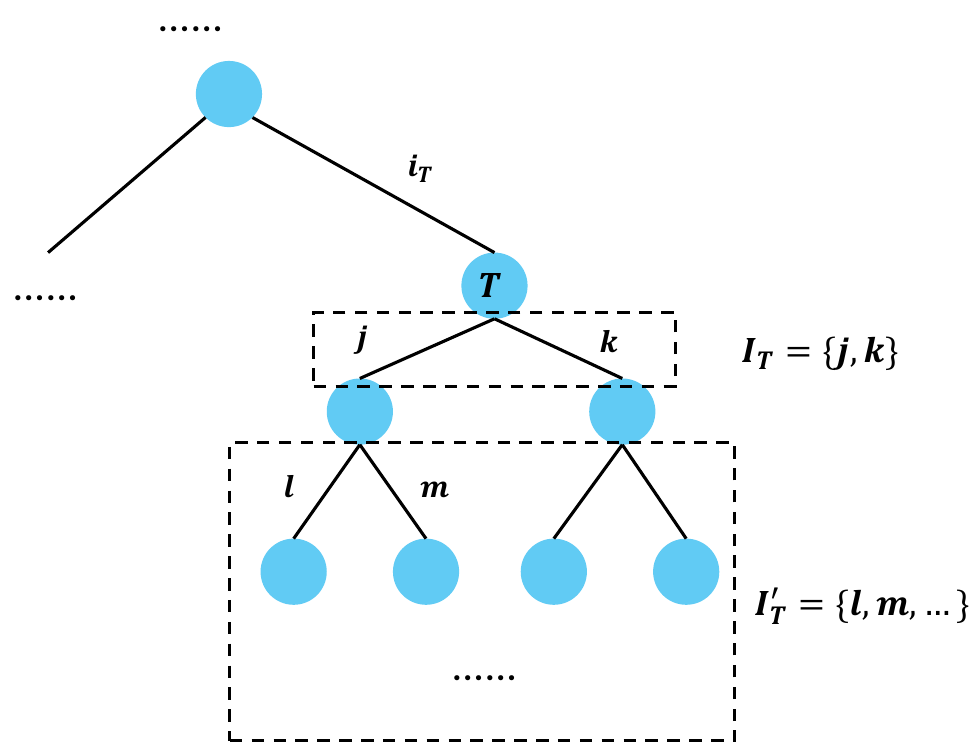}

\caption{Tree tensor contraction algorithm.}
\label{fig:tree_sampling}
\end{figure}

Our goal is to importance sample a tensor contraction with a probability distribution proportional to the product of the absolute tensor elements (see Eq. 4 in the main text):
\[
\tilde{p}^{\text{opt}}=|\boldsymbol{A}\boldsymbol{B}...|.
\]
For any tree tensor contraction, one can write it as a standard tree structure such that each tensor has a parent index (if the root tensor does not have a parent index, one can add a virtual index for it) and a few (possibly zero) children indices. Consider an arbitrary tensor $\boldsymbol{T}$, define the subtree as all tensors under this tensor (i.e. $\boldsymbol{T},$ children of $\boldsymbol{T}$, children of children of $\boldsymbol{T}$, ...). Let the parent index of $\boldsymbol{T}$ be $i_{T}$, children indices of $\boldsymbol{T}$ be $I_{T}$, and other indices of the subtree of $\boldsymbol{T}$ be $I_{T}^{\prime}$ (see Fig. \ref{fig:tree_sampling}). Let $\tilde{p}_{\text{subtree}}$ be the product of absolute tensor elements in the subtree (with the constructed similarly to $\tilde{p}^{\text{opt}}$). We define the unnormalized parent index probability $\tilde{p}(i_{T})$ and the normalized probability of children index conditioned on the parent index $p(I_{T}|i_{T})$ as:
\begin{align*}
\tilde{p}(i_{T}) & \equiv\sum_{I_{T},I_{T}^{\prime}}\tilde{p}_{\text{subtree}}(i_{T},I_{T},I_{T}^{\prime}),\\
p(I_{T}|i_{T}) & \equiv\frac{\sum_{I_{T}^{\prime}}\tilde{p}_{\text{subtree}}(i_{T},I_{T},I_{T}^{\prime})}{\sum_{I_{T},I_{T}^{\prime}}\tilde{p}_{\text{subtree}}(i_{T},I_{T},I_{T}^{\prime})}.
\end{align*}

The key observation is that, one only needs information on the (1) $\boldsymbol{T}$ tensor values, and (2) the (unnormalized) probabilities of the children indices, $\tilde{p}(i^{\prime})$ for all $i^{\prime}\in I_{T}$, to fully determine the (unnormalized) probability of the root index, $\tilde{p}(i_{T})$ and conditional probability of the children indices, $p(I_{T}|i_{T})$. The recursion relation is given by: 
\begin{equation}
\begin{aligned}\tilde{p}(i_{T}) & =\sum_{I_{T}}\tilde{p}(I_{T},i_{T}),\\
p(I_{T}|i_{T}) & =\frac{\tilde{p}(I_{T},i_{T})}{\tilde{p}(i_{T})},
\end{aligned}
\label{eq:tree_recursion}
\end{equation}
where the intermediate $\tilde{p}(I_{T},i_{T})$ is given by 
\[
\tilde{p}(I_{T},i_{T})=|T_{i_{T}I_{T}}|\prod_{i^{\prime}\in I_{T}}\tilde{p}(i^{\prime}).
\]
With these two recursive relations, one can build $\tilde{p}(i_{T})$ and $p(I_{T}|i_{T})$ for each tensor $T$ from the bottom up. Let the root index of the complete tree be $i_{\text{root}}$, then we obtain the exact decomposition of the full probability 
\[
\tilde{p}(\text{all})=\tilde{p}(i_{\text{root}})\prod_{T}p(I_{T}|i_{T}).
\]
The partition function is simply
\[
Z=\sum_{i_{\text{root}}}\tilde{p}(i_{\text{root}}).
\]
Clearly, the computational cost of constructing $\tilde{p}(i_{T})$ and $p(I_{T}|i_{T})$ from Eq. \ref{eq:tree_recursion} is the same as the size (i.e. number of elements) of $\boldsymbol{T}$.

\subsection{Proof of Eq. 5 in the main text}

Consider an importance sampling problem 
\begin{equation}
S=\sum_{i}s_{i}=\langle o_{i}=\frac{s_{i}}{p_{i}}\rangle_{i\sim p}
\end{equation}
The variance denoted as $\text{Var}$ is defined by
\begin{equation}
\text{Var}=\langle o^{2}\rangle-\langle o\rangle^{2}=\sum_{i}\frac{s_{i}^{2}}{p_{i}}-S^{2}.
\end{equation}
The relative variance is defined as 
\begin{equation}
\mathrm{RelVar}=\frac{\text{Var}}{S^{2}}=\frac{\langle o^{2}\rangle}{S^{2}}-1
\end{equation}

\begin{thm}
$\mathrm{RelVar}$ is minimized at $p_{i}=\frac{|s_{i}|}{Z}$, where $Z=\sum_{i}|s_{i}|$ normalizes the probability. The corresponding optimal variance is
\begin{equation}
\mathrm{RelVar}=\left(\frac{\sum_{i}|s_{i}|}{\sum_{i}s_{i}}\right)^{2}-1
\end{equation}
\end{thm}
\begin{proof}
By the Cauchy--Schwarz inequality, we have
\[
\langle o^{2}\rangle=\sum_{i}\frac{s_{i}^{2}}{p_{i}}=\left(\sum_{i}p_{i}\right)\left(\sum_{i}\frac{s_{i}^{2}}{p_{i}}\right)\geq\left(\sum_{i}|s_{i}|\right)^{2}=Z^{2},
\]
and it becomes an equality iff $p_{i}\propto|s_{i}|$. Thus the only choice is $p_{i}=\frac{|s_{i}|}{Z}$. Combined with the definition $S=\sum_{i}s_{i}$, this is equivalent to the given form to prove.
\end{proof}
Let $\boldsymbol{A}_{1},\boldsymbol{A}_{2},...,\boldsymbol{A}_{m}$ be tensors, and $\mathcal{C}(\boldsymbol{A}_{1},\boldsymbol{A}_{2},...,\boldsymbol{A}_{m})$ be a scalar-output tensor contractions of $\boldsymbol{A}_{1},\boldsymbol{A}_{2},...,\boldsymbol{A}_{m}$. Essentially $\mathcal{C}(\boldsymbol{A}_{1},\boldsymbol{A}_{2},...,\boldsymbol{A}_{m})$ just sums over all the indices: 
\begin{equation}
\mathcal{C}(\boldsymbol{A}_{1},\boldsymbol{A}_{2},...,\boldsymbol{A}_{m})=\sum_{I}(\boldsymbol{A}_{1}\boldsymbol{A}_{2}...\boldsymbol{A}_{m})_{I}.
\end{equation}
Thus this is equivalent to $S=\sum_{I}s_{I}$ with $s_{I}=(\boldsymbol{A}_{1}\boldsymbol{A}_{2}...\boldsymbol{A}_{m})_{I}$. Then we have
\begin{equation}
\begin{aligned}|s_{I}| & =(|\boldsymbol{A}_{1}||\boldsymbol{A}_{2}|...|\boldsymbol{A}_{m}|)_{I},\\
Z & =\sum_{I}|s_{I}|=\mathcal{C}(|\boldsymbol{A}_{1}|,|\boldsymbol{A}_{2}|,...,|\boldsymbol{A}_{m}|),\\
S & =\sum_{I}s_{I}=\mathcal{C}(\boldsymbol{A}_{1},\boldsymbol{A}_{2},...,\boldsymbol{A}_{m}).\\
\langle o^{2}\rangle & =Z^{2}=\left(\mathcal{C}(|\boldsymbol{A}_{1}|,|\boldsymbol{A}_{2}|,...,|\boldsymbol{A}_{m}|)\right)^{2}
\end{aligned}
\end{equation}
This gives our final conclusion:
\begin{cor}
(Eq. 5 in the main text)
\begin{equation}
\mathrm{RelVar}=\left(\frac{\mathcal{C}(|\boldsymbol{A}_{1}|,|\boldsymbol{A}_{2}|,...,|\boldsymbol{A}_{m}|)}{\mathcal{C}(\boldsymbol{A}_{1},\boldsymbol{A}_{2},...,\boldsymbol{A}_{m})}\right)^{2}-1
\end{equation}
\end{cor}

\subsection{Proof of Eq. 6 in the main text}
\begin{thm}
\label{thm:free_energy_thm1} Consider an importance sampling problem $S=\sum s_{i}=\langle\frac{s_{i}}{p_{i}}\rangle_{i\sim p}$. Let $\mathrm{RelVar}(p)$ be the relative variance as a function of sampling probability $p$. For two arbitrary distributions $p$ and $q$, if $p_{i}\leq Cq_{i}$ for all $i$ and fixed constant $C>0$, then
\begin{equation}
\frac{\mathrm{RelVar}(q)+1}{\mathrm{RelVar}(p)+1}\leq C.
\end{equation}
\end{thm}
\begin{proof}
We have 
\[
\mathrm{RelVar}(p)+1=\frac{1}{S^{2}}\sum_{i}\frac{s_{i}^{2}}{p_{i}}\geq\frac{1}{S^{2}}\sum_{i}\frac{s_{i}^{2}}{Cq_{i}}=\frac{1}{C}(\mathrm{RelVar}(q)+1),
\]
which is equivalent to the given form to prove.
\end{proof}
\begin{cor}
\label{cor:free_energy_thm2} (Eq. 6 in the main text) With the same notations, let $\tilde{p}$ and $\tilde{q}$ be unnormalized probabilities, and $Z_{p}=\sum_{i}\tilde{p}_{i}$, $Z_{q}=\sum_{i}\tilde{q}_{i}$ be the corresponding partition functions. If $\tilde{p}_{i}\leq\tilde{q}_{i}$ for all $i$, then
\begin{equation}
\frac{\mathrm{RelVar}(\tilde{q})+1}{\mathrm{RelVar}(\tilde{p})+1}\leq \frac{Z_{q}}{Z_{p}}.
\end{equation}
\end{cor}
\begin{proof}
Since $\tilde{p}=Z_{p}p$, $\tilde{q}=Z_{q}q$, we have $p_{i}\leq\frac{Z_{q}}{Z_{p}}q_{i}$, thus it reduces to Theorem \ref{thm:free_energy_thm1} with $C=\frac{Z_{q}}{Z_{p}}$ .
\end{proof}

\subsection{Proof of Eq. 8 in the main text}
In this section only, we use $\boldsymbol{A}\boldsymbol{B}$ for matrix product.

\begin{lem}
\label{lem:PSD_trace} Let \textbf{$\boldsymbol{P},\boldsymbol{Q}$ }be positive-semidefinite symmetric matrices, and $\boldsymbol{PQ}$ be the matrix product of $\boldsymbol{P}$ and $\boldsymbol{Q}$, then
\[
\mathrm{Tr}[\boldsymbol{PQ}]\geq0
\]
\end{lem}
\begin{proof}
Let the Cholesky decomposition of $\boldsymbol{P}$ and $\boldsymbol{Q}$ be $\boldsymbol{P}=\boldsymbol{A}\boldsymbol{A}^{T},\boldsymbol{Q}=\boldsymbol{B}\boldsymbol{B}^{T}$ , then
\begin{align*}
\mathrm{Tr}[\boldsymbol{PQ}] & =\text{Tr}[\boldsymbol{A}\boldsymbol{A}^{T}\boldsymbol{B}\boldsymbol{B}^{T}]\\
 & =\text{Tr}[\boldsymbol{B}^{T}\boldsymbol{A}\boldsymbol{A}^{T}\boldsymbol{B}]\\
 & =\text{Tr}[\left(\boldsymbol{B}^{T}\boldsymbol{A}\right)\left(\boldsymbol{B}^{T}\boldsymbol{A}\right)^{T}]\\
 & \geq0
\end{align*}
\end{proof}
\begin{lem}
\label{lem:PSD} Let \textbf{$\boldsymbol{P},\boldsymbol{Q}$ }be positive-semidefinite symmetric matrices, then
\[
\mathrm{Tr}[\boldsymbol{PQ}]\leq\mathrm{Tr}[\boldsymbol{P}]\mathrm{Tr}[\boldsymbol{Q}]
\]
\end{lem}
\begin{proof}
For a positive-semidefinite symmetric matrix $\boldsymbol{P}$ we have $\boldsymbol{P}\preceq\text{Tr}[\boldsymbol{P}]\boldsymbol{I}$, where $\boldsymbol{P}\preceq\boldsymbol{P}^{\prime}$ means $\boldsymbol{P}^{\prime}-\boldsymbol{P}$ is positive-semidefinite, and $\boldsymbol{I}$ is the identity matrix. Thus we have
\[
\text{Tr}[\boldsymbol{P}]\text{Tr}[\boldsymbol{Q}]-\text{Tr}[\boldsymbol{PQ}]=\text{Tr}\left[\left(\text{Tr}[\boldsymbol{P}]\boldsymbol{I}-\boldsymbol{P}\right)\boldsymbol{Q}\right]\geq0
\]
where Lemma \ref{lem:PSD_trace} is applied based on the fact that $\text{Tr}[\boldsymbol{P}]\boldsymbol{I}-\boldsymbol{P}$ is positive-semidefinite.
\end{proof}
\begin{lem}
Let $\boldsymbol{A},\boldsymbol{B}$ be matrices, and $\boldsymbol{AB}$ be the matrix product of $\boldsymbol{A}$ and $\boldsymbol{B}$. Then 
\begin{equation}
\Vert\boldsymbol{AB}\Vert_{2}\leq\Vert\boldsymbol{A}\Vert_{2}\Vert\boldsymbol{B}\Vert_{2}.
\end{equation}
\end{lem}
\begin{proof}
The tensor 2-norm of matrix $\boldsymbol{A}$ is equivalent to $\Vert\boldsymbol{A}\Vert_{2}=\sqrt{\text{Tr}(\boldsymbol{A}\boldsymbol{A}^{T})}=\sqrt{\text{Tr}(\boldsymbol{A}^{T}\boldsymbol{A})}$. Thus we have
\begin{align*}
\Vert\boldsymbol{AB}\Vert_{2} & =\sqrt{\text{Tr}(\boldsymbol{A}\boldsymbol{B}\boldsymbol{B}^{T}\boldsymbol{A}^{T})}\\
 & =\sqrt{\text{Tr}(\boldsymbol{A}^{T}\boldsymbol{A}\boldsymbol{B}\boldsymbol{B}^{T})}\\
 & \leq\sqrt{\text{Tr}(\boldsymbol{A}^{T}\boldsymbol{A})\text{Tr}(\boldsymbol{B}\boldsymbol{B}^{T})}\\
 & =\Vert\boldsymbol{A}\Vert_{2}\Vert\boldsymbol{B}\Vert_{2}.
\end{align*}
where Lemma \ref{lem:PSD} is used in the second last line. The inequality becomes an equality iff $\boldsymbol{A}$ and $\boldsymbol{B}$ are both rank-1 matrices, and the right singular vector of $\boldsymbol{A}$ is the same as the left singular vector of $\boldsymbol{B}$.
\end{proof}
Since contraction of two tensors is equivalent to matrix mulitplication with some re-grouping of indices, we have:
\begin{cor}
\label{cor:contraction_2norm} Let $\boldsymbol{A},\boldsymbol{B}$ be tensors, and $\mathcal{C}(\boldsymbol{A},\boldsymbol{B})$ be a tensor contraction, then
\begin{equation}
\Vert\mathcal{C}(\boldsymbol{A},\boldsymbol{B})\Vert_{2}\leq\Vert\boldsymbol{A}\Vert_{2}\Vert\boldsymbol{B}\Vert_{2}.
\end{equation}
\end{cor}
Now consider a scalar-output general tensor contraction $\mathcal{C}(\boldsymbol{A}_{1},\boldsymbol{A}_{2},...,\boldsymbol{A}_{m})$ following the Einstein summation rule, and compute it by STC with the optimal sampling probability. Temporarily using $\bar{\boldsymbol{A}}$ instead of $|\boldsymbol{A}|$ as the element-wise absolute value of $\boldsymbol{A}$ to avoid confusion with the $\Vert\cdot\Vert$ symbol, we have the following variance upper bound:
\begin{thm}
(Eq. 8 in the main text) Consider a scalar-output tensor contraction $\mathcal{C}(\bar{\boldsymbol{A}}_{1},\bar{\boldsymbol{A}}_{2},...,\bar{\boldsymbol{A}}_{m})$, then STC with the optimal sampling probabilities has variance satisfying:
\begin{equation}
\mathrm{Var}\leq\left(\prod_{i=1}^{m}\Vert\boldsymbol{A}_{i}\Vert_{2}\right)^{2}\label{eq:main_eq8}
\end{equation}
\end{thm}
\begin{proof}
The variance of STC with the optimal sampling probability is given by
\[
\text{Var}\leq\langle o^{2}\rangle=\left(\mathcal{C}(\bar{\boldsymbol{A}}_{1},\bar{\boldsymbol{A}}_{2},...,\bar{\boldsymbol{A}}_{m})\right)^{2}
\]
To compute $\mathcal{C}(\bar{\boldsymbol{A}}_{1},\bar{\boldsymbol{A}}_{2},...,\bar{\boldsymbol{A}}_{m})$, we rewrite the contraction of multiple tensors as sequential contractions of two tensors: 
\begin{align*}
\bar{\boldsymbol{A}}_{12} & =\mathcal{C}_{1,2}(\bar{\boldsymbol{A}}_{1},\bar{\boldsymbol{A}}_{2}),\\
\bar{\boldsymbol{A}}_{123} & =\mathcal{C}_{12,3}(\bar{\boldsymbol{A}}_{12},\bar{\boldsymbol{A}}_{3}),\\
 & ...\\
\bar{\boldsymbol{A}}_{123...m} & =\mathcal{C}_{123...m-1,m}(\bar{\boldsymbol{A}}_{123...m-1},\bar{\boldsymbol{A}}_{m}),
\end{align*}
which satisfies 
\[
\bar{A}_{123...m}=\mathcal{C}(\bar{\boldsymbol{A}}_{1},\bar{\boldsymbol{A}}_{2},...,\bar{\boldsymbol{A}}_{m}).
\]
Note that the above rewriting does not necessarily require the two input tensors to be connected by some indices. When they are not connected, the contraction simply becomes a tensor product, and the equations still hold. Since $\bar{A}_{123...m}$ is a scalar, we have
\[
\bar{A}_{123...m}=\Vert\bar{\boldsymbol{A}}_{123...m}\Vert_{2}.
\]
Then we sequentially apply Corollary \ref{cor:contraction_2norm}:
\begin{equation}
\begin{aligned}\Vert\bar{\boldsymbol{A}}_{123...m}\Vert_{2} & \leq\Vert\bar{\boldsymbol{A}}_{123...m-1}\Vert_{2}\Vert\bar{\boldsymbol{A}}_{m}\Vert_{2}\\
 & \leq\Vert\bar{\boldsymbol{A}}_{123...m-2}\Vert_{2}\Vert\bar{\boldsymbol{A}}_{m-1}\Vert_{2}\Vert\bar{\boldsymbol{A}}_{m}\Vert_{2}\\
 & \leq...\\
 & \leq\prod_{i=1}^{m}\Vert\bar{\boldsymbol{A}}_{i}\Vert_{2}\\
 & =\prod_{i=1}^{m}\Vert\boldsymbol{A}_{i}\Vert_{2},
\end{aligned}
\label{eq:general_STC_bound}
\end{equation}
The last step comes from the simple fact that $\Vert\boldsymbol{A}\Vert_{l}\equiv\Vert\bar{\boldsymbol{A}}\Vert_{l}$ for arbitrary $l$ and tensor $\boldsymbol{A}$. Combining all the derived conclusions, we finish the proof.
\end{proof}

\section{STC for general quantum chemistry tensor contractions}

\subsection{Variance of MP2 with uniform sampling}

We show that uniform sampling gives $O(N^{5})$ variance for the MP2 energy expression 
\begin{align*}
E & =\sum_{ijab}e_{ijab},\\
e_{ijab} & =\frac{1}{4}\frac{\left(V_{ijab}-V_{ijba}\right)^{2}}{e_{a}+e_{b}-e_{i}-e_{j}}
\end{align*}
in the limit of $N$ identical non-interacting subsystems, each with $O(1)$ size. For simplicity, let each subsystem have only one occupied orbital and one virtual orbital. In the non-interacting limit, each orbital is fully localized within one subsystem, so we use $i,j,a,b=1,2,...,N$ to indicate both the orbital and the subsystem it is in. The unbiased expectation from uniform sampling is given by 
\[
E=\left\langle o_{ijab}=N^{4}e_{ijab}\right\rangle _{i,j,a,b\sim\text{Uniform}}
\]
Clearly only when $i=j=a=b$, do we have $e_{ijab}\neq0$. Thus we can compute the variance as $\langle o^{2}\rangle-E^{2}$, where $\langle o^{2}\rangle$ is 
\begin{align*}
\langle o^{2}\rangle & =N^{8}\langle e_{ijab}^{2}\rangle_{i,j,a,b\sim\text{Uniform}}\\
 & =N^{4}\sum_{ijab}e_{ijab}^{2}\\
 & \sim O(N^{5})
\end{align*}
Since $E\sim O(N)$ is extensive, we have 
\[
\text{Var}=\langle o^{2}\rangle-E^{2}\sim O(N^{5})
\]

\subsection{Asymptotic forms of fundamental tensors for gapped systems}

We derive the asymptotic forms of the fundamental tensors $\boldsymbol{F}$ and $\boldsymbol{V}$. For a gapped system, the localized occupied or virtual orbitals $\phi_{p}$ have exponentially decaying tails around their centers $\boldsymbol{x}_{p}$: 
\[
\phi_{p}(\boldsymbol{x})\sim\exp(-|\boldsymbol{x}-\boldsymbol{x}_{p}|)
\]
Here and below, we are using $\sim\exp(-x)$ as an informal way to indicate that the decay is $\exp(-O(x))$, to avoid writing too many $O$ symbols.

For the two-electron integrals: 
\[
V_{pqrs}=\int\frac{\rho_{pr}(\boldsymbol{x}_{1})\rho_{qs}(\boldsymbol{x}_{2})}{\Vert\boldsymbol{x}_{1}-\boldsymbol{x}_{2}\Vert_{2}}d^{3}\boldsymbol{x}_{1}d^{3}\boldsymbol{x}_{2},
\]
where 
\[
\rho_{pq}(\boldsymbol{x})=\phi_{p}(\boldsymbol{x})\phi_{q}(\boldsymbol{x}).
\]
$\rho_{pq}(\boldsymbol{x})$ is exponentially small if $p$ are $q$ are distant: 
\[
\rho_{pq}(\boldsymbol{x})\sim\exp(-|\boldsymbol{x}-\boldsymbol{x}_{p}|)\exp(-|\boldsymbol{x}-\boldsymbol{x}_{q}|)\leq\exp(-|\boldsymbol{x}_{p}-\boldsymbol{x}_{q}|),
\]
where the $\leq$ relation comes from the triangle inequality. Thus $V_{pqrs}$ has exponential decaying dependence on $R_{pr}$ and $R_{qs}$. Now consider only close $(p,r)$ and $(q,s)$ pairs. Taking the pair $(p,r)$ as an example, $\rho_{pr}$ is peaked between $\boldsymbol{x}_{p}$ and $\boldsymbol{x}_{r}$ (which must be close for the pair density to be non-zero) thus we consider the single center denoted $\boldsymbol{x}_{pr}$.

In the long-range limit w.r.t. the distance between pairs, one can apply the multipole expansion: 
\begin{equation}
\int\frac{\rho_{pr}(\boldsymbol{x}_{1}-\boldsymbol{x}_{pr})\rho_{qs}(\boldsymbol{x}_{2}-\boldsymbol{x}_{qs})}{\Vert\boldsymbol{x}_{1}-\boldsymbol{x}_{2}\Vert_{2}}d^{3}\boldsymbol{x}_{1}d^{3}\boldsymbol{x}_{2}=\frac{Q_{pr}Q_{qs}}{R}+\frac{(Q_{pr}\boldsymbol{d}_{qs}-Q_{qs}\boldsymbol{d}_{pr})\cdot\hat{\boldsymbol{R}}}{R^{2}}+\frac{\boldsymbol{d}_{pr}\cdot\boldsymbol{d}_{qs}-3(\boldsymbol{d}_{pr}\cdot\hat{\boldsymbol{R}})(\boldsymbol{d}_{qs}\cdot\hat{\boldsymbol{R}})}{R^{3}}+...\label{eq:multipole}
\end{equation}
where $\boldsymbol{R}=(\boldsymbol{x}_{qs}-\boldsymbol{x}_{pr})$, $R=\Vert\boldsymbol{R}\Vert_{2}$, $\hat{\boldsymbol{R}}=\boldsymbol{R}/R$. The charge and dipole moment are 
\begin{align*}
Q_{pq} & =\int\phi_{p}(\boldsymbol{x})\phi_{q}(\boldsymbol{x})d^{3}\boldsymbol{x}=\delta_{pq}\\
\boldsymbol{d}_{pq} & =\int\boldsymbol{x}\phi_{p}(\boldsymbol{x})\phi_{q}(\boldsymbol{x})d^{3}\boldsymbol{x}
\end{align*}
Thus the leading order is $Q_{pq}$ when $p=q$, otherwise it is $\boldsymbol{d}_{pq}$. Thus the final asymptotic form is 
\begin{equation}
V_{pqrs}\sim\exp(-R_{pr})\exp(-R_{qs})R_{pq}^{-\alpha},\alpha=3-\delta_{pr}-\delta_{qs},\label{eq:asymptotic_V}
\end{equation}
where we write $R$ as $R_{pq}$, as the elements are always exponentially small in the asymptotic limit except when $R\approx R_{pq}\approx R_{ps}\approx R_{rq}\approx R_{rs}$.

In Hartree--Fock, the Fock matrix is constructed as 
\[
F_{pq}=h_{pq}+\sum_{rs}(V_{prqs}-V_{prsq})D_{rs},
\]
where $h_{pq}$ is the single-electron Hamiltonian, and $D_{rs}$ is the one-electron reduced density matrix, and both of them decay exponentially in the spatial separation of the indices. To estimate the second term, we introduce a simple equivalence relation $p\sim q$ if the given value is exponentially decaying with $R_{pq}$. It satisfies reflexivity, symmetry, and transitivity. In the expression $V_{prqs}D_{rs}$, we have $p\sim q$, $s\sim r$, and in the expression $V_{prsq}D_{rs}$, we have $p\sim s,q\sim r$ from $V$ and $r\sim s$ from $D$. So both terms give $p\sim r$, which means that 
\[
\sum_{rs}(V_{prqs}-V_{prsq})D_{rs}\sim\exp(-R_{pq})
\]
Finally, we can conclude: 
\begin{equation}
F_{pq}\sim\exp(-R_{pq}).\label{eq:asymptotic_F}
\end{equation}

\subsection{Tensor norms}

For the fundamental tensors $\boldsymbol{F}$ and $\boldsymbol{V}$, we can directly compute their asymptotic tensor $p$-norm. Let \textbf{$d$ }be the system dimension. We note that an exponential decay in $R$ integrates over $R$ to a constant, regardless of $d$, i.e. 
\[
\int\exp(-R)R^{d-1}dR\sim O(1),
\]
thus the $d$ dependence only manifests for the polynomial part. We compute the norm for general $\alpha=1,2,3$: 
\begin{align*}
\Vert\boldsymbol{F}\Vert_{p} & \sim N^{1/p},\\
\Vert\boldsymbol{V}\Vert_{p} & \sim\left(N\int_{0}^{R_{0}}R^{-p\alpha}R^{d-1}dR\right)^{1/p}\sim\left(Nf_{d-p\alpha}(R_{0})\right)^{1/p},\\
f_{\beta}(R_{0}) & =\begin{cases}
0 & \beta<0\\
\log R_{0}\sim\log N & \beta=0\\
R_{0}^{\beta}\sim N^{\beta/d} & \beta>0
\end{cases}
\end{align*}
where $R_{0}\sim N^{1/d}$ is the system lengthscale. For $p=2$, we have 
\begin{align*}
\Vert\boldsymbol{F}\Vert_{2} & \sim N^{1/2},\\
\Vert\boldsymbol{V}\Vert_{2} & \sim\begin{cases}
N^{2/3} & d=3,\alpha=1\\
(N\log N)^{1/2} & d=2,\alpha=1\\
N^{1/2} & \text{otherwise}
\end{cases}
\end{align*}
Neglecting the diagonal elements of $\boldsymbol{V}$ (i.e. considering only the $p\neq r,q\neq s$ part), both $\boldsymbol{F}$ and $\boldsymbol{V}$ always have a 2-norm that scales as $N^{1/2}$.

Combining with Eq. \ref{eq:main_eq8}, this gives the general basis-independent upper bound of the STC variance.

\subsection{Radial sign-freeness of quantum chemistry tensors}

Consider the first three leading orders in the multipole expansion Eq. \ref{eq:multipole}. For each order, the numerator is determined by $Q_{pr},Q_{qs},\boldsymbol{d}_{pr},\boldsymbol{d}_{ps}$, and the relative orientation $\hat{\boldsymbol{R}}$ between pairs $(pr),(qs)$, while the denominator only depends on the radial distance $R$ between pairs. Using an ``internal coordinate system'', where $\boldsymbol{R}_{pr},\boldsymbol{R}_{qs},\boldsymbol{R}_{pq}$ are independent variables, we can generally write 
\begin{equation}
V_{pqrs}\simeq f_{\text{sign}}(\boldsymbol{R}_{pr},\boldsymbol{R}_{qs},\hat{\boldsymbol{R}}_{pq})g_{\text{pos}}(R_{pq})\label{eq:signfree}
\end{equation}
where we introduce a new symbol $\simeq$ (not to be confused with $\sim$) defined as 
\[
x(R)\simeq y(R)\text{ if }\lim_{R\rightarrow\infty}\frac{x}{y}=1,
\]
which is a stricter relation than $\sim$, and $g_{\text{pos}}$ accounts for the long-range algebraically decaying part w.r.t. $R_{pq}$: 
\[
g_{\text{pos}}(R_{pq})\sim\text{poly}(R_{pq}^{-1}),
\]
Intuitively, this divides $V_{pqrs}$ into the product of a short-range (in $\boldsymbol{R}_{pr}$, $\boldsymbol{R}_{qs}$) part $f_{\text{sign}}$ with possible signs, and a long-range (in the radial distance $R_{pq}$) part $g_{\text{pos}}$ that is sign-free. We refer to this property as radial sign-freeness.

\subsection{Extension to higher-order tensors}

Now we extend the asymptotic forms to higher-order quantum chemistry tensors. Such higher order tensors can naturally appear from the tensor contractions of simpler tensors. For example, the following 6-index tensor implicitly appears in the (T) expression: 
\[
T_{ijkabc}^{(3)}=\sum_{f}T_{ijaf}V_{fkbc}.
\]
It has the asymptotic form 
\[
T_{ijkabc}^{(3)}\sim\exp(-R_{ia})\exp(-R_{jb})\exp(-R_{kc})R_{ij}^{-3}R_{jk}^{-2}.
\]
Generally quantum chemistry tensors have $2n$ indices. We can assign these into $n$ pairs $(\alpha_{1},\beta_{1}),(\alpha_{2},\beta_{2}),...,(\alpha_{n},\beta_{n})$, such that, in a local basis, $T_{\alpha_{1}\beta_{1},...,\alpha_{n}\beta_{n}}$ has exponential decay in the intra-pair distances $R_{\alpha_{i}\beta_{i}},1\leq i\leq n$, and algebraic decay in the inter-pair distances $R_{\alpha_{i}\alpha_{j}},1\leq i,j\leq n$. 
\begin{equation}
T_{\alpha_{1}\beta_{1},...,\alpha_{n}\beta_{n}}\sim\left(\prod_{1\leq i\leq n}\exp(-R_{\alpha_{i}\beta_{i}})\right)\left(\prod_{1\leq i<j\leq n}R_{\alpha_{i}\alpha_{j}}^{-\gamma_{ij}}\right),\label{eq:general_tensor_asymptotic}
\end{equation}
where $\gamma_{ij}\geq0$ are integer algebraic exponents.

To show that all quantum chemistry tensors have such asymptotic behavior, we only need to show (1) the fundamental tensors have such asymptotic behavior, and (2) such an asymptotic form is closed under tensor contraction, i.e. let $\boldsymbol{A},\boldsymbol{B}$ be two tensors with such asymptotic behavior, and $\mathcal{C}$ be arbitrary tensor contraction, then $\mathcal{C}(\boldsymbol{A},\boldsymbol{B})$ also has such an asymptotic form. We have already verified (1) previously. To show (2), we note that $\mathcal{C}(\boldsymbol{A},\boldsymbol{B})$ is equivalent to contracting some pair of inner indices in $\boldsymbol{A}\otimes\boldsymbol{B}$. Let $\boldsymbol{A}\otimes\boldsymbol{B}$ have indices $(\alpha_{1},\beta_{1}),(\alpha_{2},\beta_{2}),...,(\alpha_{n},\beta_{n})$, with the asymptotic form in Eq. \ref{eq:general_tensor_asymptotic}. There are two possible ways to contract a pair of indices: either the two indices are already in a pair (say $\alpha_{n}$ and $\beta_{n}$), or they are in different pairs (say $\beta_{n-1}$ and $\alpha_{n}$). The former case leads to the disappearance of such a pair in the new tensor, without changing other intra-pair or inter-pair asymptotic forms: 
\begin{align*}
T_{\alpha_{1}\beta_{1},...,\alpha_{n}\beta_{n}}^{\prime} & \equiv\sum_{\alpha_{n}=\beta_{n}}T_{\alpha_{1}\beta_{1},...,\alpha_{n}\beta_{n}}\\
 & \sim\left(\prod_{1\leq i\leq n-1}\exp(-R_{\alpha_{i}\beta_{i}})\right)\left(\prod_{1\leq i<j\leq n-1}R_{\alpha_{i}\alpha_{j}}^{-\gamma_{ij}}\right).
\end{align*}
The latter case results in a new pair $(\alpha_{n-1},\beta_{n})$, and the updated asymptotic form 
\begin{align*}
T_{\alpha_{1}\beta_{1},...,\alpha_{n-2}\beta_{n-2},\alpha_{n-1}\beta_{n}}^{\prime} & \equiv\sum_{\beta_{n-1}=\alpha_{n}}T_{\alpha_{1}\beta_{1},...,\alpha_{n}\beta_{n}}\\
 & \sim\left(\prod_{1\leq i\leq n-2}\exp(-R_{\alpha_{i}\beta_{i}})\right)\exp(-R_{\alpha_{n-1}\beta_{n}})\left(\prod_{1\leq i<j\leq n-2}R_{\alpha_{i}\alpha_{j}}^{-\gamma_{ij}}\right)\prod_{1\leq i\leq n-2}R_{\alpha_{i}\alpha_{n-1}}^{\gamma_{i,n-1}+\gamma_{i,n}}.
\end{align*}
Thus it results in a change of the $\gamma$ matrix: 
\begin{equation}
\gamma_{i,n-1}\rightarrow\gamma_{i,n-1}+\gamma_{i,n},\label{eq:gamma_update}
\end{equation}
followed by deleting the last row and column. Clearly, both cases remain in the form in Eq. \ref{eq:general_tensor_asymptotic}, thus such an asymptotic form is closed under tensor contraction.

The radial sign-freeness still holds for higher-order tensors: in the asymptotic limit, $T_{\alpha_{1}\beta_{1},...,\alpha_{n}\beta_{n}}$ can be separated as 
\begin{equation}
T_{\alpha_{1}\beta_{1},...,\alpha_{n}\beta_{n}}\simeq f_{\text{sign}}\left(\{\boldsymbol{R}_{\alpha_{i}\beta_{i}},1\leq i\leq n\},\{\hat{\boldsymbol{R}}_{\alpha_{i}\alpha_{j}},1\leq i,j\leq n\}\right)g_{\text{pos}}(\{R_{\alpha_{i}\alpha_{j}}\}),\label{eq:general_tensor_signfree}
\end{equation}
where 
\[
g_{\text{pos}}(\{R_{\alpha_{i}\alpha_{j}}\})\sim\prod_{1\leq i,j\leq n}\text{poly}(R_{\alpha_{i}\alpha_{j}}).
\]
One can similarly show that such behavior is closed under tensor contraction.

\subsection{Systematic treatment of $1/\boldsymbol{\Delta}$ tensors }\label{subsec:denominator}

The $1/\boldsymbol{\Delta}$ tensors generally appear in tensor contractions obtained from perturbation theory (for example the representative (T) expression shown in the main text). They are diagonal tensors with elements
\begin{align*}
\Delta_{ia}^{(1)} & =e_{a}-e_{i},\\
\Delta_{ijab}^{(2)} & =e_{a}+e_{b}-e_{i}-e_{j},\\
 & ...
\end{align*}
If one naively views them as tensors similar to $\boldsymbol{F}$ and $\boldsymbol{V}$, we will find that they break the Einstein summation rule, and have large tensor norms. However, we can simply view them as importance sampling weights that can be on-the-fly-computed. Now we show, for gapped systems, such treatment always leads to
\[
\exp(\Delta F)\sim O(1).
\]

Generally, following the STC protocol, let 
\[
\tilde{p}^{\text{opt}}=(1/\boldsymbol{\Delta})\boldsymbol{A}\boldsymbol{B}...
\]
be the optimal sampling probability. Viewing all (possibly multiple) $1/\boldsymbol{\Delta}$ as weights is equivalent to creating an ``ideal probability'' by removing all $1/\boldsymbol{\Delta}$ in $\tilde{p}^{\text{opt}}$. This gives
\[
\tilde{p}^{\text{ideal}}=\boldsymbol{A}\boldsymbol{B}...
\]
For gapped systems, we have $e_{a}-e_{i}\geq\text{const.}>0$, i.e. elements of \textbf{$1/\boldsymbol{\Delta}$ }are upper bounded. On the other hand, $e_{a}$ and $e_{i}$ are eigenvalues of the $\boldsymbol{F}$ tensor that follows the Eq. \ref{eq:asymptotic_F} asymptotic behavior, thus $e_{a}-e_{i}$ is also lower bounded, if the system energy is extensive. This means that, there exists system-size independent constants $C_{1}$ and $C_{2}$, such that
\[
C_{1}\leq\frac{\tilde{p}^{\text{ideal}}}{\tilde{p}^{\text{opt}}}\leq C_{2}.
\]
Therefore the partition functions satisfy
\[
\frac{Z^{\text{ideal}}=\sum\tilde{p}^{\text{ideal}}}{Z^{\text{opt}}=\sum\tilde{p}^{\text{opt}}}\leq C_{2}.
\]
The normalized probabilities thus satisfy
\[
\frac{p^{\text{ideal}}}{p^{\text{opt}}}=\frac{\tilde{p}^{\text{ideal}}}{\tilde{p}^{\text{opt}}}\times\frac{Z^{\text{opt}}}{Z^{\text{ideal}}}\geq\frac{C_{1}}{C_{2}}.
\]
From Theorem \ref{thm:free_energy_thm1}, we can now conclude that
\[
\exp(\Delta F)\leq\frac{C_{2}}{C_{1}}\sim O(1)
\]
which finishes the proof. Note that the above conclusion holds in an arbitrary basis.

\subsection{Variance of local STC with optimal sampling}

Now we prove that, for \textbf{arbitrary} tensor contractions involving \textbf{arbitrary} numbers of quantum chemistry tensors of \textbf{arbitrary} orders, local STC with optimal sampling (assuming the sampling can be performed for loopy contractions), gives $O(1)$ relative variance.

We have already shown that, the relative variance of the optimal sampling is strictly 
\begin{equation}
\mathrm{RelVar}\equiv\frac{\text{Var}}{S^{2}}=\left(\frac{\mathcal{C}(|\boldsymbol{A}|,|\boldsymbol{B}|,...)}{\mathcal{C}(\boldsymbol{A},\boldsymbol{B},...)}\right)^{2}-1.\label{eq:relvar}
\end{equation}
Now consider tensor contractions $\mathcal{C}(\boldsymbol{A},\boldsymbol{B},...)$ and $\mathcal{C}(|\boldsymbol{A}|,|\boldsymbol{B}|,...)$. We have already discussed that quantum chemistry tensor indices can be divided into pairs, yielding certain asymptotic properties. Now we consider the product of the tensors, without summing over any indices, i.e. 
\[
\boldsymbol{T}=\boldsymbol{A}\boldsymbol{B}...,
\]

The indices of \textbf{$\boldsymbol{T}$} can be paired as $(\alpha_{1},\beta_{1}),(\alpha_{2},\beta_{2}),...,(\alpha_{n},\beta_{n})$, to yield the same asymptotic behavior as in Eq. \ref{eq:general_tensor_asymptotic} and Eq. \ref{eq:general_tensor_signfree}. $\mathcal{C}(\boldsymbol{A},\boldsymbol{B},...)$ and $\mathcal{C}(|\boldsymbol{A}|,|\boldsymbol{B}|,...)$ now become 
\begin{align*}
\mathcal{C}(\boldsymbol{A},\boldsymbol{B},...) & =\sum_{\alpha_{1}\beta_{1}...\alpha_{n}\beta_{n}}T_{\alpha_{1}\beta_{1},...,\alpha_{n}\beta_{n}}\\
\mathcal{C}(|\boldsymbol{A}|,|\boldsymbol{B}|,...) & =\sum_{\alpha_{1}\beta_{1}...\alpha_{n}\beta_{n}}|T_{\alpha_{1}\beta_{1},...,\alpha_{n}\beta_{n}}|
\end{align*}
We create the following shorthand for $f_{\text{sign}}$ and $g_{\text{pos}}$: 
\begin{align*}
f_{\text{sign}}\left(\{\boldsymbol{R}_{\alpha_{i}\beta_{i}},1\leq i\leq n\},\{\hat{\boldsymbol{R}}_{\alpha_{i}\alpha_{j}},1\leq i,j\leq n\}\right) & \rightarrow f_{\text{sign}}(\beta_{1}...\beta_{n};\alpha_{1}...\alpha_{n}),\\
g_{\text{pos}}(\{R_{\alpha_{i}\alpha_{j}}\}) & \rightarrow g_{\text{pos}}(\alpha_{1}...\alpha_{n}).
\end{align*}
Using the positivity of the $g_{\text{pos}}$, we can rewrite $\mathcal{C}(\boldsymbol{A},\boldsymbol{B},...)$ and $\mathcal{C}(|\boldsymbol{A}|,|\boldsymbol{B}|,...)$ as: 
\begin{align*}
\mathcal{C}(\boldsymbol{A},\boldsymbol{B},...) & =Z_{g}\left\langle \sum_{\beta_{1}...\beta_{n}}f_{\text{sign}}(\beta_{1}...\beta_{n};\alpha_{1}...\alpha_{n})\right\rangle _{\alpha_{1}...\alpha_{n}\sim g_{\text{pos}}}\\
\mathcal{C}(|\boldsymbol{A}|,|\boldsymbol{B}|,...) & =Z_{g}\left\langle \sum_{\beta_{1}...\beta_{n}}\left|f_{\text{sign}}(\beta_{1}...\beta_{n};\alpha_{1}...\alpha_{n})\right|\right\rangle _{\alpha_{1}...\alpha_{n}\sim g_{\text{pos}}}
\end{align*}
where $Z_{g}$ is the partition function of the long-range part: 
\[
Z_{g}=\sum_{\alpha_{1}...\alpha_{n}}g_{\text{pos}}(\alpha_{1}...\alpha_{n})
\]
For any given $\alpha_{1}...\alpha_{n}$, the $f_{\text{sign}}$ part is fully short-ranged w.r.t. $\{R_{\alpha_{i}\beta_{i}}\}$, thus 
\begin{align*}
\sum_{\beta_{1}...\beta_{n}}f_{\text{sign}}(\beta_{1}...\beta_{n};\alpha_{1}...\alpha_{n}) & \sim O(1)\\
\sum_{\beta_{1}...\beta_{n}}\left|f_{\text{sign}}(\beta_{1}...\beta_{n};\alpha_{1}...\alpha_{n})\right| & \sim O(1)
\end{align*}
$f_{\text{sign}}$ depends on the system structure. Let us assume additionally that the sign contributions do not vanish as we increase the system size, or more precisely, that $|\sum_{\beta_{1}...\beta_{n}}f_{\text{sign}}(\beta_{1}...\beta_{n};\alpha_{1}...\alpha_{n})|\geq\kappa$, where $\kappa$ is a system size independent constant. Then we have 
\[
\frac{\mathcal{C}(|\boldsymbol{A}|,|\boldsymbol{B}|,...)}{\mathcal{C}(\boldsymbol{A},\boldsymbol{B},...)}\sim\frac{Z_{g}\times O(1)}{Z_{g}\times O(1)}\sim O(1).
\]
Note that the above derivation does not constrain the exact polynomial form of $g_{\text{pos}}$, and we only require it to be positive, so that it can be interpreted as a probability. Combined with Eq. \ref{eq:relvar}, now we can conclude that the relative variance of local STC with the optimal importance sampling is always $O(1)$, regardless of the tensor contraction form.

\subsection{Variance of local STC with the loop-breaking strategy }\label{subsec:local_loopy}

Now we consider an arbitrary connected, scalar-output tensor contraction $\mathcal{C}(\boldsymbol{A}_{1},\boldsymbol{A}_{2},...)$ that possibly contains loops, and the tensors $\boldsymbol{A}_{1},\boldsymbol{A}_{2},...$ follow the asymptotic forms in Eq. \ref{eq:general_tensor_asymptotic}. If we restrict the summation to off-diagonal elements, i.e. $\alpha_{i}\neq\beta_{i}$, then the leading order polynomial scaling in $\boldsymbol{V}$ (see Eq. \ref{eq:asymptotic_V}) is $R^{-3}$. We first prove that, with the off-diagonal restriction, STC with the loop-breaking strategy gives 
\[
\exp(\Delta F)\leq O(\text{polylog}(N))
\]
for arbitrary loopy tensor contractions.

Since contraction of tensors can only increase but not decrease the polynomial exponent (see Eq. \ref{eq:gamma_update}), we have $\gamma_{ij}\geq3$ for each connected pair.

\begin{figure}[tbh]
\centering 

\includegraphics[width=0.9\columnwidth]{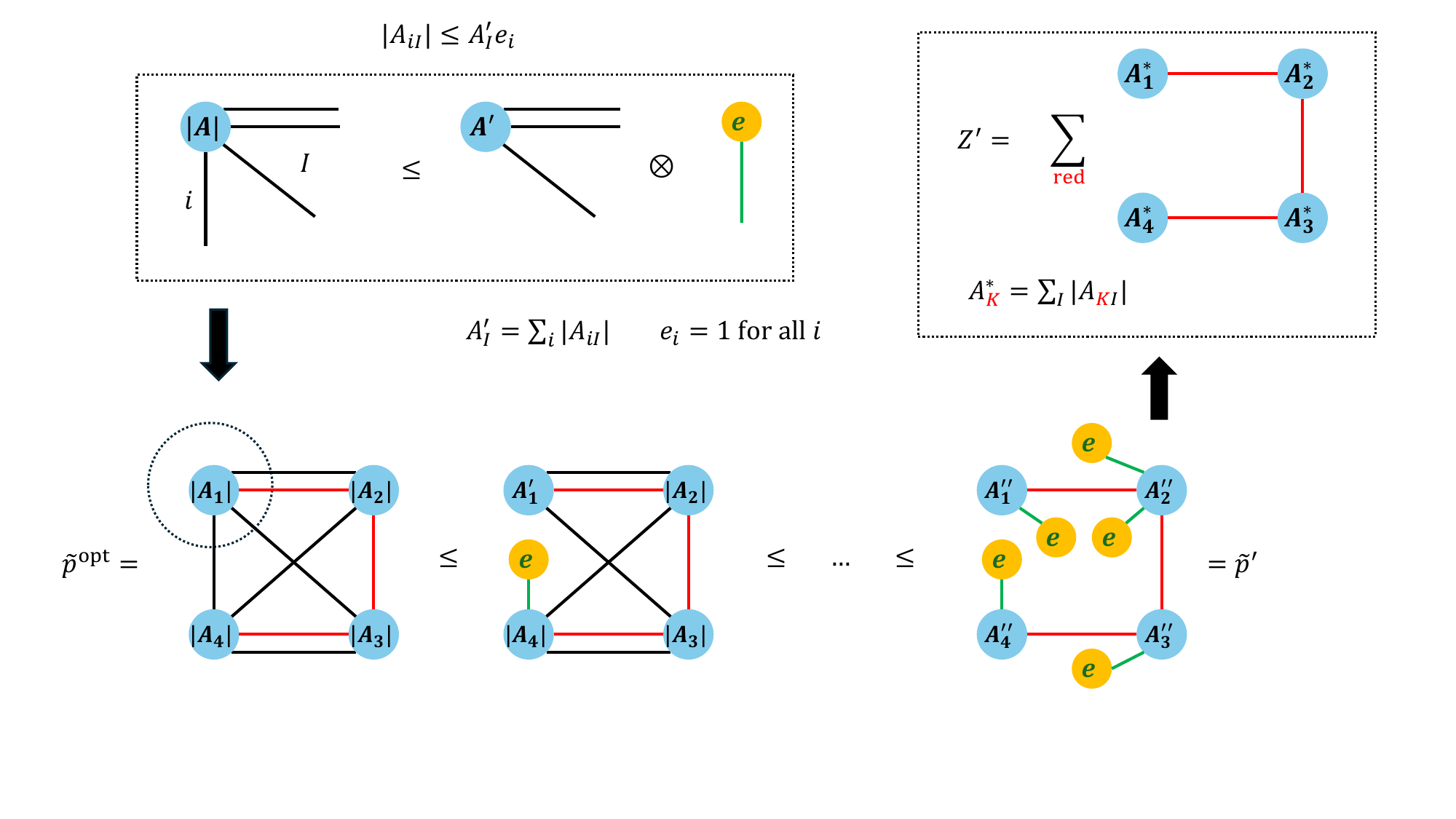}

\caption{The loop breaking strategy for general loopy tensor contractions.}

\label{fig:loop_breaking}
\end{figure}

Following the main text and previous discussions, the optimal sampling probability is
\[
\tilde{p}^{\text{opt}}=|\boldsymbol{A}_{1}\boldsymbol{A}_{2}...|.
\]
In the main text, we state the loop-breaking strategy, that is to sequentially choose some tensor $\boldsymbol{A}$, divide the indices to two groups, and apply 
\[
|\boldsymbol{A}|\leq\boldsymbol{P}\otimes\boldsymbol{Q},
\]
until all loops are broken. Then we construct the $\tilde{p}^{\prime}$ using the decomposed tensors. Now we present a simple choice of decomposition that leads to $\exp(\Delta F)=\frac{Z^{\prime}}{Z^{\text{opt}}}\leq\text{polylog}(N)$, with a schematic diagram show in Fig. \ref{fig:loop_breaking}.

For an arbitrary loopy tensor contraction, we first order the tensors in a chain as $\boldsymbol{A}_{1},...,\boldsymbol{A}_{m}$, such that $\boldsymbol{A}_{k}$ and $\boldsymbol{A}_{k+1}$ are connected by at least one index for each $1\leq k\leq m-1$. Let the index connecting $\boldsymbol{A}_{k}$ and $\boldsymbol{A}_{k+1}$ be $i_{k}$ (red indices in the figure). We will name things according to the figure, e.g. we refer to red indices and black indices. Now we recursively apply the tensor decomposition by the following procedure: choose any tensor connected to at least one black index, say tensor $\boldsymbol{A}$ and index $i$. Let $I$ be the set of all other indices of $\boldsymbol{A}$, now we apply the decomposition: 
\[
|\boldsymbol{A}|\leq\boldsymbol{A}^{\prime}\otimes\boldsymbol{e},
\]
where 
\begin{align}
A_{I}^{\prime} & =\sum_{i}|A_{iI}|,\label{eq:tensor_decomposition}\\
e_{i} & =1\text{ for all }i.\nonumber 
\end{align}
Clearly, the chosen values of $\boldsymbol{A}^{\prime}$ and $\boldsymbol{e}$ satisfy the required inequality. We also note that we can use the better decomposition:
\begin{equation}
\begin{aligned}A_{I}^{\prime} & =\max_{i}|A_{iI}|,\\
e_{i} & =1\text{ for all }i.
\end{aligned}
\label{eq:tensor_decomposition_better}
\end{equation}
although this does not lead to any change in scaling except when diagonal-elements are considered (see below). After the decomposition, we label the index $i$ as a green index, and put $\boldsymbol{A}^{\prime}$ and $\boldsymbol{e}$ back into the tensor network. We recursively apply the above decomposition until all black indices are converted to green indices. Now the tensor network becomes a tree tensor network, as the green indices only connect to a single dangling $\boldsymbol{e}$ tensor, and the red indices themselves form a chain. We choose our approximate sampling probability $\tilde{p}^{\prime}$ as the product of these new tensors $\boldsymbol{A}_{1}^{\prime\prime},...,\boldsymbol{A}_{m}^{\prime\prime}$, and the attached $\boldsymbol{e}$ tensors. Clearly $\tilde{p}^{\prime}$ supports efficient sampling (i.e. one-time cost proportional to the tensor sizes with a $O(1)$ per-sample cost). According to Eq. 6 in the main text, the relative variance using $\tilde{p}^{\prime}$ is fully determined by $\exp(\Delta F)=Z^{\prime}/Z^{\text{opt}}$. To compute $Z^{\prime}$, we contract the $\boldsymbol{e}$ tensors into the $\boldsymbol{A}^{\prime\prime}$ tensors to obtain $\boldsymbol{A}^{*}$ tensors. For any original tensor $\boldsymbol{A}$, let the set of black indices be $I$, and red indices be $K$, then $\boldsymbol{A}^{*}$ tensor is related to $\boldsymbol{A}$ by
\[
\boldsymbol{A}_{K}^{*}=\sum_{I}|\boldsymbol{A}_{KI}|.
\]
Now we estimate the magnitudes of $\boldsymbol{A}_{K}^{*}$ using the asymptotic form in Eq. \ref{eq:general_tensor_asymptotic} as well as the $\gamma\geq3$ (off-diagonal) restriction. Note that $K$ contains either one or two indices, corresponding to the two tensors at the two ends of the chain, or other tensors in the middle of the chain, respectively. Thus we can estimate the magnitudes by summing the indices in Eq. \ref{eq:general_tensor_asymptotic} until one or two indices are left. Generally, summing over an index that has a short-range connection to another index gives a $O(1)$ factor, and summing over an index that has either long-range or no connections with any other indices, gives a factor of up to 
\begin{equation}
\int_{0}^{N^{1/d}}R^{-\gamma}R^{d-1}dR\sim\begin{cases}
O(\log N) & \gamma=d=3\\
O(1) & \text{otherwise}
\end{cases}\label{eq:asymptotic_integral}
\end{equation}
where $d$ is the system dimensionality. In the worse case, the summation generates a $\text{polylog}(N)$ factor. Thus, if $\boldsymbol{A}^{*}$ has one index (i.e. at the end of the chain), then 
\begin{align*}
A_{k}^{*} & \sim\text{polylog}(N)\text{ for all }k
\end{align*}
otherwise it takes one of the following forms: 
\begin{align*}
A_{kl}^{*} & \sim R_{kl}^{-\gamma}\text{polylog}(N)\\
A_{kl}^{*} & \sim\exp(-R_{kl})\text{polylog}(N)
\end{align*}
depending on how the indices $k,l$ appear in the original tensor. With the $\gamma\geq3$ restriction, even the worst case only gives  extra $\text{polylog}(N)$ factors when summing over the $\boldsymbol{A}^{*}$ tensors to obtain $Z^{\prime}$. Thus we conclude 
\[
Z^{\prime}\leq\begin{cases}
O(N\text{polylog}(N)) & d=3\\
O(N) & \text{otherwise}
\end{cases}
\]
Since $Z^{\text{opt}}$ is at least $O(N)$, we have
\[
\exp(\Delta F)=\frac{Z^{\prime}}{Z^{\text{opt}}}\leq\begin{cases}
O(\text{polylog}(N)) & d=3\\
O(1) & \text{otherwise}
\end{cases}
\]
which proves the claim in the main text.

Now we consider the effects of diagonal elements, which may lead to $\gamma=1,2$ in Eq. \ref{eq:general_tensor_asymptotic}, and can generally result in $\text{poly}(N)$ factors in Eq. \ref{eq:asymptotic_integral}. Consider the worst case of $d=3$. For $\gamma=1,2$, Eq. \ref{eq:asymptotic_integral} becomes 
\[
\int_{0}^{N^{1/3}}R^{-\gamma}R^{2}dR\sim O(N^{1-\gamma/3}).
\]
Using the better decomposition in Eq. \ref{eq:tensor_decomposition_better}, each index pair with a $\gamma=1$ connection to others (corresponding to $V_{pqpq}$ type elements) contributes at most a $N^{2/3}$ factor, while $\gamma=2$ (corresponding to $V_{pqps}$ or $V_{pqrq})$ contributes at most a $O(N^{1/3})$ factor. One can similarily derive the results in other dimensions. Thus, a loose upper bound for a tensor contraction with $K$ index pairs, with contribution of diagonal elements included, can be obtained by assuming all polynomial connections have $\gamma=1$ decay. This gives the following loose upper bound:

\begin{equation}
\exp(\Delta F)=\frac{Z^{\prime}}{Z^{\text{opt}}}\leq\begin{cases}
O\left(N^{2K/3}\text{polylog}(N)\right) & d=3\\
O\left(N^{K/2}\text{polylog}(N)\right) & d=2\\
O\left(\text{polylog}(N)\right) & d=1
\end{cases}\label{eq:local_STC_bound}
\end{equation}

However, now we show that in many cases, the scaling can be much better than Eq. \ref{eq:local_STC_bound} with some more careful tensor decompositions.

\begin{figure}[tbh]
\centering 

\includegraphics[width=0.6\columnwidth]{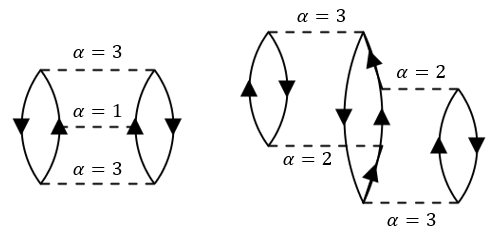}

\caption{Examples of tensor contractions (shown as Goldstone diagrams) involving diagonal blocks of $\boldsymbol{V}$, where using a careful loop-breaking strategy in local STC does not contribute a $\text{poly}(N)$ factor to $Z^{\prime}$ and $\exp(\Delta F)$. The first diagram is a third-order energy diagram in CCSD energy, while the second diagram is a fourth-order energy diagram. In the Goldstone diagram, each dashed line represents a $\boldsymbol{V}$ tensor. A $\text{poly}(N)$ factor can be generated for a $\alpha=1,2$ line, but only when cutting this line makes the diagram disconnected. In both shown diagrams, cutting all $\alpha=1,2$ lines does not make the diagram disconnected, thus we still have $Z^{\prime}\protect\leq O(N\text{polylog}(N))$, $\exp(\Delta F)\protect\leq\text{polylog}(N)$.}

\label{fig:feynman_diagram}
\end{figure}

As an example, we consider a tensor contraction $\mathcal{C}(\boldsymbol{V},\boldsymbol{A}_{1},\boldsymbol{A}_{2},...)$ involving a $\boldsymbol{V}$ tensor block with diagonal elements. Let its partition function be 
\[
Z=\mathcal{C}(|\boldsymbol{V}|,|\boldsymbol{A}_{1}|,|\boldsymbol{A}_{2}|,...).
\]
We now apply the decomposition 
\begin{align}
|V_{pqrs}| & \leq P_{pr}Q_{qs},\label{eq:cutting_Coulomb}
\end{align}
such that both $\boldsymbol{P}$ and $\boldsymbol{Q}$ are short-range (i.e. have exponential decaying). Now replace $\boldsymbol{V}$ with $\boldsymbol{P}$ and $\boldsymbol{\boldsymbol{Q}}$ in the tensor contraction, which results in $\mathcal{C}(\boldsymbol{P},\boldsymbol{Q},\boldsymbol{A}_{1},\boldsymbol{A}_{2},...)$. Note that the polynomial dacaying part of $\boldsymbol{V}$ does not exist anymore. Let the new partition function be $Z^{\prime}\geq Z$. As long as the new tensor contraction remains connected, we can now apply the same analysis above: if $\boldsymbol{A}_{1},\boldsymbol{A}_{2},...$ all satisfy $\gamma\geq3$, we have $Z^{\prime}\leq O(N\text{polylog}(N))$; otherwise each extra $\gamma=1,2$ in $\boldsymbol{A}_{1},\boldsymbol{A}_{2},...$ contributes at most $O(N^{2/3})$ and $O(N^{1/3})$ factors to $Z^{\prime}$, respectively. The above conclusion can be summarized by a simple diagrammatic argument: one can freely remove any polynomial connection without modifying the scaling, if cutting such a connection does not make the diagram disconnected. One can naturally extend this strategy for high-order tensors. Let tensor $\boldsymbol{T}$ have paired indices $(\alpha_{1},\beta_{1}),(\alpha_{2},\beta_{2}),...,(\alpha_{n},\beta_{n})$, one can divide the pairs into two groups, and apply the similar decomposition that keeps the intra-connection decay behavior.

In fact, for many of the tensor contractions appearing in perturbation theory, one can use such a strategy to construct $p^{\prime}$ with very low $Z^{\prime}$ and thus $\Delta F$. A few examples of tensor contractions that involve $\boldsymbol{V}$ diagonal element, but where one can still obtain $\exp(\Delta F)\leq\text{polylog}(N)$ with such careful decomposition strategies, are shown in Fig. \ref{fig:feynman_diagram}.

\subsection{Variance of Haar-random and canonical STC }\label{subsec:canonical}

We first provide a non-rigorous derivation, to show that, assuming the input tensors satisfy the asymptotic forms in Eq. \ref{eq:general_tensor_asymptotic}, applying STC with the loop-breaking strategy to the arbitrary contraction of $m$ tensors in the Haar-random basis gives $\Delta F\sim O(1)$.

The sketch of the derivation is that we first want to show that any tensor $\boldsymbol{T}$ after a Haar-random transformation, has elements with roughly equal magnitudes, which leads to 
\[
|\boldsymbol{T}|\approx\Vert\boldsymbol{T}\Vert_{2}\otimes_{i=1}^{n}\tilde{\boldsymbol{e}}_{i}
\]
where $\tilde{\boldsymbol{e}}_{i}$ is the normalized uniform vector in the space of index $i$, with all elements $N^{-1/2}$. Then we show this leads to $\Delta F\sim O(1)$.

We start by observing that for a Haar-random unitary $\boldsymbol{U}$, the individual elements $U_{ij}$ are very close to i.i.d. Gaussian random variables $N(0,\sigma^{2}=\frac{1}{N})$. Now we consider transforming a given tensor $\boldsymbol{T}$ with elements $T_{a_{1}...a_{n}}$ by applying the same Haar-random unitary to all indices, which gives a new $\boldsymbol{T}^{\prime}$ tensor with elements: 
\[
T_{i_{1}...i_{n}}^{\prime}=\sum_{a_{1}...a_{n}}U_{i_{1}a_{1}}...U_{i_{n}a_{n}}T_{a_{1}...a_{n}}.
\]
We write $|\boldsymbol{T}^{\prime}|$ as the sum of a uniform part and a remainder: 
\begin{align*}
|\boldsymbol{T}^{\prime}| & =c_{\boldsymbol{e}}\left(\otimes_{i=1}^{n}\tilde{\boldsymbol{e}}_{i}\right)+c_{\perp}\boldsymbol{X},\\
\boldsymbol{X} & \perp\left(\otimes_{i=1}^{n}\tilde{\boldsymbol{e}}_{i}\right)
\end{align*}
Then $c_{\boldsymbol{e}}$ is directly relate to the tensor 1-norm: 
\[
c_{\boldsymbol{e}}=N^{-n/2}\sum_{i_{1}...i_{n}}|T_{i_{1}...i_{n}}^{\prime}|.
\]

Now we treat $U_{i_{1}a_{1}},...,U_{i_{n}a_{n}}$ as i.i.d. Gaussian random variables. If $\{i_{1},i_{2},...\}$ are all distinct, the summation of each index $a_{k}$ is essentially a summation over $O(N)$ independent random variables. According to the central limit theorem, $T_{i_{1}...i_{n}}^{\prime}$ must asymptotically follow a Gaussian distribution at large $N$. Since the Haar random distribution has symmetric signs (i.e. if $\boldsymbol{U}$ is unitary, flipping the sign of one row/column also gives a unitary), we have $\langle T_{i_{1}...i_{n}}^{\prime}\rangle_{U\sim\text{Haar}}=0$, thus $T_{i_{1}...i_{n}}^{\prime}$ is a zero-centered Gaussian distribution. We can also easily find that, all different $T_{i_{1}...i_{n}}^{\prime}$ have the same variance without direct evaluation, again as long as $\{i_{1},i_{2},...\}$ are all distinct. Therefore, neglecting the $\boldsymbol{T}^{\prime}$ elements with identical indices (whose number, as a percentage of the total number of elements asymptotically becomes zero with $N$), all elements of $\boldsymbol{T}^{\prime}$ follow the same zero-centered Gaussian distribution. Note that we are not assuming these elements are independent. For a zero-centered Gaussian distribution $x\sim N(0,\sigma^{2})$, we always have
\[
\langle|x|\rangle=\sqrt{\frac{2}{\pi}}\sigma=\sqrt{\frac{2}{\pi}\langle x^{2}\rangle}
\]
Thus we can compute the expectation of $c_{\boldsymbol{e}}$ as
\[
\langle c_{\boldsymbol{e}}\rangle=N^{-n/2}\left\langle \sum_{i_{1}...i_{n}}|T_{i_{1}...i_{n}}^{\prime}|\right\rangle \geq N^{-n}\left\langle \sqrt{\frac{2}{\pi}\sum_{\text{distinct }i_{1}...i_{n}}\left\langle T_{i_{1}...i_{n}}^{\prime2}\right\rangle }\right\rangle =\left(1-O\left(\frac{1}{N}\right)\right)\left(\frac{2}{\pi}\right)^{n/2}\Vert\boldsymbol{T}^{\prime}\Vert_{2}.
\]
Thus $\boldsymbol{T}^{\prime}$ (for moderate $n$) is dominated by $\left(\otimes_{i=1}^{n}\tilde{\boldsymbol{e}}_{i}\right)$, which implies that $|\boldsymbol{T}^{\prime}|$ is almost a uniform tensor!

Now we can simply work out $\mathcal{C}(|\boldsymbol{A}_{1}|,|\boldsymbol{A}_{2}|,...,|\boldsymbol{A}_{m}|)$ with $K$ index pairs in total. For each $\boldsymbol{A}$, we can write 
\[
|\boldsymbol{A}|=\left(\frac{2}{\pi}\right)^{n}\Vert\boldsymbol{A}\Vert_{2}\otimes_{i\in I_{A}}\tilde{\boldsymbol{e}}_{i}.
\]
where $2n$ is the number of indices of $\boldsymbol{A}$. The contraction of each pair of $\tilde{\boldsymbol{e}}_{i}$ simply gives 1, i.e., $\tilde{\boldsymbol{e}}_{i}\cdot\tilde{\boldsymbol{e}}_{i}=1$. Thus we can estimate from the dominant contribution that: 
\[
\mathcal{C}(|\boldsymbol{A}_{1}|,|\boldsymbol{A}_{2}|,...)\approx\left(\frac{2}{\pi}\right)^{K}\prod_{i}\Vert\boldsymbol{A}_{i}\Vert_{2}.
\]
We can see it only differs by Eq. \ref{eq:general_STC_bound} by a constant factor $\left(\frac{2}{\pi}\right)^{K}$!

Once we establish that the $|\boldsymbol{A}|$ tensors are almost uniform tensors, it is easy to see that the loop-breaking strategy naturally gives $\exp(\Delta F)\sim O(1)$. This is simply because the uniform tensors themselves are already rank-1 tensors, thus there are in effect no loops at all.

Now we give an empirical argument that canonical STC should not be worse than Haar-random STC. For systems composed of non-interacting identical subsystems, the canonical basis become ill-defined, since the Fock matrix has $O(N)$ repeating eigenvalues, thus the canonical basis can freely switch between two limits: the local limit, and the Haar-random limit. Since we have already shown that STC leads to $\Delta F\leq\text{polylog}(N)$ in both limits (with diagonal elements neglected), we conjecture that, canonical STC also gives $\Delta F\leq\text{polylog}(N)$.

Finally we give a numerical demonstration of the scalings of Haar-random STC and canonical STC for the following two scalar-output tensor contractions: 
\begin{align*}
\mathcal{C}(\boldsymbol{V},\boldsymbol{T},\boldsymbol{T}) & =\sum_{ijkabc}V_{ijab}T_{jkbc}T_{kica},\\
\mathcal{C}(\boldsymbol{V},\boldsymbol{T},\boldsymbol{T},\boldsymbol{T}) & =\sum_{ijkabc}V_{ijab}T_{jkbc}T_{klcd}T_{lida}.
\end{align*}
We use the simplest tensor decomposition $|\boldsymbol{T}|\approx\otimes\tilde{\boldsymbol{e}}$ in the loop-breaking strategy, i.e. 
\[
|T_{jkbc}|\approx1\text{ for all }j,k,b,c.
\]
Empirically, we find the Haar-random scaling of both optimal sampling (labeled as opt) and the loop-breaking strategy (labeled as LB) gives approximately $O(N^{m})$ variance, where $m=3,4$ for the two contractions, respectively. For both contractions, the variance of Haar-random STC follows almost perfect polynomial scaling; the variance of canonical STC is not smooth, but always below the Haar-random STC.

\begin{figure}[tbh]
\centering
\includegraphics[width=0.9\columnwidth]{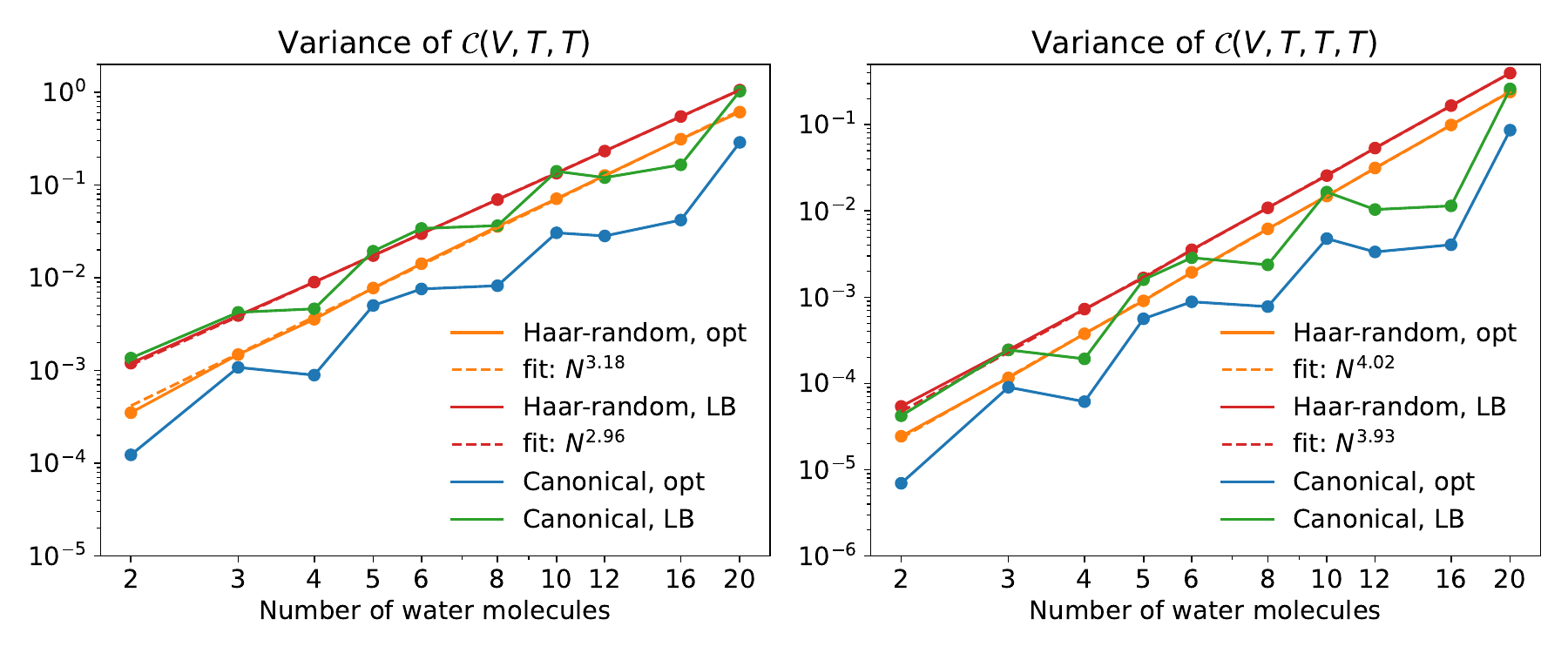}

\caption{Numerical STC variance scaling of two tensor contractions for Haar-random STC and canonical STC. Two strategies are considered, including optimal sampling (opt) and loop-breaking strategy (LB).}

\label{fig:Haar-canonical}
\end{figure}

\subsection{Summary of full STC variance scalings}

We have derived the effects of diagonal elements on variance scaling in different STC setups in the previous sections. Now we make a comprehensive summary of the upper bound of the STC relative scalings. However, the induced scaling increase from diagonal elements in diagrams strongly depends on the specific tensor contraction structure. For arbitrary-basis STC, a pair of indices that has a $\gamma=1$ connection contributes a $O(N^{1/3})$ polynomial factor in 3D systems. For local STC of loopy tensor contractions, using a naive tensor decomposition choice in the loop-breaking strategy, a pair of indices that has $\gamma=1,2$ connections contributes a $O(N^{1-\gamma/d})$ polynomial factor if $d>\gamma$. However, as mentioned in Sec. \ref{subsec:local_loopy}, for many tensor contraction structures, these factors can be removed or reduced with more careful tensor decompositions. 

For simplicity, we only show a very loose, and tensor-contraction-structure independent upper bound, and readers can refer to the above sections for tighter scaling for specific tensor contraction structures. Let the tensor contraction has $m$ tensors, $2K$ indices (i.e. $K$ pairs) in total, and let the system dimension be $d\in\{1,2,3\}$. Define

\[
f(N,K,d)=\begin{cases}
O\left(N^{K/3}\text{polylog}(N)\right) & d=3\\
O\left(\text{polylog}(N)\right) & d=2\\
O\left(1\right) & d=1
\end{cases}
\]
\[
g(N,K,d)=\begin{cases}
O\left(N^{2K/3}\text{polylog}(N)\right) & d=3\\
O\left(N^{K/2}\text{polylog}(N)\right) & d=2\\
O\left(\text{polylog}(N)\right) & d=1
\end{cases}
\]
A very loose but complete upper bound of relative variance scaling (which is the same as the computation cost to get fixed relative error in a scalar output), is shown in Table \ref{table:scaling}.

\begin{table}

\caption{Upper bound of relative variance scaling of scalar properties by STC in different bases for quantum chemistry tensor contractions, with all possibilities considered.}

\begin{tabular}{|c|c|c|c|c|}
\hline 
\multirow{1}{*}{Relative variance scaling} & Efficiently achievable & Arbitrary-basis STC & Local STC & Haar-random/canonical STC\tabularnewline
\hline 
\hline 
Optimal sampling & Tree contractions only & $O(N^{m-2})f(N,K,d)$ & $O(1)$ & $O(N^{m-2})f(N,K,d)$\tabularnewline
\hline 
Loop breaking strategy & Always & Unknown & $^{*}g(N,K,d)$ & $^{**}$$O(N^{m-2})f(N,K,d)$\tabularnewline
\hline 
\end{tabular}

$^{*}$: A very loose contraction-structure-independent upper bound. For most contractions, it can be greatly reduced by the strategy introduced in Sec. \ref{subsec:local_loopy}.

$^{**}$: Obtained from a non-rigorous estimation.

\label{table:scaling}
\end{table}

\section{STC-CCSD(T)}

\subsection{Details of the STC-CCSD algorithm}

\textbf{Density fitting}. Density fitting is a widely-used low-rank decomposition of the two-electron integrals, and is known to give small and acceptable errors for chemical quantities. In STC-CCSD, we use density fitting to reduce the memory requirements. The decomposition is given by 
\begin{equation}
V_{pqrs}\approx\sum_{\chi}P_{pr\chi}Q_{qs\chi}.\label{eq:DF}
\end{equation}
In practice, one typically chooses $P=Q$, and we write them as $R$ in such a case. As an example, we consider the contraction: 
\begin{align*}
S_{ijab} & =\sum_{cd}V_{abcd}T_{ijcd}\\
 & \approx\sum_{cd\chi}R_{ac\chi}R_{bd\chi}T_{ijcd}
\end{align*}
Although this looks like a loopy contraction, one can still achieve optimal sampling with low cost through
\begin{equation}
\tilde{p}_{ijabcd\chi}=|T_{ijcd}R_{ac\chi}R_{bd\chi}|=\tilde{p}(c,d,\chi)p(i,j|c,d)p(a|c,\chi)p(b|d,\chi).\label{eq:DF_prob}
\end{equation}
Let the number of occupied, virtual, and auxiliary orbitals be $N_{\text{o}},N_{\text{v}},N_{\text{aux}}$, the total time and memory cost of constructing all the probabilility tables and the partition function is $O(N_{v}^{2}N_{\text{aux}})+O(N_{o}^{2}N_{v}^{2})$. In practice this cost is much smaller than if using the full $V_{abcd}$ of size $O(N_{v}^{4})$.

\begin{figure}[tbh]
\centering
\includegraphics[width=0.9\columnwidth]{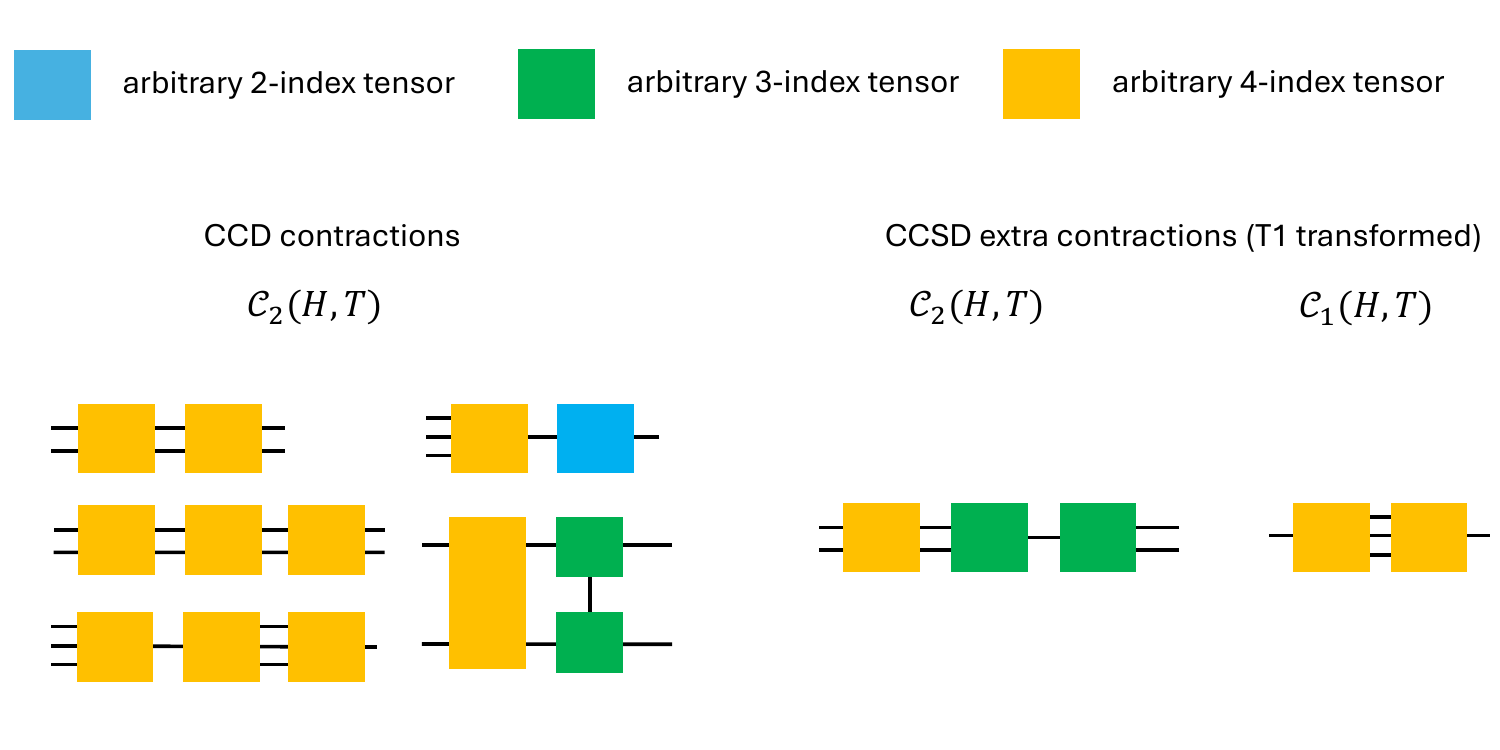}

\caption{Complete list of tensor contraction structures with cost higher than $O(N^{4})$ in STC-CCSD. }

\label{fig:CCSD}
\end{figure}


\textbf{STC-CCSD}. Before discussing the complicated CCSD equations, we first consider CCD as a simplified version of CCSD, i.e. $T_{ia}=0$ in CCSD. Let $\langle pq|rs\rangle=V_{pqrs}$, $\langle pq||rs\rangle=\langle pq|rs\rangle-\langle pq|sr\rangle$, the complete amplitude equation for CCD is 
\begin{equation}
\begin{aligned}\mathcal{C}_{\text{CCD}}(H,T) & =\langle ij||ab\rangle+P(ab)\sum_{c}f_{bc}T_{ijac}-P(ij)\sum_{k}f_{jk}T_{ikab}\\
 & +\frac{1}{2}\sum_{cd}\langle ab||cd\rangle T_{ijcd}+\frac{1}{2}\sum_{kl}\langle ij||kl\rangle T_{klab}+P(ab)P(ij)\sum_{kc}\langle kb||cj\rangle T_{ikac}\\
 & +\frac{1}{2}P(ij)P(ab)\sum_{klcd}\langle kl||cd\rangle T_{ikac}T_{ljdb}+\frac{1}{2}P(ij)\sum_{klcd}\langle kl||cd\rangle T_{ikcd}T_{ljab}\\
 & +\frac{1}{2}P(ab)\sum_{klcd}\langle kl||cd\rangle T_{klac}T_{ijdb}+\frac{1}{4}\langle kl||cd\rangle T_{ijcd}T_{klab}
\end{aligned}
\label{eq:full_CCD}
\end{equation}
where $f_{pq}=F_{pq}(1-\delta_{pq})$ is the off-diagonal part of the Fock matrix $F$. In the canonical basis, $f=0$. Following the main text, we only apply STC for those contractions with exact cost higher than $O(N^{4})$. Without differentiating between tensors with the same number of indices, there are in total 5 tensor contraction structures shown in Fig. \ref{fig:CCSD}. One structure involving 3-index tensors comes from applying density fitting to the $\sum_{cd}\langle ab||cd\rangle T_{ijcd}$ term. Now we consider CCSD. The extra tensor contractions in $\mathcal{C}_{\text{CC}}^{(2)}(\boldsymbol{H},\boldsymbol{T})$ can generally be obtained by $T_{ijab}\rightarrow T_{ia}T_{jb}$ for one or two $\boldsymbol{T}^{(2)}$ tensors in Eq. \ref{eq:full_CCD}, and keeping the connected diagrams. This could in principle generate contractions involving up to 5 tensors, which could lead to higher scaling of the canonical STC-CCSD. However, one can again can use the density fitting decomposition, and ``absorb'' the $\boldsymbol{T}^{(1)}$ tensors into the density fitting tensor. One example is 
\begin{align*}
S_{ijab} & =\sum_{klcd}V_{klcd}T_{id}T_{la}T_{jkbc}\\
 & =\sum_{klcd\chi}R_{kc\chi}R_{ld\chi}T_{id}T_{la}T_{jkbc}\\
 & =\sum_{kc\chi}\left(\sum_{ld}R_{ld\chi}T_{id}T_{la}\right)R_{kc\chi}T_{jkbc}.
\end{align*}
Such absorption of $\boldsymbol{T}^{(1)}$ tensors into the density fitting tensors has $O(N^{4})$ cost, and thus can be done deterministically without increasing the final deterministic scaling. One can apply a similar strategy to $\mathcal{C}_{\text{CC}}^{(1)}(\boldsymbol{H},\boldsymbol{T})$. The extra contractions in CCSD are shown in Fig. \ref{fig:CCSD}. Clearly there is only one contraction type involving a loop, which is the $S_{ijab}=\sum_{cd\chi}R_{ac\chi}R_{bd\chi}T_{ijcd}$ term (and some other contractions with the same structure), but we have already given the exact decomposition in Eq. \ref{eq:DF_prob}. Thus, all contractions in CCSD can be performed by optimal sampling, if STC is performed for the tensor-output contractions. For simplicity, for contractions with loopy structures, but which can still be efficiently sampled as in the case of Eq.~\ref{eq:DF_prob}, we will say they ``effectively'' have a tree structure.

\textbf{Loop-breaking strategy}. One can also perform STC by viewing the next-iteration's energy expression as a scalar-output contraction, which leads to a lower energy variance. The full CCSD energy expression is 
\[
E_{\text{CC}}=\frac{1}{4}\sum_{ijab}\langle ij||ab\rangle(T_{ijab}+T_{ia}T_{jb}-T_{ib}T_{ja}).
\]
Again, since $\boldsymbol{T}^{(2)}$ in practice dominates both the energy and its variance, we only consider applying the loop-breaking strategy to $\boldsymbol{T}^{(2)}$ here. Since $\boldsymbol{S}=-\mathcal{C}_{\text{CC}}^{(2)}(\boldsymbol{H},\boldsymbol{T})$ gives $S_{ijab}=\Delta_{ijab}T_{ijab}$, the related energy expression is essentially 
\[
E=-\frac{1}{4}\sum_{ijab}\left(A_{ijab}\mathcal{C}_{2}(\boldsymbol{H},\boldsymbol{T})_{ijab}\right),
\]
where we define an extra $\boldsymbol{A}$ tensor as 
\[
A_{ijab}=-\frac{\langle ij||ab\rangle}{4\Delta_{ijab}}.
\]
Adding such an extra $\boldsymbol{A}$ tensor in the tensor contraction generally makes the contraction loopy. Thus we apply the loop breaking strategy by decomposing the tensor\textbf{ $\boldsymbol{A}$}. We have already mentioned above that all terms in $\mathcal{C}_{\text{CC}}^{(2)}(\boldsymbol{H},\boldsymbol{T})_{ijab}$ have (or effectively have) tree structures. When adding the decomposed tensors of $\boldsymbol{A}$, they are independently connected to the original structure, thus all structures still (effectively) remain tree structures. Therefore, the tree sampling algorithm can be directly used. Naively, one can apply the decomposition in Eq. \ref{eq:tensor_decomposition_better} for local STC, or the uniform decomposition $A_{ijab}\approx1$ for canonical STC. In practice, with a careful choice of the decomposition, one can get a lower energy variance. However, for different contractions, the required decomposition structures can be different. Here we list some of the required decompositions to break loops in different contractions: 
\begin{align*}
|A_{ijab}| & \approx M_{ij}M_{ab}\\
|A_{ijab}| & \approx M_{ia}M_{jb}\\
|A_{ijab}| & \approx M_{ij}M_{a}M_{b}\\
 & ...
\end{align*}
To simplify the notation, let any decomposed tensor be $M_{I}$, where $I$ is some subset of $\{i,j,a,b\}$. Then we empirically choose 
\begin{equation}
M_{I}=\left(\sum_{\{i,j,a,b\}/I}A_{ijab}^{2}\right)^{\frac{1}{2}\gamma}\label{eq:CCSD_loop_breaking}
\end{equation}
and use $\gamma=\frac{1}{2}$.

Finally we use the example of 
\[
S_{ijab}=\sum_{klcd}\langle kl||cd\rangle T_{ijcd}T_{klab},
\]
which is the last term in the CCD equations, to show how STC is exactly applied. We first write the next-iteration energy expression for this term: 
\begin{align*}
E & =\sum_{ijab}A_{ijab}S_{ijab}\\
 & =\sum_{ijklabcd}A_{ijab}\langle kl||cd\rangle T_{ijcd}T_{klab}
\end{align*}
Now the unnormalized optimal sampling probability is 
\[
\tilde{p}_{ijklabcd}^{\text{opt}}=|A_{ijab}\langle kl||cd\rangle T_{ijcd}T_{klab}|.
\]
We apply the loop-breaking strategy: 
\[
|A_{ijab}|\approx M_{ij}M_{ab}
\]
where $M_{ij}$ and $M_{ab}$ are constructed through Eq. \ref{eq:CCSD_loop_breaking}. Now the unnormalized approximate probability is 
\[
\tilde{p}_{ijklabcd}^{\prime}=|M_{ij}M_{ab}\langle kl||cd\rangle T_{ijcd}T_{klab}|
\]
which now becomes a tree structure. Thus we can rewrite 
\[
\tilde{p}_{ijklabcd}^{\prime}=\tilde{p}(ij)p(cd|ij)p(kl|cd)p(ab|kl).
\]
and compute the involved probability tables $\tilde{p}(ij),p(cd|ij),p(kl|cd),p(ab|kl)$ with $O(N^{4})$ cost using the tree tensor contraction algorithm described in Sec. \ref{subsec:tree}. Note that the decomposition is not unique, as one can write a tree structure starting from any tensor as the root. Now one can sample $i,j,k,l,a,b,c,d\sim\tilde{p}_{ijklabcd}^{\prime}$ with $O(1)$ cost. Each set of sampled indices generates an unbiased sample for both $\boldsymbol{S}$ (one can divide by $\boldsymbol{\Delta}$ to get the updated $\boldsymbol{T}^{(2)}$) and the next-iteration energy corresponding to this term: 
\begin{align*}
\boldsymbol{S} & =\left\langle \frac{\langle kl||cd\rangle T_{ijcd}T_{klab}}{\tilde{p}_{ijklabcd}^{\prime}}\boldsymbol{e}_{ijab}\right\rangle _{i,j,k,l,a,b,c,d\sim\tilde{p}^{\prime}}\\
 & =\left\langle \text{sgn}\left[\langle kl||cd\rangle T_{ijcd}T_{klab}\right]\frac{1}{M_{ij}M_{ab}}\boldsymbol{e}_{ijab}\right\rangle _{i,j,k,l,a,b,c,d\sim\tilde{p}^{\prime}}\\
E & =\left\langle \text{sgn}\left[\langle kl||cd\rangle T_{ijcd}T_{klab}\right]\frac{A_{ijab}}{M_{ij}M_{ab}}\right\rangle _{i,j,k,l,a,b,c,d\sim\tilde{p}^{\prime}}
\end{align*}
where $\text{sgn}(x)=\frac{x}{|x|}$.

\textbf{Optimized implementation}: We now describe the ``optimized'' implementations mentioned in the main text. There are two major changes. First, we use exact computations for part of the $O(N^{5})$ tensor contractions, specifically the contraction $\sum_{kc}V_{jkbc}T_{ikac}$ in the last term of the second line of Eq. \ref{eq:full_CCD}, which can be simplified by density fitting. Other $O(N^{5})$ terms are still evaluted by STC. The reason for the different treatments is that, since $\boldsymbol{T}$ tensors can be labelled as first order in the interaction perturbation, higher-order terms with multiple $\boldsymbol{T}$ tensors have much small values and thus much smaller variances, which favors the STC treatment; while the leading order term dominates the contribution, and in the system ranges we test, exact contraction is cheaper for our desired accuracy. Second, we evaluate $\mathcal{C}_{CC}^{(l=1,2)}(\boldsymbol{H},\boldsymbol{T})$ without the terms involving $\boldsymbol{F}$ tensors (i.e. the second and third terms of Eq. \ref{eq:full_CCD}), and exactly transform the resulting tensor to the canonical basis, apply the element-wise multiplication with $1/\boldsymbol{\Delta}$ in the canonical basis, and finally transform back to the local basis. The basis transformation results in a $O(N^{5})$ cost, but makes the iterative convergence faster and reduces the STC variance.

\textbf{Orbital localization}: Motivated by the theoretical variance analysis, in local STC-CCSD, we perform joint orbital localization of the occupied, virtual, and auxiliary orbitals by minimizing the tensor norms. Similar but slightly different strategies are used for the pure implementation and optimized implementation. In the pure implementation, we minimize the 1-norm of the $R_{ia\chi}$ block of the density fitting tensors, i.e., with the loss function
\[
L_{\text{optimized}}\propto\sum_{ia\chi}|R_{ia\chi}|,
\]
and the Rprop optimizer (a gradient-based optimizer). Each iteration has a $O(N^{4})$ deterministic computational cost, and we consistently find $\sim50$ iterations are enough to obtain stably localized orbitals in all the tested systems. Note that this localization scheme is specially designed for STC, so one should not expect it to generate physically meaningful local orbitals comparable to the well-known Foster-Boys localization or Pipek-Mezey localization methods.

In the pure implementation, since the off-diagonal part of the $\boldsymbol{F}$ is involved in the STC we add an extra term in the loss function to minimize the off-diagonal magnitudes of the $\boldsymbol{F}$ tensor (the occupied part and virtual part are independently added to the loss function). Let $\boldsymbol{F}_{\text{diag}}$ and $\boldsymbol{F}_{\text{offdiag}}$ be the diagonal and off-diagonal part of $\boldsymbol{F}$ (either the occupied or virtual block), and $\boldsymbol{F}_{\text{eff}}=|\boldsymbol{F}_{\text{diag}}^{-1}\boldsymbol{F}_{\text{offdiag}}|$, where $|\cdot|$ is the element-wise absolute value. The extra term is
\[
L_{F}\propto\left(\frac{1}{N}\text{Tr}[\boldsymbol{F}_{\text{eff}}^{k}]\right)^{1/k}.
\]
The complete loss function of the pure implementation is
\[
L_{\text{pure}}=\alpha L_{\text{optimized}}+\beta(L_{F,\text{occ}}+L_{F,\text{vir}}),
\]
with some empirical tunable coefficients $\alpha,\beta$.

\textbf{DIIS}: We employ a noise-resistant version of DIIS to speedup the iterative convergence. It differs from the normal DIIS by a diagonal regularization term in the overlap matrix between solutions (i.e. $\boldsymbol{T}$ tensors) at different iterations. 

\textbf{Average over different iterations}: After reaching convergence (which means that the energies should then fluctuate unbiasedly around the true CCSD energy), we run a few extra iterations, and average the energies in different iterations as the final STC-CCSD energy. We find that energies in different iterations have very weak statistical correlations. This allows us to increase the target error by $m$ times (which means less samples are needed), and average $m^{2}$ extra iterations, which effectively gives the same target error. We use $m=2$ in all calculations reported in the main text, including the benzene energy statistics (Fig. 4 in the main text), which verifies that such an ``effective'' target error is still in good agreement with the true error statistics. When combined with STC-(T), the averaged $\boldsymbol{T}$ ampitudes of the extra iterations are used for the following STC-(T) calculation.

\subsection{Variance of STC-CCSD}

Now we consider the variance of STC-CCSD. For simplicity we neglect all $O(\text{polylog}(N))$ factors. For canonical STC-CCSD, we have already mentioned that the next-iteration energy expression involves up to 4 tensors. Thus, with the diagonal contributions of $\boldsymbol{V}$ neglected, the variance is at most $O(N^{4})$. When the diagonal contribution is considered, each tensor with $V_{pqpq}$ elements contributes an extra $O(N^{1/3})$ factor (in addition to the $O(N)$ factor from the off-diagonal contributions) in the worst 3D case. However, $V_{pqpq}$ only appears in contractions with 3 tensors. An example is the term $S_{ijab}=\sum_{cd}T_{ijab}V_{abcd}$ shown above (when considering the energy expression $E=\sum_{ijab}S_{ijab}A_{ijab}$ there are 3 tensors $\boldsymbol{T},\boldsymbol{V},\boldsymbol{A}$). Thus we conclude that canonical STC-CCSD has $O(N^{4})$ absolute energy variance or $O(N^{2})$ relative energy variance. 

For local-CCSD, we first show that introducing the density fitting does not change the variance scaling of local STC, as it preserves a similar asymptotic structure to the original $\boldsymbol{V}$. Physically, the density fitting approximation comes from 
\[
\phi_{p}(\boldsymbol{x})\phi_{r}(\boldsymbol{x})\approx\sum_{\chi}P_{pr\chi}\phi_{\chi}(\boldsymbol{x}).
\]
We have mentioned earlier that the $p,r$ indices are short-ranged connected, thus density fitting effectively just ``groups'' the two short-range connected indices, and remains local, thus after density fitting, the resulting tensors retain all their long-range asymptotic algebraic behaviors. From the tensor perspective, this means with density fitting some index pairs $(\alpha_{i},\beta_{i})$ of a general tensor $T_{\alpha_{1}\beta_{1},...,\alpha_{n}\beta_{n}}$ are replaced by the auxiliary basis index $\chi_{i}$, with the same long-range behaviors. As all $\text{poly}(N)$ factors in local STC come from the long-range algebraic form, density fitting does not change the variance scaling in local STC.

Now consider the case without density fitting. For simplicity, we first consider the CCD expressions in Eq. \ref{eq:full_CCD}. As proved in Sec. \ref{subsec:denominator}, extra $\text{poly}(N)$ variance scalings only possibly appear when some tensors have $\gamma=1,2$ decay. In CCD, only the three terms in the second line of Eq. \ref{eq:full_CCD} have $\gamma=1$. However, all of them lead to a next-iteration energy diagram exactly the same as the left one shown in Fig. \ref{fig:feynman_diagram}, thus in fact they do not contribute extra $\text{poly}(N)$ factors. Thus, all contractions in CCD have $O(N^{2})$ absolute energy variance, or $O(1)$ relative energy variance.

For CCSD, additional $V_{iabc}$ or $V_{iajk}$ blocks (corresponding to $1/R^{2}$ asymptotic decay) appear. However, such blocks only appear in combination with the $\boldsymbol{T}^{(1)}$ tensors. As we mentioned earlier, the $\boldsymbol{T}^{(1)}$ tensors are absorbed into the density fitting tensors with an $O(N^{4})$ preprocessing cost. In terms of the asymptotic behavior, it is effectively equivalent to absorbing the $\boldsymbol{T}^{(1)}$ tensors into these $V_{iabc}$ or $V_{iajk}$ blocks. After all $\boldsymbol{T}^{(1)}$ tensors are absorbed, all tensor contractions now have the same structure as the CCD tensor contractions in Eq. \ref{eq:full_CCD}.

Combining the above analysis, we conclude that the diagonal elements of $\boldsymbol{V}$ do not contribute to extra polynomial factors in the variance, and local STC-CCSD gives $O(N^{2})$ absolute energy variance or $O(1)$ relative energy variance.

\subsection{Convergence of iterative STC-CCSD equations}

We consider a single-variable model to mimic the nonlinear CCSD equations: 
\[
x_{k+1}=x_{0}+ax_{k}+bx_{k}^{2}
\]
where $a$ and $b$ are i.i.d. random variables with mean and variance $\mu_{a,b}$ and $\sigma_{a,b}^{2}$ in all iterations to mimic the ``STC sampling''. If there is no quadratic term, i.e. the equation is 
\[
x_{k+1}=x_{0}+ax_{k}
\]
then we have an iterative equation for the mean $\mu_{k}=\langle x_{k}\rangle$ and variance $\sigma_{k}^{2}=\langle x_{k}^{2}\rangle-\langle x_{k}\rangle^{2}$: 
\begin{align*}
\mu_{k+1} & =x_{0}+\mu_{a}\mu_{k}\\
\sigma_{k+1}^{2} & =\sigma_{a}^{2}\mu_{k}^{2}+(\mu_{a}^{2}+\sigma_{a}^{2})\sigma_{k}^{2}
\end{align*}
Clearly, the equation for the mean is unbiased, and the variance converges if 
\begin{equation}
\mu_{a}^{2}+\sigma_{a}^{2}=\langle a^{2}\rangle<1.\label{eq:cond1_convergence}
\end{equation}
Since $\sigma_{a}^{2}$ mimics the stochastic sampling error, we have 
\begin{equation}
\sigma_{a}^{2}\propto\frac{1}{N_{\text{samples}}}\label{eq:cond2_convergence}
\end{equation}
Combining Eq. \ref{eq:cond1_convergence} and Eq. \ref{eq:cond2_convergence}, we can see there exists some critical point $N_{\text{critical}}$, such that the variance of stochastic iteration converges if $N_{\text{samples}}>N_{\text{critical}}$, and does not converge otherwise.

With the quadratic term, the iterative mean equation becomes: 
\begin{align*}
\mu_{k+1} & =x_{0}+\mu_{a}\mu_{k}+\mu_{b}(\mu_{k}^{2}+\sigma_{k}^{2})
\end{align*}
which is not unbiased anymore, and the deviation is proportional to $\sigma_{k}^{2}$. As long as $\sigma_{k}^{2}$ converges, we have 
\[
\text{bias}\sim(\text{statistical error})^{2}.
\]

\begin{figure}[tbh]
\centering
\includegraphics[width=0.6\columnwidth]{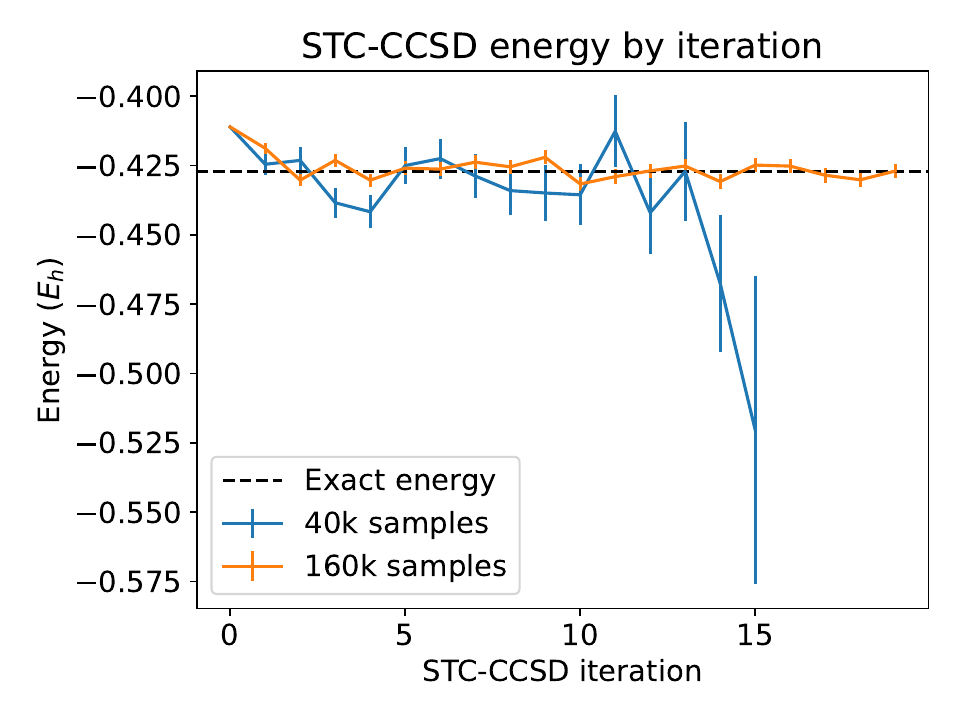}

\caption{Convergence behavior of STC-CCSD, benchmarked on water dimer/cc-pVDZ.}

\label{fig:convergence}
\end{figure}

Finally, we give a numerical demonstration of this critical behavior, shown in Fig. \ref{fig:convergence}.

\subsection{Variance of the representative tensor contraction in STC-(T)}

We now derive the variance scaling of the representative tensor contraction in STC-(T) shown in the main text. The representative tensor contraction is
\[
\mathcal{C}_{\text{(T)}}(H,T)=-\frac{1}{4}\sum_{ijkabc}\frac{1}{\Delta_{ijkabc}^{(3)}}\left(\sum_{f}T_{ijaf}V_{fkbc}\right)^{2}.
\]
Following Sec. \ref{subsec:denominator}, we can always start from the ideal sampling probability by removing the $1/\Delta_{ijkabc}^{(3)}$ part, with only an $O(1)$ extra factor of $\exp(\Delta F)$, assuming the system is gapped. The rest part is simply a normal tensor contraction of 4 tensors, following the Einstein summation rule, and all 2-norms are $O(\sqrt{N})$ following the asymtotic forms in Eq. \ref{eq:asymptotic_V}. (the involved $\boldsymbol{T}$ and $\boldsymbol{V}$ blocks have $1/R^{3}$ and $1/R^{2}$ decay, respectively). Thus, using the canonical STC algorithm with the loop-breaking strategy, we get a conjectured $O(N^{4})$ or $O(N^{2})$ cost  for fixed absolute or relative energy error, respectively (see Sec. \ref{subsec:canonical}).

Now we show the approach introduced in the main text leads to a provable cost upper bound of $O(N^{4})$ or $O(N^{2})$ for fixed absolute or relative energy error, respectively. Following the main text, we define the tensor $\boldsymbol{T}\cdot\boldsymbol{V}$ with elements  $(\boldsymbol{T}\cdot\boldsymbol{V})_{ijkabc}=\sum_{f}T_{ijaf}V_{fkbc}$. Following Sec. \ref{subsec:denominator}, the ideal probability is then
\[
\tilde{p}_{ijkabc}^{\text{ideal}}=\left|(\boldsymbol{T}\cdot\boldsymbol{V})_{ijkabc}\right|^{2}=\left(\sum_{f}T_{ijaf}V_{fkbc}\right)^{2}.
\]
The approximate probability introduced in the main text is
\[
\tilde{p}_{ijkabc}^{\prime}=N(\boldsymbol{T}^{2}\cdot\boldsymbol{V}^{2})_{ijkabc}=N\sum_{f}T_{ijaf}^{2}V_{fkbc}^{2}
\]
We now prove that: (1) $\tilde{p}_{ijkabc}^{\prime}\geq\tilde{p}_{ijkabc}^{\text{ideal}}$ for any $i,j,k,a,b,c$, and (2) $\frac{Z^{\prime}}{Z^{\text{ideal}}}\sim O(N)$ assuming a gapped system. The former one can be simply derived from the Cauchy-Schwarz inequality:
\[
\tilde{p}_{ijkabc}^{\prime}=\left(\sum_{f}1^{2}\right)\left(\sum_{f}T_{ijaf}^{2}V_{fkbc}^{2}\right)\geq\left(\sum_{f}T_{ijaf}V_{fkbc}\right)^{2}=\tilde{p}_{ijkabc}^{\text{ideal}}
\]
To show the latter one, we recall that the Einstein summation is invariant to unitary transformation applied to any indices. Here $Z^{\text{ideal}}$ is a Einstein summation, while $Z^{\prime}$ is partially Einstein --- index $i,j,k,a,b,c$ appears twice, while $f$ appears four times, thus $Z^{\prime}$ is invariant to unitary transformations applied on $i,j,k,a,b,c$. Now define two new tensors $\boldsymbol{T}\cdot\boldsymbol{T}$ and $\boldsymbol{V}\cdot\boldsymbol{V}$:
\begin{align*}
(\boldsymbol{T}\cdot\boldsymbol{T})_{fg} & =\sum_{ija}T_{ijaf}T_{ijag},\\
(\boldsymbol{V}\cdot\boldsymbol{V})_{fg} & =\sum_{kbc}V_{fkbc}V_{gkbc,}
\end{align*}
where $f,g$ are virtual indices. Since it follows the Einstein summation rule, they are valid tensors, one can compute its value in the local basis, and then transform the result to the canonical basis later. Now let all indices to be local basis indices, following the results for the asymptotic forms, $\boldsymbol{T}$ has leading order algebraic decay like $R^{-3}$, and $\boldsymbol{V}$ (note that $k\neq c$) has leading order algebraic decay like $R^{-2}$ 
\begin{align*}
(\boldsymbol{T}\cdot\boldsymbol{T})_{fg} & \sim\exp(-R_{fg})\int\left(R^{-3}\right)^{2}dV\sim\exp(-R_{fg})\\
(\boldsymbol{V}\cdot\boldsymbol{V})_{fg} & \sim\exp(-R_{fg})\int\left(R^{-2}\right)^{2}dV\sim\exp(-R_{fg})
\end{align*}
Note that these scalings do not depend on the system dimension (up to 3D) due to the fast decay of the integrand. It is already enough for us to estimate $Z^{\text{ideal}}$ in the local basis as 
\[
Z^{\text{ideal}}=\sum_{ijkabc}\tilde{p}_{ijkabc}^{\text{ideal}}=\sum_{fg}(\boldsymbol{T}\cdot\boldsymbol{T})_{fg}(\boldsymbol{V}\cdot\boldsymbol{V})_{fg}\sim O(N)
\]
For $Z^{\prime}$, the current information is not enough yet, because $Z^{\prime}$ must be computed in the canonical basis for index $f$. Now we estimate the eigenvalues of the two matrices from the well-known inequality, that the spectral radius (i.e., largest eigenvalue $\lambda_{\text{max}}$ by magnitude) of a matrix is bounded by any induced operator $p$-norm defined in Eq. \ref{eq:op_norm}. Specifically consider the operator $\infty$-norm, which is given by 
\[
\Vert\boldsymbol{M}\Vert_{\text{op},\infty}=\max_{i}\sum_{j}|M_{ij}|
\]
For $M_{fg}\sim\exp(-R_{fg})$, it is clearly $O(1)$. Thus 
\[
\lambda_{\text{max}}\leq\Vert\boldsymbol{M}\Vert_{\text{op},\infty}\sim O(1)
\]
Now we transform $\boldsymbol{T}\cdot\boldsymbol{T}$ and $\boldsymbol{V}\cdot\boldsymbol{V}$ to the canonical basis, which does not change their eigenvalues. Since the diagonal matrix element is upper bounded by the largest eigenvalue, we now have: $(\boldsymbol{T}\cdot\boldsymbol{T})_{ff}\sim(\boldsymbol{V}\cdot\boldsymbol{V})_{ff}\sim O(1)$. Note that such conclusion holds in any basis, including the canonical basis. We immediately have
\[
Z^{\prime}=\sum_{ijkabc}\tilde{p}_{ijkabc}^{\prime}=N\sum_{f}(\boldsymbol{T}\cdot\boldsymbol{T})_{ff}(\boldsymbol{V}\cdot\boldsymbol{V})_{ff}\sim O(N^{2})
\]
Thus we conclude: 
\[
\exp(\Delta F)\sim\frac{Z^{\prime}}{Z^{\text{ideal}}}\sim O(N).
\]
According to Eq. 6 in the main text, we can now conclude that the variance of such a representative tensor contraction in STC-(T) is upper-bounded by 
\[
\text{Var}\leq O(N^{3})
\]

\subsection{The complete STC-(T) algorithm and variance}

The full closed-shell (T) expression is
\[
E_{(T)}=-\frac{1}{3}\sum_{ijkabc}\left[\frac{W_{ijkabc}\mathcal{R}(W+\frac{1}{2}Q)_{ijkabc}}{\Delta_{ijkabc}^{(3)}}\right],
\]
where 
\begin{align*}
W_{ijkabc} & =\mathcal{P}(\tilde{W})_{ijkabc},\\
Q_{ijkabc} & =\mathcal{P}(\tilde{Q})_{ijkabc},\\
\tilde{W}_{ijkabc} & =\sum_{f}T_{ijaf}V_{fkbc}-\sum_{m}T_{imab}V_{jkmc}\\
\tilde{Q}_{ijkabc} & =V_{ijab}T_{kc}\\
\mathcal{P}(X)_{ijkabc} & =X_{ijkabc}+X_{jkibca}+X_{kijcab}+X_{kjicba}+X_{jikbac}+X_{ikjacb}\\
\mathcal{R}(X)_{ijkabc} & =4X_{ijkabc}+X_{ijkbca}+X_{ijkcab}-2X_{ijkcba}-2X_{ijkbac}-2X_{ijkabc},\\
\Delta_{ijkabc}^{(3)} & =e_{a}+e_{b}+e_{c}-e_{i}-e_{j}-e_{k}
\end{align*}
Here we give a full derivation that extends the algorithm shown in the main text for a representative term, to the full expression. Readers only interested in the final algorithm but not derivations can directly go to Eq. \ref{eq:(T)_full_prob} and Eq. \ref{eq:(T)_full_expr}.

Note that the $W+\frac{1}{2}Q$ term essentially divides the (T) energy into two parts. Since $\boldsymbol{T}^{(1)}$ comes from higher order perturbation contributions than $\boldsymbol{T}^{(2)}$, in practice $W$ is much more important than $Q$. Thus to estimate the variance, we focus on the first term
\[
E_{1}=-\frac{1}{3}\sum_{ijkabc}\left[\frac{W_{ijkabc}\mathcal{R}(W)_{ijkabc}}{\Delta_{ijkabc}^{(3)}}\right].
\]
One can expand the permutations, and rewrite: 
\[
E_{1}=-\sum_{\sigma,\tau=1}^{6}M_{\sigma\tau}\sum_{ijkabc}\frac{W_{ijk\sigma(abc)}W_{ijk\tau(abc)}}{e_{a}+e_{b}+e_{c}-e_{i}-e_{j}-e_{k}}
\]
where $\sigma,\tau$ are 6 rank-3 permutations, i.e., $\sigma(abc)=\{abc,bca,cab,cba,bac,acb\}$. $M_{\sigma\tau}$ is a fixed $6\times6$ symmetric matrix encoding the coefficients. Note that if one (1) only keeps $\sigma=\tau$, (2) keep only the non-crossing term in $\mathcal{P}$, and (3) keep only the first part of $\tilde{W}_{ijkabc}$, it now reduces to the representative contraction we show in the main text. Now we extend of the method shown in the main text to compute $E_{1}$, and show that it has the same cost and variance scaling. A conceptually simple way is to diagonalize the $M$ matrix as 
\[
M_{\sigma\tau}=\sum_{\mu=1}^{6}\lambda_{\mu}C_{\sigma}^{\mu}C_{\tau}^{\mu}.
\]
Then we can rewrite: 
\[
E_{1}=-\sum_{\mu=1}^{6}\lambda_{\mu}\left[\sum_{ijkabc}\frac{\left(\sum_{\sigma}C_{\sigma}^{\mu}W_{ijk\sigma(abc)}\right)^{2}}{\Delta_{ijkabc}^{(3)}}\right].
\]
and compute the contributions corresponding to $\mu=1,...,6$ separately, then add them together. (In fact, $M$ is positive-semidefinite, thus one does not need to worry about sign cancellations). Now for any $\mu$, following the same idea shown in the main text, treating $\frac{1}{\Delta_{ijkabc}^{(3)}}$ as extra weights on the indices $ijkabc$, the ideal sampling probability is 
\[
\tilde{p}^{\text{ideal}}=\left(\sum_{\sigma}C_{\sigma}^{\mu}W_{ijk\sigma(abc)}\right)^{2}
\]
Now we sequentially apply the Cauchy-Schwarz inequality:
\begin{align*}
\tilde{p}_{ijkabc}^{\text{ideal}} & =\left(\sum_{\sigma=1}^{6}C_{\sigma}^{\mu}W_{ijk\sigma(abc)}\right)^{2}\\
 & \leq6\left[\sum_{\sigma=1}^{6}\left(C_{\sigma}^{\mu}\right)^{2}\right]\mathcal{P}_{abc}\left[W_{ijkabc}^{2}\right]\\
 & =6\mathcal{P}_{abc}\left[W_{ijkabc}^{2}\right]\\
 & \leq36\mathcal{P}\mathcal{P}_{abc}\left[\tilde{W}_{ijkabc}^{2}\right]\\
 & \leq72\mathcal{P}\mathcal{P}_{abc}\left[\left(\sum_{f}T_{ijaf}V_{fkbc}\right)^{2}+\left(\sum_{m}T_{imab}V_{jkmc}\right)^{2}\right]\\
 & \leq72\mathcal{P}\mathcal{P}_{abc}\left[N_{v}\sum_{f}T_{ijaf}^{2}V_{fkbc}^{2}+N_{o}\sum_{m}T_{imab}^{2}V_{jkmc}^{2}\right]
\end{align*}
where $\mathcal{P}$ has already been defined above, and
\[
\mathcal{P}_{abc}(X)_{ijkabc}=\sum_{\sigma=1}^{6}X_{ijk\sigma(abc)}.
\]
In the derivation of the above inequality, we used the Cauchy-Schwarz inequality four times, one for $\sum_{\sigma=1}^{6}$, one for 6 permutations in $W$, one for the two parts in $\tilde{W}$, and one for $\sum_{f}$ and $\sum_{m}$. We also used $\sum_{\sigma}C_{\sigma}^{2}=1$, which comes from the fact that the eigenvectors are normalized. One can now see that the upper bound is already independent of $\mu$. Now we construct 
\begin{equation}
\begin{aligned}\tilde{p}_{ijkabc}^{\prime} & =\mathcal{P}\mathcal{P}_{abc}\left[\tilde{p}_{1}^{\prime}+\tilde{p}_{2}^{\prime}\right]\\
 & \equiv\mathcal{P}\mathcal{P}_{abc}\left[N_{v}\sum_{f}T_{ijaf}^{2}V_{fkbc}^{2}+N_{o}\sum_{m}T_{imab}^{2}V_{jkmc}^{2}\right],
\end{aligned}
\label{eq:(T)_full_prob}
\end{equation}
where $\tilde{p}_{1}^{\prime}$ and $\tilde{p}_{2}^{\prime}$ represent the first and second terms, respectively. Note that $\tilde{p}_{1}^{\prime}$ is exactly the same as the $\tilde{p}^{\prime}$ shown in the main text, which supports efficient sampling and gives $\exp(\Delta F)\sim O(N)$. Since $\tilde{p}_{2}^{\prime}$ has the same structure, the same conclusion also holds for $\tilde{p}_{2}^{\prime}$. $\mathcal{P}\mathcal{P}_{abc}$ simply contributes a multiplication factor of $36=6\times6$. Thus for $\tilde{p}^{\prime}$ we also have $\exp(\Delta F)\sim O(N)$. Thus the final deterministic cost, and sampling variance have the same scaling with the shown representative tensor contraction term in the main text.

We have already shown in the main text that $\tilde{p}_{1}^{\prime}$ (and $\tilde{p}_{2}^{\prime}$ similarly) supports efficient sampling (i.e. $O(1)$ per-sample cost using the alias method after constructing the probability tables) since it has a tree structure. We now show that $\tilde{p}^{\prime}$ also supports efficient sampling in two steps. To sample from $\tilde{p}_{1}^{\prime}+\tilde{p}_{2}^{\prime}$, one just needs to draw samples from $\tilde{p}_{1}^{\prime}$ with probability $\frac{Z_{1}^{\prime}}{Z_{1}^{\prime}+Z_{2}^{\prime}}$, and draw samples from $\tilde{p}_{2}^{\prime}$ with probability $\frac{Z_{2}^{\prime}}{Z_{1}^{\prime}+Z_{2}^{\prime}}$, thus sampling from $\tilde{p}_{1}^{\prime}+\tilde{p}_{2}^{\prime}$ is also efficient. To sample from $\tilde{p}^{\prime}=\mathcal{P}\mathcal{P}_{abc}\left[\tilde{p}_{1}^{\prime}+\tilde{p}_{2}^{\prime}\right]$, we just need to uniformly shuffle the indices from the 36 combinations of $\mathcal{P}\mathcal{P}_{abc}$ after sampling from $\tilde{p}_{1}^{\prime}+\tilde{p}_{2}^{\prime}$, therefore this is again efficient.

To extend the above algorithm to the complete (T) expression, we just add the $Q$ part to the sampling weights as 
\begin{equation}
E_{(T)}=-\frac{1}{3}Z^{\prime}\left\langle \frac{W_{ijkabc}\mathcal{R}[W+\frac{1}{2}Q]_{ijkabc}}{\Delta_{ijkabc}^{(3)}\tilde{p}_{ijkabc}^{\prime}}\right\rangle _{ijkabc\sim\tilde{p}^{\prime}}.\label{eq:(T)_full_expr}
\end{equation}
The combination of Eq. \ref{eq:(T)_full_prob} and Eq. \ref{eq:(T)_full_expr} now gives the complete STC-(T) algorithm.

Finally, we introduce a useful trick to improve the STC-(T) with negligible extra cost. When combining STC-CCSD and STC-(T), the $\boldsymbol{T}$ tensors obtained from STC-CCSD are stochastic. Note that STC-(T) is essentially a quadratic form of $\boldsymbol{T}$, i.e. 
\[
E_{(T)}=\mathcal{C}_{(T)}(\boldsymbol{T},\boldsymbol{T}),
\]
where $\mathcal{C}_{(T)}$ has a bi-linear form. Denote the above STC-(T) algorithm as $\mathcal{C}_{\text{STC}}(\boldsymbol{T})=\mathcal{C}_{\text{STC}}(\boldsymbol{T},\boldsymbol{T})$, which has a stochastic output. Even if $\boldsymbol{T}$ is unbiased, the statistical error of $\boldsymbol{T}$ leads to a bias 
\[
\left\langle \mathcal{C}_{\text{STC}}(\boldsymbol{T},\boldsymbol{T})\right\rangle \neq\mathcal{C}_{(T)}(\langle\boldsymbol{T}\rangle,\langle\boldsymbol{T}\rangle)
\]
Although this bias is again 
\[
\text{bias}\sim(\text{stastical error})^{2},
\]
the prefactor can sometimes lead to non-negligible errors. We employ a strategy to get two $\boldsymbol{T}$ tensors, $\boldsymbol{T}_{1}$, $\boldsymbol{T}_{2}$, from different STC-CCSD iterations. If they are uncorrelated, then $\mathcal{C}_{\text{STC}}(\boldsymbol{T}_{1},\boldsymbol{T}_{2})$ is now an unbiased estimator: 
\[
\langle\mathcal{C}_{\text{STC}}(\boldsymbol{T}_{1},\boldsymbol{T}_{2})\rangle=\mathcal{C}_{(T)}(\langle\boldsymbol{T}\rangle,\langle\boldsymbol{T}\rangle).
\]
To implement $\mathcal{C}_{\text{STC}}(\boldsymbol{T}_{1},\boldsymbol{T}_{2})$ without modifying the original algorithm, we use 
\begin{align*}
\mathcal{C}_{\text{STC}}(\boldsymbol{T}_{1},\boldsymbol{T}_{2}) & =\mathcal{C}_{\text{STC}}(\boldsymbol{T}_{+},\boldsymbol{T}_{+})-\mathcal{C}_{\text{STC}}(\boldsymbol{T}_{-},\boldsymbol{T}_{-})\\
 & =\mathcal{C}_{\text{STC}}(\boldsymbol{T}_{+})-\mathcal{C}_{\text{STC}}(\boldsymbol{T}_{-})
\end{align*}
where 
\begin{align*}
\boldsymbol{T}_{+} & =\frac{1}{2}\left(\boldsymbol{T}_{1}+\boldsymbol{T}_{2}\right),\\
\boldsymbol{T}_{-} & =\frac{1}{2}\left(\boldsymbol{T}_{1}-\boldsymbol{T}_{2}\right).
\end{align*}